\documentclass[aps,nofootinbib,longbibliography,superscriptaddress,
tightenlines,notitlepage,prd]{revtex4-1}

\usepackage{amsmath,amsthm,latexsym,amssymb,amsfonts}
\usepackage{graphicx,color,natbib}
\usepackage[export]{adjustbox}
\usepackage[colorlinks=true]{hyperref}

\begin{document}

\title{Coarse graining spin foam quantum gravity - a review}

\author{Sebastian Steinhaus}
\email{sebastian.steinhaus@uni-jena.de}
\affiliation{Theoretisch-Physikalisches Institut, Friedrich-Schiller-Universit\"at Jena,\\ Max-Wien-Platz 1, 07743 Jena, Germany}

%\author[\firstAuthorLast ]{\Authors} %This field will be automatically populated
%\address{} %This field will be automatically populated
%\correspondance{} %This field will be automatically populated

%\extraAuth{}% If there are more than 1 corresponding author, comment this line and uncomment the next one.
%\extraAuth{corresponding Author2 \\ Laboratory X2, Institute X2, Department X2, Organization X2, Street X2, City X2 , State XX2 (only USA, Canada and Australia), Zip Code2, X2 Country X2, email2@uni2.edu}

\begin{abstract}

%%% Leave the Abstract empty if your article does not require one, please see the Summary Table for full details.
In quantum gravity, we envision renormalization as the key tool for bridging the gap between microscopic models and observable scales. For spin foam quantum gravity, which is defined on a discretisation akin to lattice gauge theories, the goal is to derive an effective theory on a coarser discretisation from the dynamics on the finer one, coarse graining the system in the process and thus relating physics at different scales.

In this review I will discuss the motivation for studying renormalization in spin foam quantum gravity, e.g. to restore diffeomorphism symmetry, and explain how to define renormalization in a background independent setting by formulating it in terms of boundary data. I will motivate the importance of the boundary data by studying coarse graining of a concrete example and extending this to the spin foam setting. This will naturally lead me to the methods currently used for renormalizing spin foam quantum gravity, such as tensor network renormalization, and a discussion of recent results. I will conclude with an overview of future prospects and research directions.

\end{abstract}

\maketitle

\section{A brief introduction to spin foam quantum gravity} \label{sec:Intro}

Spin foam quantum gravity \cite{Perez:2012wv,carlobook} is a promising approach to quantum gravity closely related to loop quantum gravity \cite{thomasbook}. The aim is to define the path integral for gravity in a non-perturbative and background independent fashion, that is without any reference to a fixed background space-time or structure.

The starting point of spin foam models is the Plebanski-Holst formulation of general relativity \cite{Plebanski:1977zz}, in which gravity is formulated as constrained topological BF theory \cite{Baez:1999sr}. To formulate this theory as a path integral, one introduces a lattice as a regulator, more precisely a 2-complex, in order to truncate the number of degrees of freedom. On this 2-complex, which is a collection of vertices, edges and faces, the topological theory is first discretized and quantized. This is in close analogy to 3D (topological) gravity, where this formulation gives rise to the Ponzano-Regge model \cite{PR,Freidel:2004vi,Barrett:2008wh}, a well-defined model of 3D quantum gravity defined on a triangulation.

However, gravity in 4D is not topological, which requires the implementation of so-called simplicity constraints. In the continuum they serve the role to break the too many symmetries of the theory and reduce the $B$-field in BF theory to a simple 2-form, reducing the action to the familiar Holst action \cite{Holst:1995pc}. In spin foam quantum gravity, one derives such constraints for the discretisation of the classical $B$-field, so-called bivectors. In 4D, bivectors are assigned to 2D faces, e.g. triangles, and encode their geometry. The constraints ensure that these bivectors are simple, i.e. they can always be written as a wedge product of two vectors. Geometrically these vectors span two edges of a triangle. Different versions of these discrete constraints agree for single, classical building blocks, e.g. a 4-simplex, such that they correspond to different discretisations. However, their implementation in the quantum theory, which leads to restrictions on the variables of the theory, generically results in different models with starkly different dynamics. Two examples are the Barrett-Crane (BC) model \cite{Barrett:1997gw,Barrett:1999qw} and the Engle-Pereira-Rovelli-Livine / Freidel-Krasnov (EPRL / FK) model \cite{Engle:2007qf,Engle:2007wy,Freidel:2007py}. The former strongly implements a condition on bivectors, which significantly reduces the degrees of freedom of the model. This was criticized \cite{Alesci:2007tx,Alesci:2007tg} and motivated the development of the EPRL / FK model, in which constraints are implemented weakly, i.e. at the level of expectation values. Despite these insights, in remains an open question whether these constraints are sufficient to recover general relativity in a continuum limit. The hope is that coarse graining / renormalization can shed a light on this intriguing question.

\begin{figure}[h!]
\begin{center}
  \includegraphics[width=0.6\textwidth]{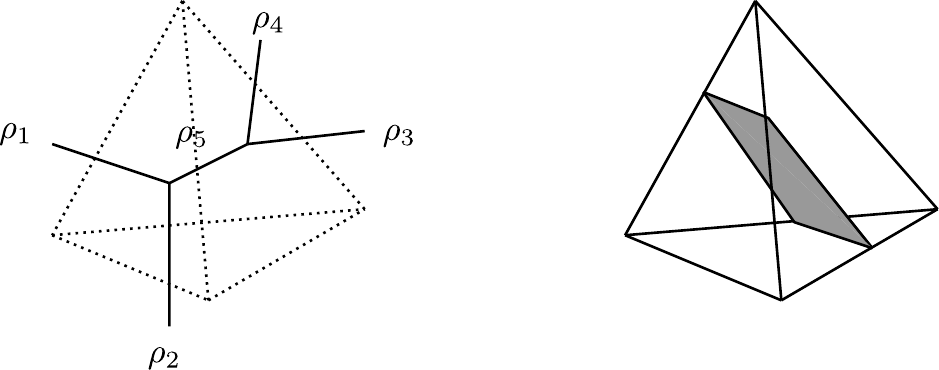}
  \caption{ \label{fig:intertwiner}
  {Left: A 4-valent intertwiner, dual to a tetrahedron, expanded in an orthonormal basis. The shape of a tetrahedron is not determined by the areas of its four triangles. Right: The intermediate representation gives the area of a parallelogram in the center of the tetrahedron. Its corners are located on the center points of the edges of the tetrahedron according to the split of the intertwiner.}
  }
\end{center}
\end{figure}

Despite these differences, all spin foam models are written in a similar form. The 2-complex, which is frequently dual to a triangulation, is coloured with group theoretic data: each face $f$ carries an irreducible representation $\rho_f$ of the underlying symmetry group ($\text{Spin}(4)$ for Riemannian, $\text{SL}(2,\mathbb{C})$ for Lorentzian signature), while each edge $e$ carries an intertwiner $\iota_e$, an invariant tensor in the tensor product of representation spaces associated to faces meeting at an edge. These data encode the geometry of the spin foam: in 4D, each edge is dual to a 3D polyhedron, which has as many faces as (dual) faces that share this edge. Then, the areas of these faces are given by the associated representations $\rho_f$. However, this does not determine the shape of the polyhedron uniquely, see e.g. a tetrahedron. A flat tetrahedron is uniquely determined by specifying its six edge lengths, whereas it only has four faces. Thus, the areas of these faces alone do not fix the shape of the tetrahedron. Part of the information on the shape is stored in the intertwiner, which can be expanded into an orthonormal basis using group representation theory. For a tetrahedron, its dual 4-valent intertwiner can be split into two 3-valent ones, where the new link carries again a group representation labeling the basis element. Geometrically this representation gives the area of a parallelogram spanned by the midpoints of the edges of the tetrahedron \cite{Baez:1999tk,Baez:1999sr}, see also fig. \ref{fig:intertwiner}. However, due to the uncertainty principle the shape cannot be fully specified since area operators associated with intersecting faces do not commute and thus cannot be diagonalized simultaneously. Note that coherent intertwiners can be defined that are sharply peaked on the geometry of a classical polyhedron\footnote{These coherent intertwiners, called Livine-Speziale intertwiners \cite{Livine:2007vk} (see also sec. \ref{sec:restricted}), are given by a tensor product of coherent $\text{SU}(2)$ states, which are sharply peaked on the outward pointing normals of the faces of the polyhedron. If these normals times their respective areas sum up to zero, these data uniquely define a convex polyhedron by Minkowski's theorem \cite{Minkowski,Bianchi:2010gc}.}.

\begin{figure}[h!]
\begin{center}
  \includegraphics[width=0.2\textwidth]{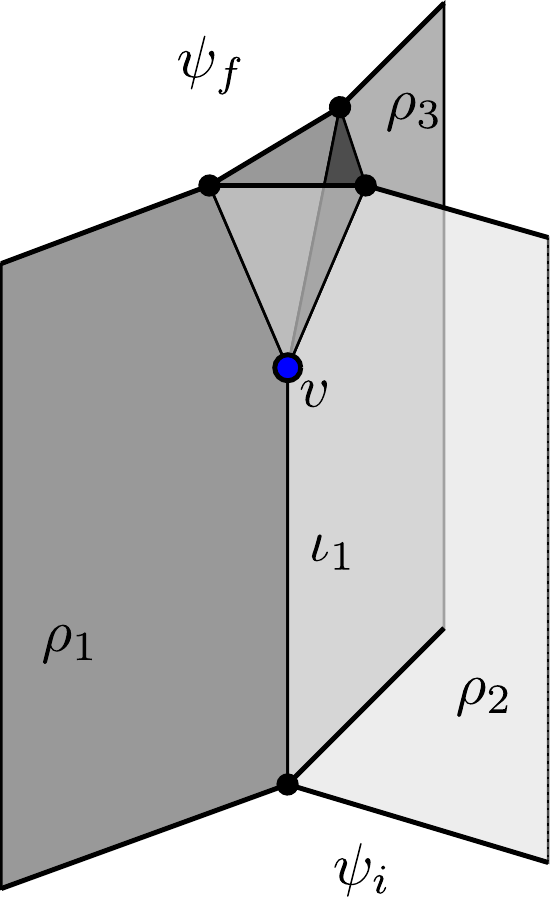}
  \caption{ \label{fig:quantum-spacetime}
  A spin foam in 3D with boundary. {The boundary of this spin foam is made up of an initial and a final graph, which carry states $\psi_i$ and $\psi_f$ respectively. The links of a graph carry representations $\rho_f$, whereas the nodes carry intertwiners $\iota_e$. For $\psi_i$, these are $\rho_i$, $i=1,2,3$, and $\iota_1$. The data for $\psi_f$ is omitted for clarity of the figure. The 2-complex interpolates between these two graphs and evolves the states. Thus, in the 2-complex, representations are associated with faces and intertwiners with edges. Several edges meet at a vertex $v$, here shown in blue. The vertex shown here is dual to a 3D tetrahedron: four edges, each dual to a triangle, are glued together to form a tetrahedron.}
  }
\end{center}
\end{figure}

From these 3D polyhedra a 4D geometry is built at the vertices of a 2-complex. At such a vertex, several edges and faces meet, indicating how 3D polyhedra are glued together to form a 4D geometry. If two edges meet at the same vertex and are part of the same face, their dual 3D polyhedra are glued along the shared face. Crucially, since the representation associated to the face determines its area, it is ensured that the face has the same area in both polyhedra. From the group theoretic data, this `gluing' is performed by contracting the intertwiners according to the combinatorics of the 2-complex, which essentially amounts to a spin network evaluation\footnote{This spin network can be obtained by drawing a 3-sphere around a vertex in the 2-complex. Edges, corresponding to intertwiners, intersect the sphere at nodes. Faces, intersect the sphere at links, connecting the nodes, determining how to contract the intertwiners. }. The resulting number is known as the vertex amplitude $\mathcal{A}_v$, i.e. the amplitude of the spin foam model assigned to the discrete 4D geometry dual to the vertex $v$ with configuration $\{\rho_f,\iota_e\}$. Similarly, we assign local amplitudes to the $\mathcal{A}_e$ and $\mathcal{A}_f$ to the edges $e$ and faces $f$ respectively. The former ensures that intertwiners are normalized, while the latter corresponds to the dimension of the representation $\rho_f$. See \cite{Perez:2012wv} for more details of the derivation.
%In 4D, $\rho_f$ gives the area of a 2D face, whereas $\iota_e$ encodes the shape of the dual 3D polyhedron. These building blocks are then glued together according to the combinatorics of the 2-complex to form a discrete 4D geometry. The theory then associates an amplitude to each configuration $\{\rho_f,\iota_e\}$ by locally assigning amplitudes $\mathcal{A}_v$, $\mathcal{A}_e$ and $\mathcal{A}_f$ to the vertices $v$, edges $e$ and faces $f$ of the 2-complex.
Eventually, the path integral is defined as a sum over all these configurations:
\begin{equation} \label{eq:sf_partition}
Z = \sum_{\{\rho_f,\iota_e\}} \prod_f \mathcal{A}_f \prod_e \mathcal{A}_e \prod_v \mathcal{A}_v \; .
\end{equation}
Crucially, these geometric building blocks and amplitudes are derived from general relativity formulated as a constrained topological field theory. In case the spin foam has a boundary (see fig. \ref{fig:quantum-spacetime}), it serves as an amplitude functional mapping states from its boundary Hilbert space into the complex numbers. This concept will be crucial in this review.

At the level of a a few simplices, these models are well explored. A well understood result across models, which furthermore underlines the relation to general relativity, is the asymptotic expansion of the vertex amplitude dual to a 4-simplex \cite{Barrett:1998gs,Conrady:2008mk,Barrett:2009gg,Kaminski:2017eew,Liu:2018gfc}. In these works the vertex amplitude is investigated for coherent intertwiners, which are sharply peaked on the geometry of classical polyhedra. Then the vertex amplitude can be written as an integral over several copies of the symmetry group. This integral is then evaluated via a stationary phase approximation by uniformly scaling up all representations. Hence it is commonly referred to as large-$j$ limit, in which the amplitude is generically proportional to the cosine of the Regge action \cite{Regge:1961px,Barrett:2018ybl}, a discretisation of classical gravity. Physically, this amplitude should be valid for 4-simplices of mesoscopic and even macroscopic size. In recent years, numerical calculations of the vertex amplitude beyond the asymptotic expansion have seen promising progress \cite{Dona:2017dvf,Dona:2019dkf}.

Beyond a single building block, the dynamics is less explored, in particular how the choice of the 2-complex impacts the results of the theory. Indeed, a priori the theory itself does not specify how the 2-complex should be chosen. In this review we take a viewpoint that is akin to lattice gauge theory, and regard the 2-complex as a regulator, a particular choice to truncate the number of degrees of freedom of the theory. As such, physics must not depend on this choice and it is must be removed eventually, e.g. in a continuum limit, in order to derive consistent results. One route towards such a limit lies in coarse graining: By coarse graining, i.e. defining an effective coarse amplitude from a collection of fine ones, we readily relate two theories defined on two different regulators. Moreover, by coarse graining we gain insight into the dynamics of a collection of building blocks and learn which configurations are more relevant on a coarser scale. In short, the aim is to derive a family of amplitudes to assign to different regulators, which reproduce the same physics (at least approximately). This defines a renormalization group flow of amplitudes \cite{Dittrich:2013xwa,Dittrich:2014ala}. It is the purpose of this article to review the progress of this approach and outline how it helps turning spin foam quantum gravity into a computational formalism.

This review is structured as follows: in section \ref{sec:challenges} we start by outlining the most pressing challenges faced by spin foam quantum gravity and how these are addressed by coarse graining. Section \ref{sec:disc_diffeo} discusses the issue of restoring diffeomorphism invariance in the discrete as well as the typical appearance of non-local interactions under coarse graining, which is one motivation for the consistent boundary formulation outlined in section \ref{sec:embedding}. Section \ref{sec:coarse_graining} reviews two numerical methods to perform such coarse graining algorithms, namely tensor network renormalization and restricted path integral models. In section \ref{sec:directions} we conclude with several interesting future research directions.

\section{Key challenges in spin foam quantum gravity} \label{sec:challenges}

Before explaining renormalization in spin foam models and its progress over the last decade, it is crucial to first discuss the key challenges spin foam quantum gravity is facing and how renormalization plays a vital role in overcoming them.

\subsection{Fate of diffeomorphism symmetry}

Diffeomorphism symmetry, as the fundamental symmetry of general relativity, is deeply intertwined with the dynamics of gravity. It implies that physics must not depend on the choice of coordinates and only diffeomorphism invariant (Dirac) observables are physically meaningful \cite{Dirac,Henneaux:1992ig}. Moreover, this symmetry forbids a choice of a preferred or fixed background space-time. Conversely, the complexity of this symmetry is a root of the difficulty for defining a theory of quantum gravity; spin foam models are no exception.

While spin foam quantum gravity embraces the concept of background independence, the introduced regulator, frequently a 2-complex dual to a triangulation, generically breaks (a discrete remnant of) diffeomorphism symmetry \cite{Dittrich:2008pw,Bahr:2009ku,Bahr:2015gxa}, often called a vertex translation symmetry \cite{Bahr:2011uj}. There exist instances, where this symmetry is preserved in the discrete, where the discretisation perfectly reflects the continuum dynamics, or the symmetry can be restored iteratively via coarse graining. We explain this in detail in section \ref{sec:disc_diffeo}. For spin foams to be a viable theory of quantum gravity, diffeomorphism symmetry must be restored, at least approximately, in order to derive reliable physical predictions. There exists strong evidence that the amplitudes of the system can be systematically improved via coarse graining \cite{Bahr:2009qc,Dittrich:2012qb,Dittrich:2012jq}, such that the symmetry is broken less. Naturally the question arises whether this procedure converges to a fixed point, which would automatically imply an independence of the chosen regulator. Due to the non-local nature of diffeomorphism symmetry and in order to find a theory with propagating degrees of freedom, we conjecture such a fixed point to lie on a phase transition of second order. There it would be possible to take the continuum (or rather refinement limit) of the theory.

\subsection{Discretisation (in)dependence}

Closely related to diffeomorphism symmetry is the question of discretisation (in)dependence. Generically the results computed in spin foam models will depend sensitively on the chosen regulator, e.g. the number of simplices and subsequently the number of degrees of freedom. Moreover, there is no input from the theory itself which regulator to choose. However, in order to have a viable theory, it is imperative to find the same results no matter which discretisation is chosen, at least to an approximation.

In the research community, there exist two complementary paths addressing this question \cite{Rovelli:2010qx}. On the one hand, there is the approach to solve discretisation dependence by summing over all possible regulators, e.g. triangulations. This summation over triangulations (and topologies) is most holistically formulated in terms of group field theories \cite{Freidel:2005qe,Oriti:2013aqa}, which are quantum field theories formulated on several copies of a Lie group. The fields represent atoms of space-time, e.g. tetrahedra, whose interaction terms describe how 4D objects are formed, e.g. five tetrahedra glued together to form a 4-simplex. From this formalism, spin foam amplitudes arise as Feynman diagrams in a perturbative expansion. As for all quantum field theories, it must be shown that this theory is renormalizable, e.g. via perturbative or non-perturbative methods, see \cite{Carrozza:2016vsq} for a review.

On the other hand, we discuss the refinement approach \cite{Dittrich:2013xwa,Dittrich:2014ala} in this review, where we interpret the triangulation as a regulator to truncate the number of degrees of freedom, similar to the lattice in lattice field theories. The idea to overcome discretisation dependence is by assigning different amplitudes to different discretisations in such a way that the results agree. One example is to derive coarse amplitudes from fine ones via coarse graining. In this way, we are relating theories across different discretisations. The goal is to derive such relations for all possible discretisations, which is equivalent to a complete renormalization group trajectory. Again, this is similar to lattice field theory, where one also assigns different theories to different lattices, parametrized by coupling constants.% In contrast to spin foams however, these lattices come with an intrinsic scale given by the lattice constant.

\subsection{Computability}

The choice of a discretisation (and thus appropriate) amplitude also enters, at least partially, in another key challenge for spin foams, namely their computability. To be more precise, by computability we refer to two interconnected issues. On the one hand, there is the challenge to compute the fundamental spin foam amplitudes for a single building block, e.g. a 4-simplex. While this is well-studied and explored in the semi-classical regime \cite{Barrett:1998gs,Conrady:2008mk,Barrett:2009gg,Kaminski:2017eew,Liu:2018gfc}, in particular using coherent states and stationary phase approximation, computing a vertex amplitude in the quantum regime, e.g. for small spins, can only be done numerically. However, in recent years there has been significant progress in computing these amplitudes, e.g. for the EPRL / FK in Euclidean and the more challenging Lorentzian signature \cite{Dona:2017dvf,Dona:2018nev,Dona:2019dkf}.

Renormalization and coarse graining become important at the stages when we calculate amplitudes or observables for mulitple vertices / larger triangulations. Even if we have an efficient way of calculating spin foam amplitudes (or can access the relevant amplitudes from a database), summing over the various degrees of freedom remains a difficult task for such a high dimensional configuration space\footnote{Monte Carlo methods are only of limited use, since spin foam models are proper quantum amplitudes, i.e. complex-valued and highly oscillatory.}. However, if we assume that the full RG trajectory of the system is known, we can use the discretisation to our advantage and perform the same calculation on a much coarser spin foam with appropriately adapted amplitudes. Alternatively and more realistically, one can envision coarse graining the system first, essentially evaluating it in parts, deriving an effective theory on a coarser regulator from a finer one. On this coarse theory, expectation values of coarse observables can efficiently computed. This method is already realized nowadays in tensor network renormalization techniques \cite{Levin,GuWen,vidal-TNR}, see e.g. \cite{Cunningham:2020uco}. Note that the existence of a continuum limit is not assumed, rather we assume that coarse graining can be performed without severe truncations.

\subsection{Uniqueness, phase diagram and continuum limit}

Discretizing a continuum theory is generically not a unique process, take the 1D non-relativistic particle in a non-trivial potential as an example. There exist many choices how to discretize the potential, which all result in different dynamics. However, the expectation is that, no matter the choice, we reobtain the original continuum physics in a suitable  continuum limit (or approximate it well in a fine discretisation). This is even more severe in the case where the continuum theory possesses a symmetry, like reparametrisation or diffeomorphism invariance, which is broken in general in the discrete \cite{Bahr:2009qc,Bahr:2011uj}.

These topics, uniqueness of the theory, universality and the continuum limit remain open questions in spin foam quantum gravity. Modern spin foam models are frequently derived by starting from topological BF theory and then imposing simplicity constraints in the discrete \cite{Perez:2012wv}. The latter procedure is not unique, where e.g. the well-developed EPRL / FK model imposes the linear simplicity constraints weakly \cite{Engle:2007wy,Freidel:2007py}. Some effects on different choices of (implementations of) simplicity constraints can be found in the literature \cite{Alesci:2007tx}, however a phase diagram differentiating different universal dynamics is missing, and with it potential hints for a continuum limit and UV-completion of the theory.

These key challenges are deeply intertwined with one another and can be addressed by a coarse graining / renormalization scheme. In the following we review how our understanding of these connections developed over time and what the role of coarse graining is.

\section{Restoring diffeomorphism symmetry in the discrete} \label{sec:disc_diffeo}

Regge calculus \cite{Regge:1961px} is a discretisation of general relativity. In it the differentiable manifold is replaced by a $D$-dimensional triangulation, whose edge lengths are the dynamical degrees of freedom. Crucially, Regge calculus does not refer to coordinates of vertices of the triangulation and is solely formulated in terms of their distances. Hence it is manifestly coordinate free.
Each of the $D$-simplices is internally flat, i.e. its $D+1$ vertices can be embedded into $\mathbb{R}^D$. Curvature is distributional and located on $(D-2)$-sublimplices, so-called hinges. To each of these hinges in the bulk one associates a deficit angle $\epsilon_h$, which is the difference between the sum all dihedral angles of simplices meeting at this hinge and $2\pi$. This is nicely visualized in $d=2$: Several triangles meet at a single vertex. If their angles located at this vertex sum up to $2\pi$, it is flat and can be drawn on a piece of paper. However, if the deficit angle differs from $0$, e.g. if $\epsilon_h > 0$, we can no longer embed this collection of triangles into $\mathbb{R}^2$ and observe positive curvature around that vertex. Note that the edge lengths are the only dynamical variables, as the dihedral angles are given as functions of the edge lengths.

In addition to making no reference to coordinates, in some instances Regge calculus possesses additional symmetries in the discrete linked to diffeomorphism invariance \cite{Rocek:1982fr}. One such example is 3D Regge calculus for $\Lambda = 0$: its equations of motion state that all deficit angles $\epsilon_e = 0$ in the bulk, for all boundary data, describing a theory that glues piecewise flat tetrahedra in a flat way. Thus, it perfectly matches the continuum solution. Moreover, the Regge action is invariant under vertex translations, i.e. moving a vertex and accordingly changing the edge lengths it is connected to. One such example is the 4-1 Pachner move: If we place an additional vertex in the centre of a tetrahedron, we can freely choose three edge lengths connecting it to the vertices of the coarse tetrahedron. The fourth is then fixed uniquely by the equations of motion. This symmetry is reflected by nulleigenvalues of the matrix of second derivates of the action.

\begin{figure}[h!]
\begin{center}
\includegraphics[width=0.6\textwidth]{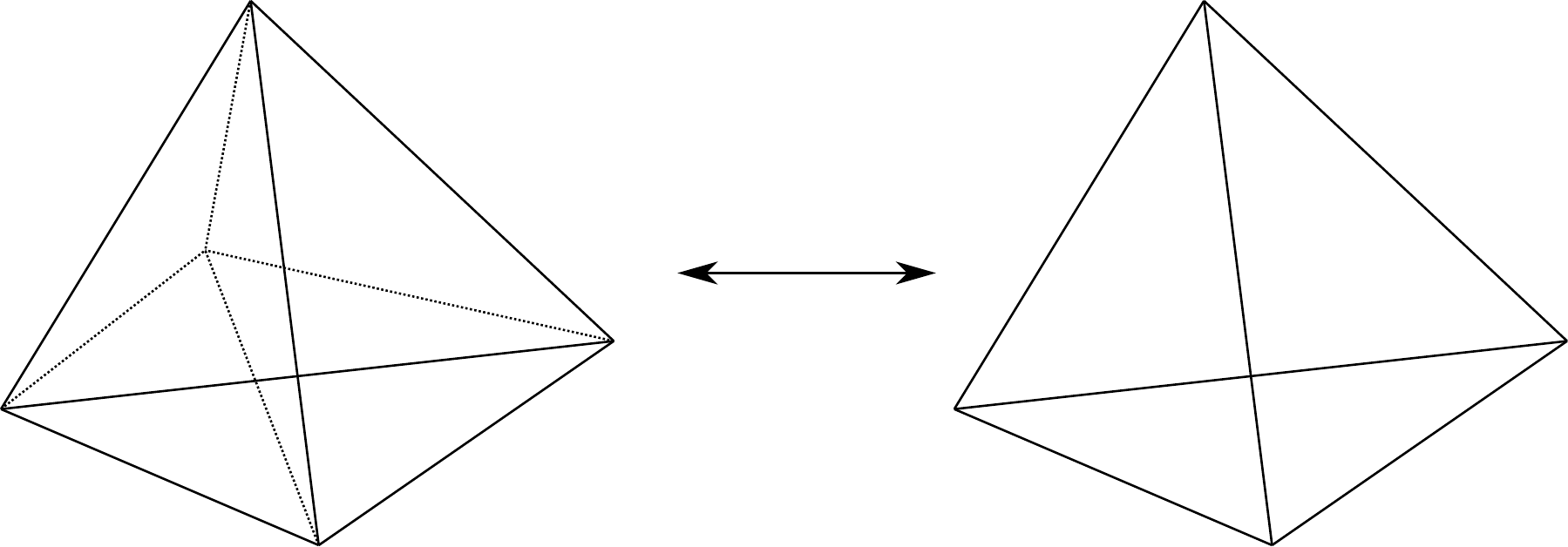}
\caption{\label{fig:Pachner_move}
The 4-1 Pachner move in 3D: the 3D Regge action is invariant under this move. Moreover, the configuration on the left hand side possesses a vertex translation symmetry. Three of the edge lengths connecting the inner vertex to the remaining ones can be chosen freely; the fourth one is then uniquely determined.
}
\end{center}
\end{figure}

Moreover, the 3D Regge action itself is invariant under such Pachner moves, i.e. local changes of the triangulation. This renders it triangulation independent, since any triangulation of a manifold can be related to any other triangulation of the same manifold by a consecutive application of Pachner moves \cite{pachner,pachner1}. This is not surprising, since 3D gravity is topological, i.e. has no local degrees of freedom. Nevertheless, we are convinced that triangulation independence and diffeomorphism symmetry in the discrete, in the form of a vertex translation symmetry, are closely related also beyond topological theories. Diffeomorphism symmetry is deeply entangled with the dynamics of general relativity. When perfectly realized in a discrete system, by fully capturing the continuum dynamics, it is irrelevant whether we consider a coarse or a fine discretisation. Thus the theory is discretisation independent. Invoking the invariance under vertex displacements, we can imagine this by moving vertices on top of each other, effectively removing them. Conversely, achieving discretisation independence by finding a fixed point of a coarse graining flow, e.g. on a second order phase transition, does not necessarily imply that diffeomorphism symmetry is restored, yet this conjecture is supported by several examples that we outline below.

A nice example how coarse graining can improve an action (or amplitudes in the quantum case) is again 3D Regge calculus with a non-vanishing cosmological constant. Due to the cosmological constant, the equations of motion state that deficit angles do not vanish. Moreover, the theory is not triangulation independent and the vertex translation symmetry is broken. In \cite{Bahr:2009qc}, Bahr and Dittrich device a coarse graining scheme for the triangulation: On a refined triangulation, subdividing large edges into small ones, they solve the equations of motion for the small edges and define an effective action for the remaining large ones. This procedure converges to a fixed point action, which describes Regge calculus for constantly curved tetrahedra. On this fixed point, deficit angles vanish, the theory is triangulation independent (by definition) and the vertex displacement symmetry is restored. Indeed, this improved discretisation / action encodes the continuum solution in the discrete, thus implementing a discrete remnant of diffeomorphism symmetry. Moreover, since it correctly captures the continuum dynamics, no information or accuracy is lost when using coarse triangulations. An analogous quantum version is the Turaev-Viro spin foam model \cite{Turaev:1992hq}, defined as a quantum deformed Ponzano-Regge model \cite{PR}.

There exist several instances where the continuum solution is pulled back to the discrete setting, where the discrete theory possesses a vertex translation symmetry. One example is 4D Regge calculus \cite{Rocek:1982fr}, when the boundary data allow for flat solutions in the bulk, or the 1D quantum parametrised (an)harmonic oscillator \cite{Bahr:2011uj}. In general we cannot guess these solutions, but with coarse graining methods we can construct or at least approximate them well. However, the examples that we discuss here are either topological or one-dimensional, and hence it is possible to retain a local description. For higher dimensional, interacting theories, non-local interactions appear, which can be a stumblestone for coarse graining methods.

\subsection{Non-localities}

Before explaining non-localities or non-local interactions, we first need to state what a local theory is in this context. In most discrete theories, we associate variables to parts of the discretisation, e.g. in spin foam models we assign irreducible representations $\rho_f$ to faces $f$ of the dual complex and intertwiners $\iota_e$ to edges $e$. In Regge calculus, we assign edge lengths $l_e$ to the edges $e$ of a triangulation. We define this theory to be local if the partition functions is given by a product of amplitudes assigned to (sub)simplices or if the action is given as a sum over actions assigned to (sub)simplices. Moreover, the action and amplitude for each (sub)simplex only depend on those variables attached to (sub)simplices contained in the (sub)simplex. Spin foam models are an example for such local theories, since the partition function is given via a local assignment of vertex, edge and face amplitudes, see eq. \eqref{eq:sf_partition}. Similarly, the Regge action can be written as a sum over contributions associated to the $D$-simplices of the triangulation.

When we apply coarse graining methods to such interacting, i.e. non-topological, theories, it is highly unlikely that this local form of the theory can be preserved. There exist several examples in the literature where this has been shown in the past. In \cite{Dittrich:2011vz}, 4D Regge calculus was linearized around a flat background solution and the perturbations of the edge length integrated over. The question is whether it is possible to find a path integral measure that is invariant under Pachner moves. However, when integrating out these degrees of freedom, one picks up a non-local factor that cannot be written as a local product. In \cite{Dittrich:2014rha} it is shown that said factor is related to a condition whether the six vertices involved in the Pachner move lie on a 3-sphere. Moreover, these articles reveal that the 4D Regge action itself is not invariant under Pachner moves. In a similar vain, \cite{Banburski:2014cwa} studies Pachner moves in 4D holomorphic spin foam models \cite{Dupuis:2011fz}. The advantage of these models is that Pachner moves can be explicitly computed. Again, the resulting amplitude is non-local, in the sense that the resulting expression cannot be written as a assignment of local amplitudes.

To illustrate this point further let us consider the concrete example of the 2D Ising model.

\subsubsection{Ising model as an example}

There exist plenty of ways to coarse grain discrete systems. A straightforward example is the 2D Ising model subject to a simple decimation procedure, where one simply sums over ``every other'' spin to derive an effective model on a larger scale.

We consider the Ising model defined on a regular 2D lattice with vanishing external magnetic field. There are only nearest neighbour interactions, i.e. an Ising spin $\sigma_i \in \{-1,1\}$ only interacts with its direct neighbours. Then we can write the partition function as product of weights associated to the edges of the lattice:
\begin{equation}
  Z_{\text{Ising}} = \sum_{\{\sigma_i\}} \prod_e \exp(\beta \, \sigma_{s(e)} \, \sigma_{t(e)}) \; ,
\end{equation}
where $\beta$ is the inverse temperature, and $s(e)$ / $t(e)$ denote the source and target of the edge $e$\footnote{The choice of orientation is fiducial, but allows for a short-hand notation.}. Note that the system has a global $\mathbb{Z}_2$ symmetry, it remains invariant if all spins are flipped.

\begin{figure}[h!]
\begin{center}
\includegraphics[width=0.95\textwidth]{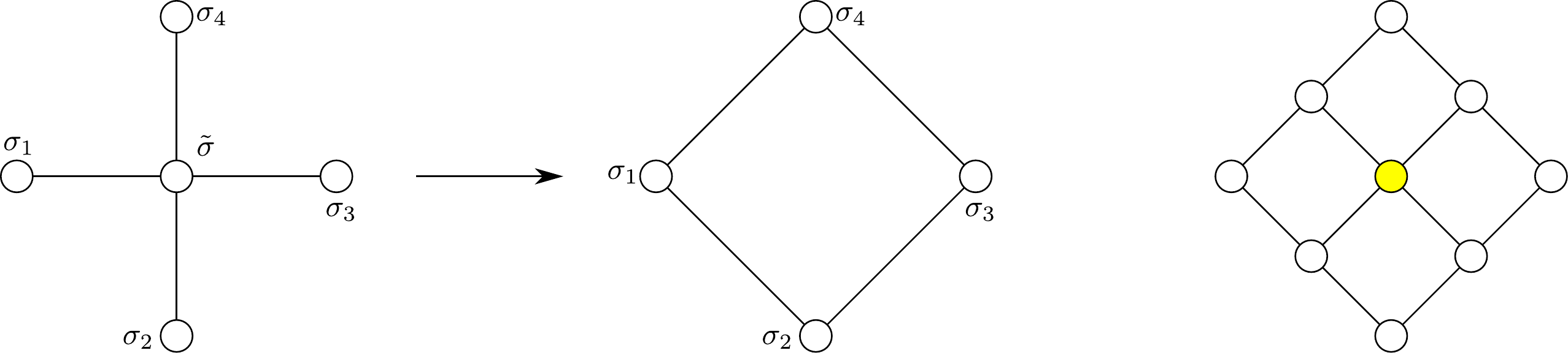}
\caption{ \label{fig:Ising_decimation}
Left: Decimating an Ising spin in the 2D model results in non-local interactions among all four spins the decimated spin is connected to. Right: In the next step, decimating one spin, here in yellow, would result in an amplitude non-locally depending on eight spins.
}
\end{center}
\end{figure}

We implement a decimation procedure by summing over every other spin, essentially evaluating the partition function in parts. In order to derive the new effective amplitude of the system, it is sufficient to consider four Ising spins $\sigma_1,\dots,\sigma_4$ that all connect to another Ising spin $\tilde{\sigma}$, see fig. \ref{fig:Ising_decimation}. The four coarse spins sit on the corner of a coarser square rotated by $45^\circ$ with spin $\tilde{\sigma}$ in the center of the square. We obtain:
\begin{equation}
  \sum_{\tilde{\sigma}} \exp(\beta \, \tilde{\sigma} \, (\sigma_1 + \sigma_2 + \sigma_3 + \sigma_4)) \; = \; 2 \cosh(\beta \, (\sigma_1 + \sigma_2 + \sigma_3 + \sigma_4) .
\end{equation}
Clearly this expression is not of the same form as the original action, in particular is not written in terms of $\mathbb{Z}_2$ group multiplications. Nevertheless the remaining spins still satisfy the global $\mathbb{Z}_2$ symmetry. Thanks to this global symmetry, this expression can only take three different values depending on the configurations of the four spins $\{\sigma_i\}$; either all spins are aligned, one is not aligned with the others or we have two pairs of aligned spins. To express this again in terms of spin interactions, we make the most general ansatz of four spin interactions compatible with the global $\mathbb{Z}_2$ symmetry:
\begin{equation}
  \mathcal{A}(\sigma_1,\sigma_2,\sigma_3,\sigma_4) :=  \exp(a \, (\sigma_1 \sigma_2 + \sigma_2 \sigma_3 + \sigma_3 \sigma_4 + \sigma_1 \sigma_4) +
   b \, (\sigma_1 \sigma_3 + \sigma_2 \sigma_4) + c \, \sigma_1 \sigma_2 \sigma_3 \sigma_4 + d ) \; .
\end{equation}
$a$ is the parameter for nearest neighbour interactions, $b$ for next-to-nearest neighbour interactions, $c$ for a four spin interaction and $d$ is a constant. We can compare these equations directly for each configuration:
\begin{align}
  \exp(4a + 2b + c + d) & = 2 \cosh(4 \beta) & \text{all }\sigma_i = \pm 1 \nonumber \\
  \exp(-c + d) & = 2 \cosh(2 \beta) & \sigma_i = \sigma_j = \sigma_k = -\sigma_l \nonumber \\
  \exp(-2b + c + d) & = 2 & \sigma_i = \sigma_j = -\sigma_k = -\sigma_l \nonumber \\
  \exp(-4a + 2b + c + d) & = 2 & \sigma_i = -\sigma_j = \sigma_k = -\sigma_l \; ,
\end{align}
where we denote a cyclic order $i,j,k,l$ around the square. Here we have four equations for four unknown parameters, which we can straightforwardly solve. We leave deriving the solution to the interested reader.

The coarse grained amplitude is notably different than the initial one. While we find again nearest neighbour interactions, new non-local interactions appear as well. From this new form it is not obvious how to return to the original expression. Moreover, it is not clear how to iterate the procedure without approximations: decimating one spin alone results in non-local interactions among eight spins, some of which ought to be decimated as well, see again fig. \ref{fig:Ising_decimation}. Nevertheless, already this simple example hints towards a resolution: the non-localities arise since we attempt to express the coarse grained dynamics in terms of the old degrees of freedom and building blocks. Yet we can still write the partition function as a local product of amplitudes associated to rotated squares, where the non-local interactions are completely contained within these locally assigned amplitudes. In the next section we introduce this change of perspective more concretely and discuss the concept of generalized boundary data\footnote{Note that the discussion of the Ising model is primarily to give an intuitive example on how non-local interactions arise under coarse graining. Since the 2D Ising model is solved analytically \cite{PhysRev.65.117}, it is ideal to test the capabilities and feasibility of new coarse graining approaches. See e.g. \cite{RevModPhys.86.647} for an overview of real-space renormalization techniques in statistical physics.}.

\section{Change of concept: generalized boundaries and amplitude maps} \label{sec:embedding}

The vital insight to arrive at a practical coarse graining scheme for spin foams is the following: instead of pertaining the original degrees of freedom and building blocks, e.g. simplices, and allowing more and more complicated, non-local interactions among them, we work with locally interacting amplitudes, which allow for more general and complex boundary data. The non-locality is still present, yet contained within the amplitudes and expressed as interactions of these boundary data. Thus, the complexity of the boundary data controls the non-locality preserved under coarse graining and the complexity of the amplitude. Truncating the boundary data allows us to introduce controllable approximations, while the partition function is still written as a local assignment of amplitudes. Thus we can iterate a coarse graining procedure that only needs to consider few building blocks at a time.

As a path integral approach, spin foam quantum gravity is already phrased in this language, as amplitude functionals for certain boundary states. Take a spin foam on a 2-complex $\Gamma_b$ with boundary $b$. Since the 2-complex is discrete, namely a collection of faces, edges and vertices, its boundary $b$ is also discrete, namely a graph, with nodes and links. The complexity of this boundary depends on the number of nodes and links. To each of these boundaries $b$ one associates a boundary Hilbert space $\mathcal{H}_b$, whose complexity again depends on the complexity of the boundary. A spin foam model for said two complex then acts as an amplitude functional $\mathcal{A}_b$ mapping states $\psi_b \in \mathcal{H}_b \rightarrow \mathbb{C}$.

The vital difference is that we allow for more general building blocks, in particular with more complex boundary data and thus boundary Hilbert spaces. When using Pachner moves, one integrates out bulk degrees of freedom while keeping the boundary unchanged. In 4D, when performing a $4-2$ move, one integrates out bulk variables and derives one effective amplitude for two glued 4-simplices, prescribed by the same boundary data. However splitting this effective amplitude into two, one assigned to each building block, is not straightforward due to the previously mentioned non-local interactions. Instead, we allow for more general building blocks with more complicated boundary data. That way, we still have local assignments of amplitudes to building blocks, which in turn interact locally with neighbouring ones. In turn, non-localities still arise, yet they are contained in each building block and captured by more complex boundary data.

While this picture recasts the problem of arising non-localities, three immediate challenges arise. Firstly, iterating this procedure leads to more and more complicated building blocks, whose amplitudes are still given by the fine boundary degrees of freedom. From this perspective we have not achieved a derivation of coarse scale physics, since the dynamics are still expressed in terms of fine scale degrees of freedom. Secondly, in order to define a renormalization group flow it is crucial to compare amplitudes after each coarse graining step. And thirdly, deriving amplitudes with more and more complex boundary data quickly becomes unfeasible, independent whether one is using analytical or numerical techniques, as one can already see for the 2D Ising model.

Hence, the next vital ingredient for a coarse graining scheme is the introduction of variable transformations, that map a collection of fine boundary degrees of freedom to a collection of coarse effective degrees of freedom. More precisely, we want to map states on a fine boundary Hilbert space $\mathcal{H}_{b'}$ on $b'$ to a coarse boundary Hilbert space $\mathcal{H}_{b}$ on $b$. In the next section, we will explain the idea behind this concept and its physical interpretation. To do so, we work in the opposite direction and explain how to add degrees of freedom using embedding maps.

\subsection{Embedding maps and the notion of vacuum}

As outlined above, a key idea of any renormalization procedure is to compare and relate theories defined on different scales. Given two spin foam amplitudes $\mathcal{A}_b$ and $\mathcal{A}_{b'}$, which are functionals for the Hilbert spaces $\mathcal{H}_b$ and $\mathcal{H}_{b'}$ respectively, these amplitudes can only be compared for the same physical processes. That is, given a state $\psi_b$ in the coarse Hilbert space $\mathcal{H}_b$ one must represent $\psi_b$ in the Hilbert space $\mathcal{H}_{b'}$. Then, each states can be evaluated with their respective amplitude and the results compared. For this purpose one defines so-called embedding maps:
\begin{equation}
  \iota_{b' b} : \mathcal{H}_b \hookrightarrow \mathcal{H}_{b'} \; .
\end{equation}
For this to work, the boundary $b$ must be embeddable into the boundary $b'$, denoted as $b \prec b'$. Thus, the boundaries $b$ form a partially ordered set. In case that two boundaries $b$ and $b'$ cannot be directly related, i.e. $b$ cannot be embedded into $b'$, one embeds both into a common refinement $b''$, written as $b \prec b''$ and $b' \prec b''$.

Hence, the goal is the following: given a state $\psi_b$ in a coarse Hilbert space $\mathcal{H}_b$, we want to define an equivalence class of states in all finer, more complex Hilbert spaces $\mathcal{H}_{b'}$ in order to readily compare the associated amplitude functionals. This equivalence class of states is defined as follows: given two states $\psi_b \in \mathcal{H}_b$ and $\phi_{b'} \in \mathcal{H}_{b'}$,
\begin{equation}
  \psi_b \sim \phi_{b'} \; \iff \; \iota_{b'' b}(\psi_b) = \iota_{b'' b'}(\phi_{b'}) \; \forall \; b'' \text{s.t. } b\prec b'' \text{and } b' \prec b'' \; .
\end{equation}
For this condition to be well-defined, the embedding maps need to satisfy a consistency condition, referred to as cylindrical consistency:
\begin{equation}
\iota_{b'' b'} \circ \iota_{b' b} = \iota_{b'' b} \; \text{with} \; b \prec b' \prec b'' \; .
\end{equation}
Essentially, it should not matter whether a state is directly embedded into a fine boundary $b''$ or via an (or any other) intermediate boundary $b'$. Given these conditions and relations, one can (at least formally) define a continuum Hilbert space via an inductive limit: $\mathcal{H} := \overline{\cup_{b} \mathcal{H}_b / \sim}$.

Beyond this formal definition, the action of embedding maps is best understood in the following way. As illustrated before, they serve the purpose of representing a coarse state in a finer Hilbert space, which can encode more complex configurations. Hence, embedding maps specify how and in which {\it state} degrees of freedom are added. Moreover, they thus define an inner product allowing us to compare states across Hilbert spaces. Since the information of the coarse state ought to be unchanged, these new degrees of freedom are added in a {\it vacuum state} prescribed by the embedding map. These concepts are familiar in the kinematical Hilbert space of loop quantum gravity \cite{thomasbook} expressed in terms of spin network functions, where new degrees of freedom are added in the Ashtekar-Lewandowski vacuum \cite{ashtekar-lewan1,ashtekar-lewan2}, which describes no space. In contrast, a dual BF representation \cite{Dittrich:2014wpa,Dittrich:2014wda,Bahr:2015bra} constructed in the last few years adds degrees of freedom that are peaked on flat connections. However, this notion of vacuum does not imply that this is a physical vacuum. Both examples given above are kinematical vacua in 4D gravity, i.e. they do not satisfy diffeomorphism and Hamiltonian constraints.

\subsection{Renormalization group flow of amplitudes}

Once given such a choice of embedding maps, these can be readily used to compare spin foam amplitudes. Again, given two amplitudes $\mathcal{A}_b$ and $\mathcal{A}_{b'}$, a state $\psi_b \in \mathcal{H}_b$ and an embedding map $\iota_{b'b}$, we compare both amplitudes:
\begin{equation}
  \mathcal{A}_b (\psi_b) \overset{?}{=} \mathcal{A}_{b'} (\iota_{b' b}(\psi_b) ) =: \mathcal{A}'_{b}(\psi_b) \; .
\end{equation}
Due to the embedding map, we define an effective amplitude $\mathcal{A}'_b$ for the coarse Hilbert space $\mathcal{H}_b$ from the fine one $\mathcal{A}_{b'}$ for $\mathcal{H}_{b'}$. If performed for all possible states in $\mathcal{H}_{b'}$, we obtain the coarse grained amplitude. Thus, embedding maps, which specify how to add degrees of freedom under refinement of states, serve as coarse graining maps for amplitudes. There, they specify how to define effective degrees of freedom. Consequently, since $b'$ can capture more information that $b$, embedding maps also encode how to truncate degrees of freedom.

\begin{figure}[h!]
  \begin{center}
  \includegraphics[width=0.95\textwidth]{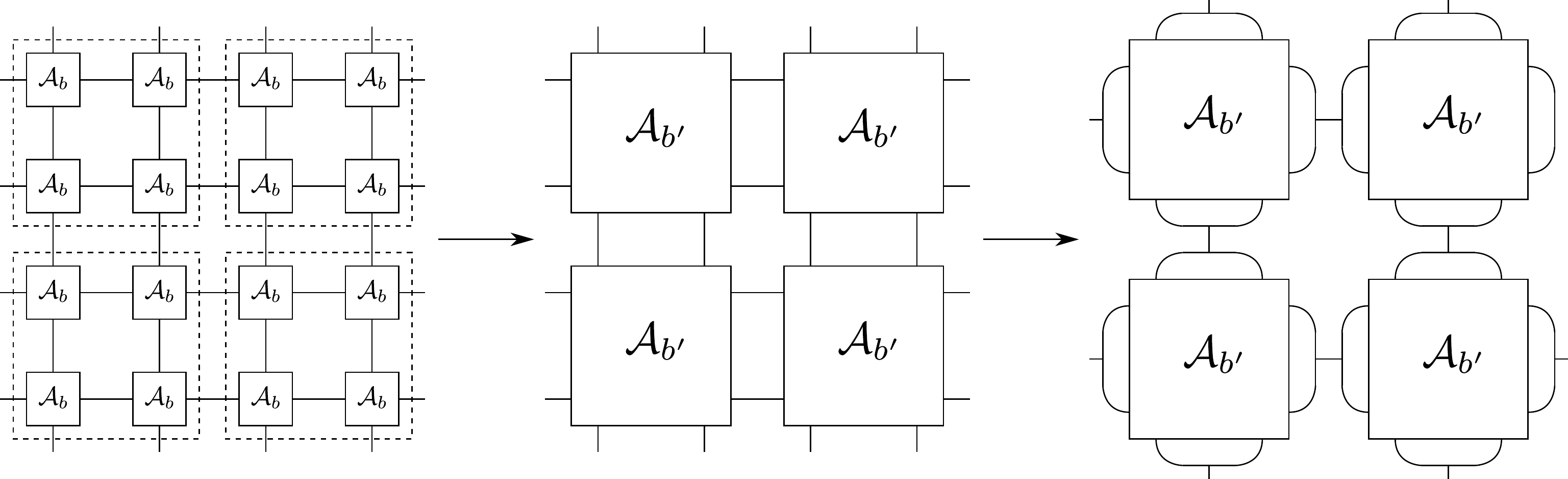}
  \caption{\label{fig:coarse_graining_gen}
  The basic steps of coarse graining: blocking of amplitudes, summing over fine degrees of freedom and introduction of embedding maps to define an effective amplitude for the original Hilbert space $\mathcal{H}_b$.
  }
\end{center}
\end{figure}

To summarize, a class of embedding maps defines a coarse graining / renormalization group flow of amplitudes, formulated with respect to their boundary. This flow is given by the following equation:
\begin{equation}
  \mathcal{A}'_b = \mathcal{A}_{b'} \circ \iota_{b'b} \; .
\end{equation}
To showcase the implications for the system as a whole, it is instructive to consider the partition function of the system. For simplicity, we assume that we can write it as a collection of amplitudes $\mathcal{A}_b$\footnote{
E.g. this is the case for spin foam models defined on 2-complexes with regular, i.e. hypercubic, combinatorics. Due to the regularity, the system can be written purely as a collection of vertex amplitudes.
}:
\begin{equation}
  Z = \sum_{j_b} \prod_{b} \mathcal{A}_b(j_b) = \sum_{j_{b'}} \prod_{b'} \left[ \sum_{j_b \in \text{bulk of } b'} \prod_{b \supset b'} \mathcal{A}_b(j_{b'},j_b) \right] =: \sum_{j_{b'}} \prod_{b'} \mathcal{A}_{b'}(j_{b'})
\end{equation}
The original partition function is given as a product of amplitudes $\mathcal{A}_b$ assigned to building blocks with boundary Hilbert space $\mathcal{H}_b$. This Hilbert space is spanned by an orthonormal basis with labels $j_b$. In the second equality, we perform a blocking of amplitudes, e.g. of 16 vertex amplitudes for hypercubic combinatorics. The degrees of freedom $j_b$ are split into two groups: one group makes up the boundary degrees of freedom $j_{b'}$ of the blocked amplitudes, while the other are block ``bulk'' degrees of freedom $j_b$ and are summed over. The latter part is then summarized as the fine amplitude $\mathcal{A}_{b'}$. See fig. \ref{fig:coarse_graining_gen}.

As the final step, we implement the embedding maps (or rather coarse graining maps) to derive the effective amplitude $\mathcal{A}'_b$ for the original Hilbert space.
\begin{equation}
  Z = \sum_{j_{b'}} \prod_{b'} \mathcal{A}_{b'}(j_{b'}) \approx \sum_{j_b} \prod_b \left[ \sum_{j_{b'}} \iota_{b' b}(j_{b'},j_b) \mathcal{A}_{b'}(j_{b'}) \right] =: \sum_{j_b} \prod_b \mathcal{A}'_b(j_b)
\end{equation}

In this context, the embedding maps serve as variable transformations and truncations, see again fig. \ref{fig:coarse_graining_gen}. Indeed, this inclusion of embedding maps necessarily alters the partition function of the system, and we must ensure that we can still draw reliable conclusions about the original system. Thus, the combined embedding maps from neighbouring amplitudes should be close to the identity on the respective Hilbert space, in the sense that we only truncate irrelevant degrees of freedom. For example this is realized in tensor network renormalization, where these embedding maps are unitary as we explain in section \ref{sec:TNW}.

\begin{figure}[h!]
  \begin{center}
  \includegraphics[width=0.5\textwidth]{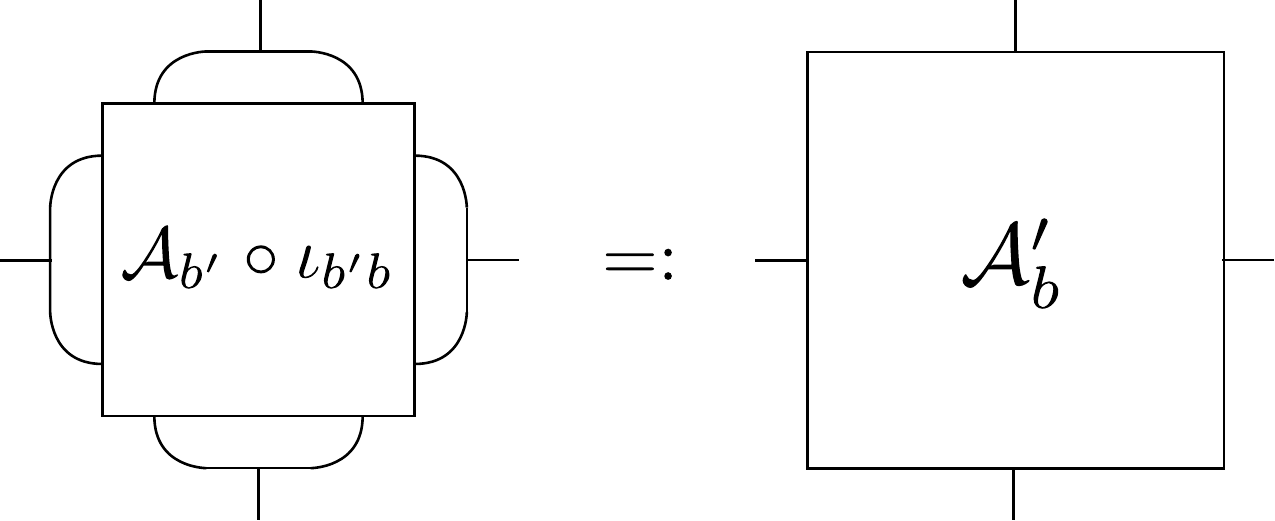}
  \caption{\label{fig:effective_amplitude}
  Definition of the effective amplitude $\mathcal{A}'_b$ defined as the concatination $\mathcal{A}_{b'} \circ \iota_{b'b}$.
  }
\end{center}
\end{figure}

Essentially, with this coarse graining procedure, we achieve two goals. On the one hand, we evaluate the partition function in parts, purely from local considerations of a subset of amplitudes. This is computationally efficient and makes calculations more accessible. On the other hand, we derive an effective theory on a coarser lattice, with less degrees of freedom, from a theory defined on a finer lattice. That way, we relate two theories on two different regulators by assigning different amplitudes to different lattices. Thus, the renormalization group flow of spin foam quantum gravity is defined as a family of amplitudes assigned to a family of 2-complexes / discretisations:
\begin{equation}
  \mathcal{A} \; \rightarrow \; \mathcal{A}' \; \rightarrow \; \mathcal{A}'' \; \rightarrow \; \dots \; .
\end{equation}
Before we discuss the consequences of this renormalization group flow in detail, it is important to discuss the role of the embedding / coarse graining and how they ought to be chosen.

\subsection{Dynamical embedding maps and the physical vacuum}

The embedding maps play a pivotal role in the renormalization group flow. Generically, different embedding maps result in different flows, since they determine how the effective coarse degrees of freedom depend on the fine ones. Thus, if one fixes the embedding maps a priori, this choice must be carefully checked. Instead, it is vital that these maps are directly determined from the dynamics encoded in the amplitudes themselves \cite{Dittrich:2012jq,Dittrich:2013xwa,Dittrich:2014ala}.

This reasoning is intuitive to follow, e.g. consider the Ising model. The effective degrees of freedom most suitably describing the coarse dynamics are sensitive to the temperature and significantly differ between low and high temperature. Hence, one size does not fit all: fixing embedding maps a priori can seriously distort the RG flow and give wrong results. Thus, we are convinced that dynamical embedding maps are vital for a successful coarse graining scheme. We explain how to implement this in practice in section \ref{sec:TNW} on tensor network renormalization. For now, we discuss its implications and physical interpretation.

As discussed above, embedding maps are prescriptions how and in which state degrees of freedom are added under refinement. In particular they allow to relate and identify states across Hilbert spaces, defining an inner product and notion of vacuum. When these embedding maps are chosen dynamically, i.e. with respect to the dynamics encoded in the amplitude, the new degrees of freedom are added in a {\it dynamical} vacuum state. %This concept fits well into a theory of quantum gravity.

General relativity is a totally constrained theory, i.e. in a canonical formulation its evolution is not governed by a Hamiltonian but rather by a sum of constraints, namely the diffeomorphism and Hamiltonian constraints. These constraints are generators of gauge transformations, which implies that evolution in gravity amounts to gauge transformations. This is known as the infamous problem of time \cite{time-anderson2}. For a quantum theory, the goal is to find the physical Hilbert space, i.e. the space of all states that are annihilated by all the constraints. Therefore, given an initial physical state, evolution by the constraint operators leave the physical state unchanged. Following this insight, dynamical embedding maps in a quantum gravity theory should be {\it physical} embedding maps that add degrees of freedom in the {\it physical} vacuum under refinement. Thus, such embedding maps do not add new information to the state and represent the same physical state on a finer boundary Hilbert space.

In the context of path integral formalisms of gravity, this insight is particularly intriguing: since evolution in the canonical formulation is governed by constraint operators, the path integral merely imposes the constraints, projecting out kinematical degrees of freedom and leaving physical states unchanged \cite{Dittrich:2013xwa}. In short, the path integral serves as a projector onto the physical Hilbert space. Indeed, this insight is one of the original motivation behind constructing spin foam models in the attempt to define a riggin map / physical inner product for kinematical states of loop quantum gravity  \cite{thomasbook}.

Following this line of thought, spin foams themselves are dynamical embedding maps. Consider a spin foam evolving an initial state to a final state, where these states are defined on different boundaries\footnote{
For an in-depth derivation of a canonical formalism for phase spaces and Hilbert spaces of varying dimension / complexity and the appearance of pre- / post-constraints, see \cite{Dittrich:2013jaa,Hoehn:2014fka,Hoehn:2014wwa}.
}. When interpreting the spin foam as a map from one Hilbert space to another, instead of as an amplitude functional, it is by definition an embedding map. Moreover, if the spin foam acts as a projector onto the physical Hilbert space, concatenated spin foams still act as a projector, implying path independence of evolution. Conversely, this is interpreted as first evolving to an intermediate state, thus cylindrical consistency conditions of embedding maps are satisfied. Additionally, the projector property implies that this evolution is independent of the choice 2-complex / discretisation, and it would mark a fixed point of the renormalization group flow:
\begin{equation}
  \mathcal{A}_b = \mathcal{A}_{b'} \circ \iota_{b'b} \; .
\end{equation}
This implies that assigning the same amplitude to all discretisations gives the same results.

All of the conditions mentioned above are highly non-trivial and rely on a perfect implementation of diffeomorphism symmetry in the discrete. Indeed, this assumption is hidden in the projector property of the path integral / spin foam, which implies an implementation of diffeomorphism and Hamiltonian constraints. Path / discretisation independence follow immediately and underline the strong connection of diffeomorphism symmetry and discretisation independence. In fact, this construction would be a realization of the perfect action program \cite{Bahr:2009qc} for quantum gravity and would imply that the dynamics of quantum gravity are solved non-perturbatively and pulled back onto the discrete.

Unsurprisingly, these conditions are not met by spin foam models: spin foam amplitudes do not act like projectors \cite{Freidel:2005qe,Thiemann:2013lka} and explicitly break diffeomorphism symmetry \cite{Dittrich:2008pw,Bahr:2015gxa}. Furthermore, it is unlikely that these conditions can be perfectly realized without approximations in full generality. Thus, the goal of the coarse graining scheme is to iteratively improve spin foam amplitudes in order to well approximate the ideal solution. We discuss this in the next section.

\subsection{Lessons from the RG flow}

The idea behind the coarse graining method outlined above is that it  allows us to iteratively improve the amplitudes, such that the conditions mentioned above are approximately implemented. Furthermore we can tackle discretisation dependence of the theory. The expectation is that on a second order phase transition the regulator can be removed and diffeomorphism symmetry is restored. Let us explain this step by step.

Firstly, it is straightforward to recognize that the renormalization group flow addresses the question of discretisation dependence and choice of 2-complex. By deriving an effective amplitude $\mathcal{A}_b$ from coarse graining $\mathcal{A}_{b'}$, we directly relate theories on 2-complexes defined from building blocks with boundaries $b$ and $b'$ respectively. This information is vital for any discrete theory: it states that we perform (approximately) the same calculation on $b$ when using $\mathcal{A}_b$ as on $b'$ when using $\mathcal{A}_{b'}$. In particular, following this prescription, it does not matter whether we calculate a coarse observable on the coarse or the fine lattice. Thus, we account for the discretisation dependence of the theory and ensure at the same time that the results are reliable. Indeed, understanding this behaviour is indispensable when trying to make contact with experiments. %Moreover, the situation is similar to other approaches relying on lattices as regulators, e.g. lattice gauge theory. There one also performs simulations on different lattices, which are labeled by the lattice constant $a$, and changes the coupling constant depending on the scale.

Understanding the lattice dependence of the theory is an important step towards determining when and how the regulator can be removed entirely. To this end, one has to study the whole coarse graining flow, that is choose an initial amplitude (e.g. given by a choice of parameters) and follow the flow until it reaches a fixed point. Within a certain approximation, e.g. restricting to finite dimensional boundary Hilbert spaces, this generically happens. These attractive fixed points frequently describe topological theories\footnote{This is expected for disretisation independent theories with finitely many degrees of freedom. Examples are the Ponzano-Regge model in 3D \cite{PR}, or more generically BF theory in any dimension \cite{Baez:1999sr}.}, where a continuum limit can be trivially taken. However, these theories do not describe gravity in four dimensions, since they lack propagating degrees of freedom. Furthermore, these attractive fixed points denote the phases of the theory.

All initial amplitudes, e.g. all amplitudes from a certain region in parameter space, that flow under coarse graining to the same attractive fixed point lie in the same phase. These theories have the same dynamics on sufficiently coarse grained discretisations and thus lie in an universality class and share qualitative features, e.g. in expectation values of observables. An example would be the strong coupling phase in lattice gauge theory, in which one expects the Wilson loop operator to satisfy an area law. Frequently models possess multiple phases with phase transitions separating them. We are particularly interested in phase transitions of second order.

In standard lore, second order phase transitions are characterized by a diverging correlation length. This implies that degrees of freedom infinitely far away are correlated and, thus, infinitely many degrees of freedom are relevant for the dynamics. Moreover, right on the phase transition, the system is scale-invariant, i.e. physics do not change with scale. Therefore, on a second order phase transition one can take the continuum limit to arrive at a continuum theory with propagating degrees of freedom.

We expect the same to hold for second order phase transitions in spin foam models, with a slightly different interpretation: Background independent theories lack an absolute length scale. Still, the regulator allows us to define a combinatorial distance. Essentially, the idea is to define a distance between vertices of the 2-complex, by counting the number of vertices one has to pass in order to reach the other one. If they are directly connected by an edge, this distance would be one\footnote{This idea is inspired by a similar concept in Causal Dynamical Triangulations \cite{Loll:2019rdj}, where the distance between two vertices of the triangulation is given by the minimal number of links between them, albeit with the notable difference that each length has a specified length assigned to it. This concept is used e.g. to measure the geodesic distance between two vertices \cite{Ambjorn:2004qm}.}. Then, on a second order phase transition, degrees of freedom that are infinitely ``far'' away with regard to the lattice are correlated and the combinatorial correlation ``length'' diverges. Furthermore, the notion of scale invariance is consequently replaced by a discretisation independence / invariance, fixed point equations for the amplitudes are satisfied and a continuum / refinement limit can be taken. The implications of constructing the theory on this fixed point must be stressed: Due to the discretisation independence, calculations can be performed in the continuum or on any discretisation, giving the same results. This exactly corresponds to the idea of perfect action, and thus solving the coarse graining flow corresponds to solving the theory on all lattices.

Nevertheless, two caveats must be observed: firstly, finding such a second order phase transition (if it exists) does not guarantee that the corresponding theory is a correct theory of quantum gravity. Secondly, if infinitely many degrees of freedom become relevant, truncated coarse graining schemes can only approximate the desired theory to a certain order, as it is the case in other renormalization schemes.

In the next section we discuss the notion of scale in more detail.

\subsection{Background independence and the interpretation of scale}

Before we continue with reviewing how to coarse grain in practice, it is crucial to discuss the notion of ``scale'' - or the lack thereof. As a background independent approach, one cannot assign a scale to a spin foam since one sums / superimposes all possible geometries (allowed by a certain 2-complex). Thus, we use the 2-complex itself, here formulated via the boundary of the amplitudes, to order degrees of freedom according to a relative scale. Consequently, we replace the familiar notions of ultraviolet (UV) and infrared (IR) by ``fine'' and ``coarse'' respectively. Following this perspective, we are not integrating out short scale degrees of freedom under coarse graining. Instead, we sum over finer discrete degrees of freedom and define effective coarse degrees of freedom, which encode (superpositions of) geometries of different scales.

Alternatively, one can introduce a specific scale in this coarse graining procedure via boundary states. That is, we do not consider the entire Hilbert space, but only a specific state because we are interested in studying a transition of geometries. Then, this fixed boundary states introduces a physical scale via the encoded 3D geometry, e.g. implemented in the restricted path integral formalism, see section \ref{sec:restricted}.

A further comment on the coarse graining scheme is in order: here we purely formulate it in terms of the boundary discretisation and its associated boundary Hilbert space, not in terms of the bulk. From a practical perspective these questions are less important, e.g. if one assumes regular combinatorics such that the coarse graining procedure can be straightforwardly iterated; then both the boundary and bulk form totally ordered sets. Nevertheless, there is a proposal by Bahr \cite{Bahr:2014qza} for formulating the coarse graining scheme in the bulk, essentially by defining embedding maps for 2-complexes.

\section{Coarse graining methods} \label{sec:coarse_graining}

For the rest of this article, let us focus on numerical methods that allow us to realize this coarse graining method in practice and review the results.

\subsection{Tensor network renormalization methods} \label{sec:TNW}

Tensor network methods originate in the fields of quantum information and condensed matter and aim at efficiently studying quantum many body systems. In this review we focus on tensor network renormalization methods\footnote{
There also exist tensor network methods that aim at constructing specific states of many body quantum systems, e.g. matrix product states (MPS), projected entangled pair states (PEPS) \cite{Orus_2014} or multi-scale entanglement renormalization ansatz (MERA) \cite{mera}. Their goal is to efficiently represent a small subspace of an exponentially large Hilbert space, containing the ground state.
} \cite{Levin,GuWen,vidal-TNR}, a numerical algorithm for coarse graining discrete systems. To this end the partition function of the system is represented as a contraction of a tensor network. In the context of this article, a tensor $T_{abc\dots}$ is best understood as the amplitude assigned to a region. The boundary data of these amplitudes are represented as indices of the tensor, which is graphically represented as a vertex with as many legs as it has indices. The partition function is then rewritten as a contraction of tensors:
\begin{equation}
Z = \sum_{a_1 b_1 c_1 \dots} T_{a_1 b_1 c_1 d_1} T_{c_1 b_2 c_2 d_2} \dots = \text{Tr}(T \, T \dots T) \; .
\end{equation}
Graphically, each identified and contracted index is represented by connecting the respective tensor indices. Thus, the partition function is represented by a collection of tensors connected to one another in a local fashion, a tensor network, see fig. \ref{fig:Tensor_network}.

So far, this is merely a rewriting of the original system. The goal is to locally manipulate the tensors in order to rewrite the partition function as a coarser tensor network, see again fig. \ref{fig:Tensor_network}. This may require truncations / approximations for which the error can be estimated. There exist several tensor network schemes, yet they all have a series of steps in common that we illustrate for a concrete example.

\begin{figure}[h!]
\begin{center}
  \includegraphics[width=0.65\textwidth]{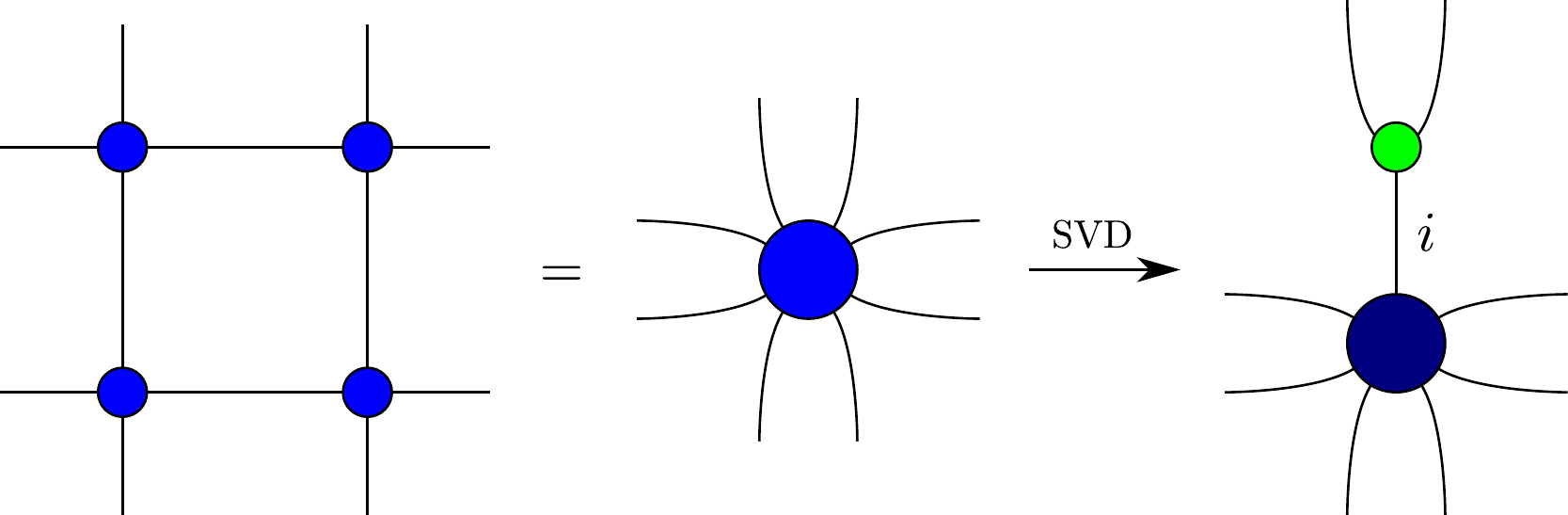}
  \caption{\label{fig:Emb_map_tensor}
  The embedding maps are computed from the contracted tensor with multiple indices. It is split apart into two tensors, connected by a new effective edge labeled by the singular values indicating the relevance of the degree of freedom. The green, three-valent tensor then serves as an embedding / coarse graining map for the fine degrees of freedom.
  }
\end{center}
\end{figure}

For simplicity, take a 2D quadratic tensor network. One step present in all tensor networks is an explicit summing of degrees of freedom, referred to contraction of indices. In our network we group together four tensors $T$ and sum over their shared indices and obtain a new tensor $\tilde{T}$, which has twice as many indices, yet the network remains local, see fig. \ref{fig:Emb_map_tensor}.
\begin{equation}
  \tilde{T}_{a_1 a_2 b_1 b_2 \dots d_1 d_2} := \sum_{i j k l} T_{b_1 i l a_1} T_{l k d_1 a_2} T_{b_2 c_1 j i} T_{j c_2 d_2 k} \; .
\end{equation}
We observe an immediate issue: if the original tensor had an index range of $\chi$, called the bond dimension, the new tensor has a range of $\chi^2$. Thus, while we evaluate the partition function in steps, we cannot continue indefinitely without truncations / approximations. To implement those, dynamical variable transformations are derived from the $\tilde{T}$ via a singular value decomposition. That way, we define effective coarse degrees of freedom as functions of the fine ones. Crucially the effective degrees of freedom are derived from the dynamics encoded in the tensors. This works as follows:

\begin{figure}[h!]
\begin{center}
  \includegraphics[width=0.6\textwidth]{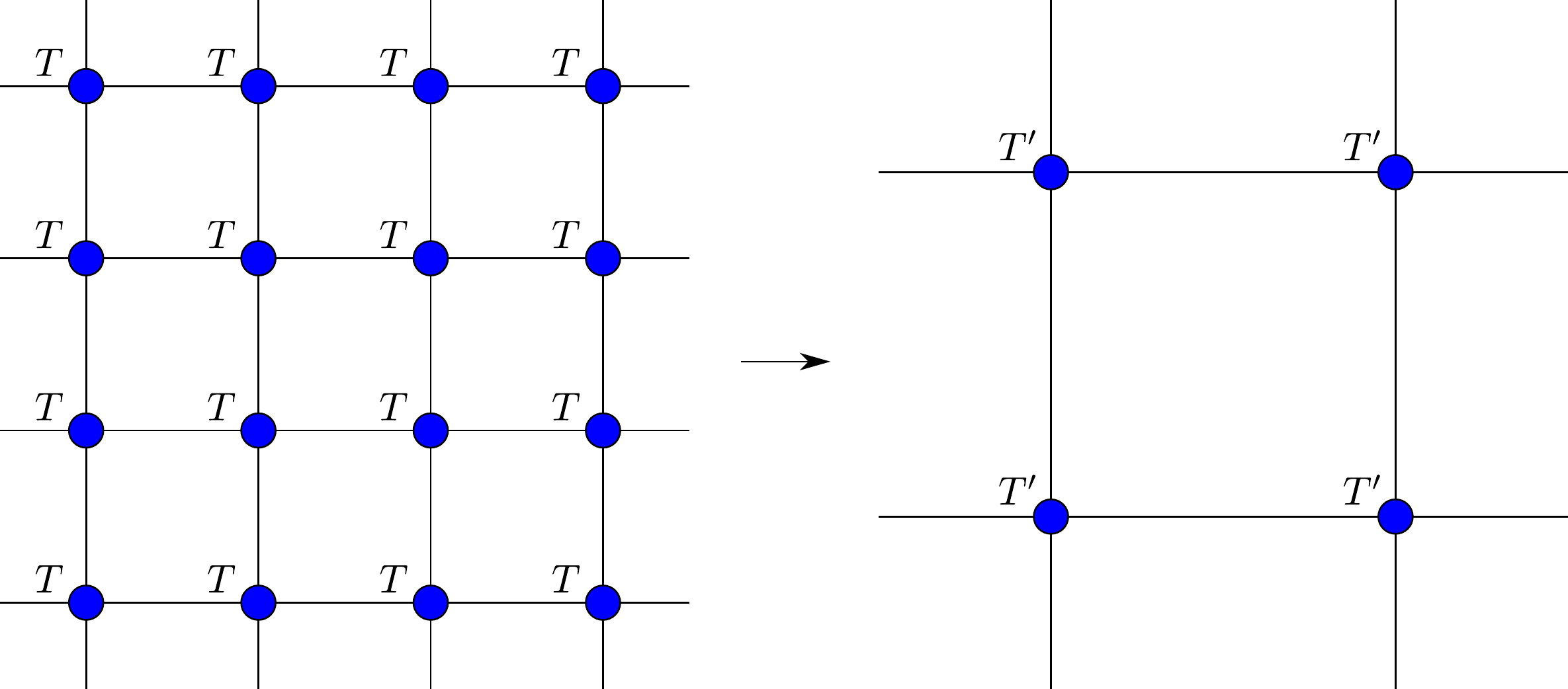}
  \caption{\label{fig:Tensor_network}
  The basic idea of tensor network renormalization: write the partition function as a contraction of a tensor network of tensors $T$ and then locally manipulate the tensors, such that the same partition is approximated by a coarser network of effective tensors $T'$. This defines a flow in ``tensor space''.
  }
\end{center}
\end{figure}

Given the new tensor $\tilde{T}$, we intend to map two indices into an effective one. To do so, we split $\tilde{T}$ in two, separating the strands $a_1,a_2$ from all other variables. This is generically not possible unless the tensor factorises. To split the tensor, we first rewrite it in terms of a matrix $M_{AB}$, where index $A = \{a_1,a_2 \}$ and $B$ contains all remaining indices. On this matrix we perform a singular value decomposition:
\begin{equation}
  \tilde{T}_{(a1,a2),(b_1,b_2,\dots,d_1,d_2)}=: M_{AB} = \sum_{i = 1}^{\chi^2} U_{A,i} \;  \lambda_i \; (V)^\dagger_{i,B} \; .
\end{equation}
The matrices $U$ and $V$ are unitary and contain the left and right singular vectors of $M$. $\lambda$ is a diagonal matrix of singular values, where $\lambda_1 \geq \lambda_2 \geq \dots \geq \lambda_{\chi^2} \geq 0$. See right side of fig. \ref{fig:Emb_map_tensor}.

\begin{figure}[h!]
\begin{center}
  \includegraphics[width=0.65\textwidth]{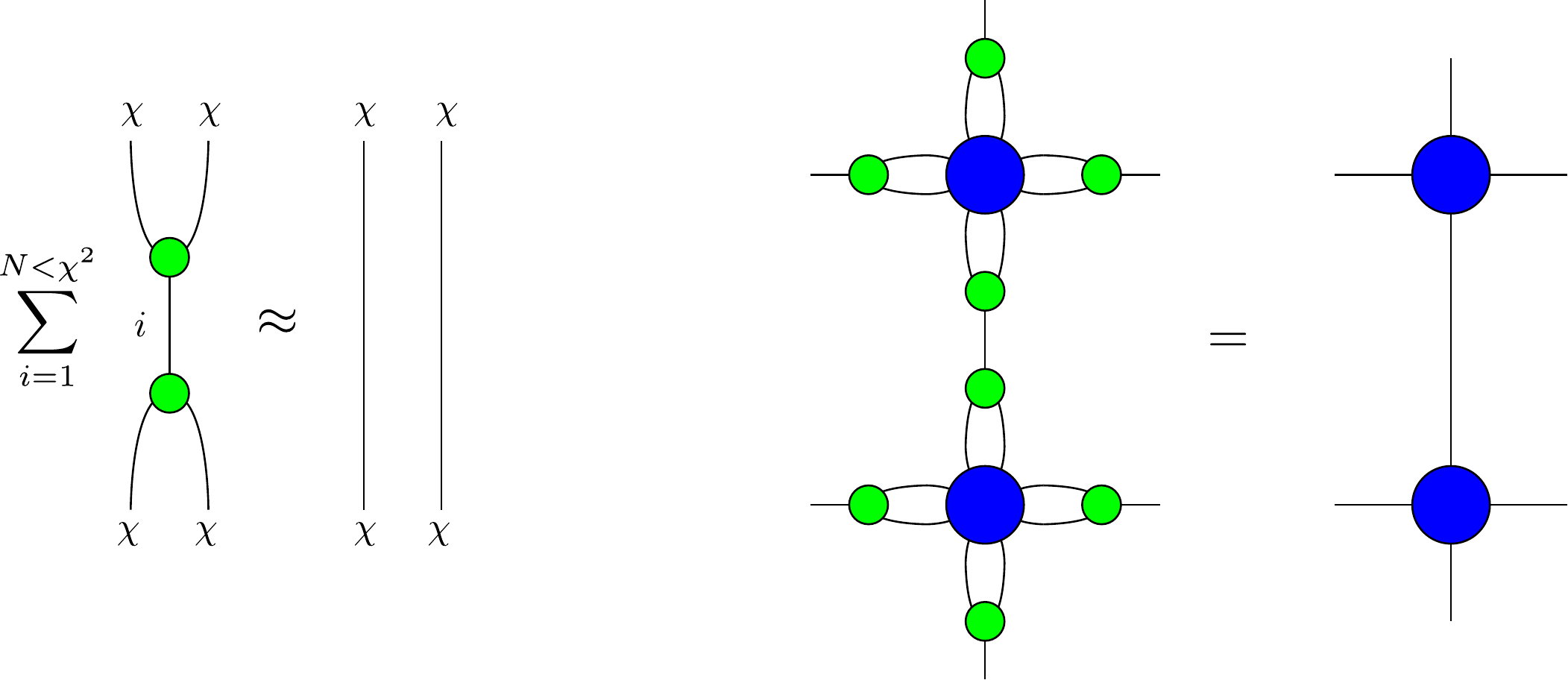}
  \caption{\label{fig:New_tensor}
  Left: In case the singular values for $i > N$ are negligible, the truncated maps $U^\dagger U$ approximate the resolution of identity well, such that inserting them inbetween the pair of indices barely changes the partition function. %For each pair of indices that are to be combined into an effective coarse degree of freedom, one introduces a pair of maps $U^\dagger U$, which change the partition function as little as possible.
  Right: Starting from the tensors in the middle of fig. \ref{fig:Emb_map_tensor}, we insert the truncated resolution of identity for each pair of edge. Then we obtain the new effective tensor by contracting the previous tensor with its respective embedding maps.
  %The new effective tensor is then defined as the contraction with these embedding maps.
  }
\end{center}
\end{figure}

If we translate the matrix indices $A$ and $B$ back into the original tensor indices, we see that the singular value decomposition allows to write the tensor $\tilde{T}$ as the contraction of a three-valent and a seven-valent tensor, where the summed index $i$ labels the singular values. The three-valent tensor $U$ encodes the desired variable transformation, translating the degrees of freedom $a_1,a_2$ into an effective coarse degree of freedom / index $i$. This transformation is exact, since $i$ (generically) has a range of $\chi^2$. Since $U$ is a unitary matrix, we can introduce resolutions of identity $U U^\dagger$ in the partition function (see left of fig. \ref{fig:New_tensor}) without changing it and sum over the indices $a_1,a_2$ as well as the indices $c_1,c_2$ on the opposite site. Then we repeat this procedure for the remaining indices to obtain a new effective tensor $T'$, see right side of fig. \ref{fig:New_tensor}.

Hence, we define a new effective tensor, yet its index range is still $\chi^2$, and we cannot continue to iterate this procedure without truncations. The singular value decomposition allows us to implement this truncation in an optimal way. Since all singular values are positive semi-definite and ordered in size, $\frac{\lambda_i}{\lambda_1}$ indicates how significant $i$ is with respect to the most significant one, $i=1$. Indeed, we can approximate the rank $\chi^2$ matrix $M$ by a rank $d$ matrix by ignoring all $\lambda_i$ with $i > d$. Crucially, in terms of the least squared error, this matrix is the best rank $d$ approximation of the matrix $M_{AB}$. Whether this is a good approximation can be readily inferred from the size of the singular values. Truncating the degrees of freedom $i$ directly translates into truncations on the variable transformations $U$ and the new tensor $T'$ respectively. The accuracy of the simulations are then determined by the bond dimension, i.e. the number of degrees of freedom kept in each iteration.

The algorithm briefly sketched above is deliberately chosen to showcase that tensor network renormalization provides a concrete realization of the spin foam coarse graining scheme. Firstly, it blocks together tensors and exactly sums over their internal degrees of freedom. Secondly, the singular value decomposition provides dynamical variable transformations that fulfill the role of dynamical embedding maps, and, moreover, allow for efficient and controllable truncations. One way to check whether these approximations are justified is to gradually increase the bond dimension, i.e. the number of kept singular values, to see whether the properties of the system depend on this choice, e.g. the position of a phase transition in parameter space. In case the results converge, the approximation is sufficiently good and one can extrapolate to infinite bond dimension. However, there exist situations in which no truncation should be implemented, in particular on second order phase transitions \cite{vidal-TNR}. There, one observes that more and more singular values are relevant the closer one tunes towards the phase transition, such that one would require an infinite bond dimension or equivalently infinitely fine boundary data. This is expected, since one models a highly non-local system by locally gluing amplitudes.% Thus, there is significant effort to apply these methods to 4D spin foam models of quantum gravity. We will review the studied models and results below.

We would like to highlight some general advantages and disadvantages of tensor network methods, as well as modifications to the method that so far are not applied to models of quantum gravity. Compared to other numerical methods, like Monte Carlo methods, tensor network algorithms do not suffer from the sign problem; the algorithms are perfectly applicable to quantum (oscillating) amplitudes, like in spin foams. The reason is that tensor networks do not rely on (random) sampling methods for a large system, but usually focus on all possible configurations of an amplitude for one building block (tensor). However, this leads to one of their disadvantages: in order to save all configurations of a tensor, it must have a finite dimensional boundary Hilbert space. Moreover, the numerical costs, both in terms of computational time and memory usage, scale with the dimension of the boundary Hilbert space. In particular for lattice gauge theories and spin foams in higher dimensions, this requires extensive optimization to make the numerical simulations feasible.

Before reviewing results in quantum gravity (related) models, we would like to highlight a few methods from the tensor network community. One key modification is called entanglement filtering \cite{GuWen,vidal-TNR,Hauru:2017jbf}. It removes entanglement between short-scale degrees of freedom, which would otherwise get promoted to larger scales and lead to unphysical fixed points in the renormalization group flow. Other modifications aim at including Monte Carlo methods into tensor network algorithms, e.g. for contracting tensor indices \cite{Ferris_2012} or to sample over the probability distribution of coarse degrees given by the singular values \cite{ferris2015unbiased}.

In the following, we first review works on tensor network renormalization applied to 2D analogue models. There we focus on the introduction of symmetry preserving methods that use the symmetry of the system to label the effective degrees of freedom with the original variables. In the second part, we discuss how to apply these methods to 3D lattice gauge theories and spin foam models, which require an efficient description of the model given by so-called decorated tensor networks.

\subsubsection{Analogue spin foam models in 2D}

By 2D analogue spin foam models, sometimes also called spin net models, we mean spin systems with a global symmetry. The typical example is the Ising model (with vanishing external magnetic field) that has a global $\mathbb{Z}_2$ symmetry. Typically these models are written in terms of group variables {color{red} $g_v \in G$} assigned to the vertices of the lattice, which only interact with their nearest neighbours expressed in ``edge weights'' $\omega_e(g_{s(e)} g_{t(e)}^{-1})$. In order to work with finite dimensional Hilbert spaces we restrict $G$ to be a finite group (or quantum group later on) \cite{Bahr:2011yc}. The partition function of the system is given by:
\begin{equation}
  Z = \sum_{g_v} \prod_e \omega_e(g_{s(e)} g_{t(e)}^{-1}) \; .
\end{equation}
To be invariant under the global symmetry, i.e. an element $h \in G$ acting on all vertices at once, these edge weights must satisfy
\begin{equation}
  \omega_e(h \, g \, h^{-1}) = \omega_e(g) \; \forall \; h \in G \; .
\end{equation}
Thus $\omega_e$ are class functions and
Since the function $\omega_e$ are invariant under conjugation, each one can be expanded via Peter Weyl's theorem \cite{liegroups} into a sum over irreducible representations $\rho$ of the character $\chi_\rho$ of $G$:
\begin{equation}
  \omega_e(g) = \sum_{\rho} \tilde{\omega}_\rho \, \chi_\rho(g) \; .
\end{equation}
$\tilde{\omega}_\rho$ stands for the group Fourier transform of the edge weight $\omega_e$.
Performing this for all edges and expanding the characters as a trace of representation matrices, the expression factorizes over all group elements $g_v$, such that the group integrations / summations can be performed analytically:
\begin{equation} \label{eq:projector}
\mathbf{P}_v (\{ \rho_e \}_{e \supset v}) := \sum_{g_v} \bigotimes_{e} {\rho_e(g_v)}^{m_e}_{n_e} \; .
\end{equation}
$\mathbf{P}_v$ denotes the Haar projector of the group $G$, i.e. the projector onto the invariant subspace. We suppress its many indices for clarity of the notation. After performing all group integrations / summations, the partition functions reads:
\begin{equation}
 Z = \sum_{\rho_e} \prod_e \tilde{\omega}_{\rho_e} \prod_v {\mathbf{P}_v} (\{ \rho_e \}_{e \supset v})
\end{equation}
Note that the indices of the Haar projectors $\mathbf{P}_v$ are contracted with projectors on neighbouring vertices. For more details on these models and their relation to spin foam models (with finite groups) see \cite{Bahr:2011yc}.

The expansion sketched here is completely analogous to the derivation of the spin foam representation familiar from spin foam literature. Thus, while the dimensionality is lower and the dynamics simpler, the dynamical ingredients - irreducible representations $\rho$ and projectors onto the invariant subspace $\mathbf{P}$ / intertwiners $\iota$ - are the same as for spin foam models. Moreover, it is expected that these 2D spin systems share statistical properties with the 4D gauge theories of the same group \cite{Kogut:1979wt}. As a final point, these models can be related to peculiar spin foams that only possess two vertices and many edges \cite{Dittrich:2013voa}.

\begin{figure}[h!]
\begin{center}
\includegraphics[width=0.7\textwidth]{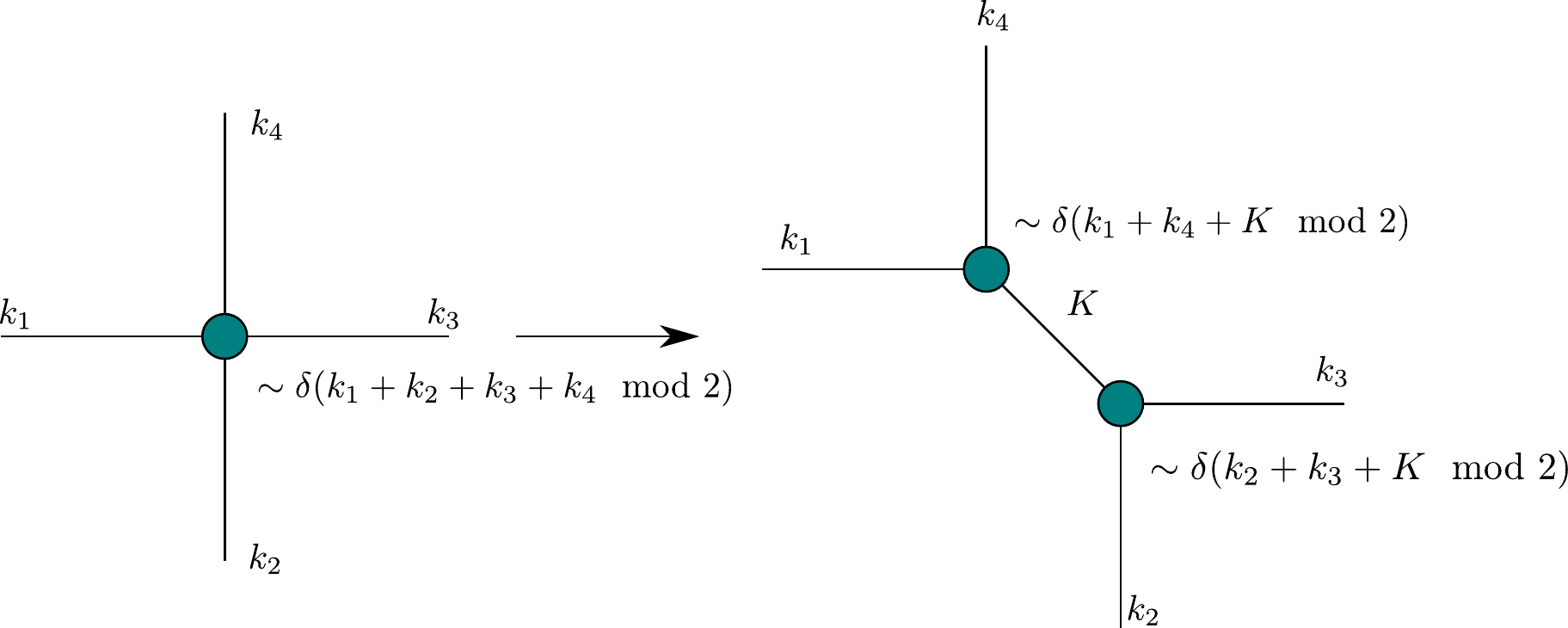}
\caption{\label{fig:Symmetry_pres}
By explicitly preserving the symmetry of the tensor, we assign an irreducible representation to the new effective edge, thus preserving also the original theory space. Here shown for the $\mathbb{Z}_2$ / Ising model case.
}
\end{center}
\end{figure}

Hence these models represent ideal test cases for applying tensor network renormalization to spin foam models and derive first hints for the RG flow of the full theory. Fortunately, the translation of the partition function into tensor network language is straightforward: the projectors $\mathbf{P}$ are essentially tensors, whose variables are the irreducible representations $\rho_e$ on the edges. Just the weights $\tilde{\omega}_{\rho_e}$ need be split per edge via a squareroot. While tensor network algorithms can be readily applied, it is vital to consider the symmetries encoded in $\mathbf{P}$: the irreducible representations $\rho_e$ meeting at the vertex $v$ must satisfy the coupling rules, i.e. they must couple to the trivial representation to satisfy gauge invariance. These restrictions can be used to optimize tensor network renormalization methods in two ways: firstly, by only storing and summing over configurations allowed by the coupling rules, the memory cost and numerical cost for the singular value decomposition and index contractions can be drastically reduced. For Abelian models this is straightforward, since all representations are one-dimensional and the projector $\mathbf{P} = \delta^{(G)}(\sum_{e \supset v} k_e)$ (modulo orientation of the edges). Thus, under splitting of a tensor, e.g. to define the variable transformations / embedding map, one defines an intermediate representation for the new effective edge satisfying the coupling rules for both tensors. This new representation will be the label of the effective degrees of freedom and thus allows us to explicitly preserve the symmetry, see fig. \ref{fig:Symmetry_pres}. Moreover, ordering the entries of the matrix according to the intermediate representation turns the matrix into a block diagonal form. Thus, the algorithm can be further optimized by performing a singular value decomposition individually for each block\footnote{A SVD of a $p \times q$ matrix with $q > p$ scales with $p^2 q$ in terms of computational time. Thus it is beneficial to perform multiple decompositions of smaller matrices.}.

For non-Abelian models, a further comment is necessary. As sketched in equation \eqref{eq:projector}, the tensor possesses ``magnetic'' indices $m,n$ per edge in addition to the irreducible representation $\rho$. More precisely, for fixed $\rho$, each edge carries the vector space $V_\rho \otimes V_{\rho^\ast}$, where $\rho^\ast$ denotes the dual representation to $\rho$. Since $V_{\rho^\ast}$ can be identified with the dual vector space $(V_\rho)^\ast$, we label each edge with a single representation. For tensor networks these magnetic indices pose a significant challenge: if we were to include them, the size of the tensor would render the simulations unfeasible. Fortunately, the dependence on these indices is entirely encoded in representation theory of the group $G$ and does not change under coarse graining. Essentially these indices get ``pre-contracted'' \cite{Dittrich:2013uqe,Dittrich:2013voa}, which is accounted for by $G$ recoupling symbols in the coarse graining equations. From these equations one can read off another feature of non-Abelian models: the renormalization group equations for the representations $\rho$ and $\rho^\ast$ are decoupled. Thus, it is possible that under coarse graining the effective edges will carry representations $(\rho,\rho')$ with $\rho' \neq \rho^\ast$.

Mentioning the channels $(\rho,\rho')$ is a good keyword to explain the flow as well as the approximation scheme. Due to the explicit symmetry preservation, the renormalized tensors are expressed in terms of the same variables as the original theory (with a slightly more general theory space). Thus, instead of directly comparing all entries of the tensors, we study the coarse graining flow by considering the singular values per channel $(\rho,\rho')$. This is completely sufficient to characterize the flow and read off different fixed points. Then, in order to determine which degrees of freedom are more relevant, one needs to compare the singular values from all channels and truncate accordingly. This can then result in a higher multiplicity of the same representation labels and thus more general boundary data, which improves the accuracy of the simulation. However, in most cases \cite{Dittrich:2013uqe,Dittrich:2013voa,Dittrich:2016tys} a simple scheme is used, where only the largest singular value per block is kept. While this is a strong simplification, it is sufficient to identify interesting phases, labeled by attractive fixed points of the flow, while keeping the simulations feasible, in particular for studies on quantum groups $\text{SU}(2)_\text{k}$ \cite{Dittrich:2013voa} and $\text{SU}(2)_\text{k} \times \text{SU}(2)_\text{k}$ \cite{Dittrich:2016tys}.

The articles \cite{Dittrich:2011zh,Dittrich:2011av} investigate spin net models for the Abelian finite groups $\mathbb{Z}_q$, for so-called cut-off models. The starting point are the edge weights in the zero-temperature limit with $\tilde{\omega}_k = 1 \; \forall k \in \{0,\dots,q \}$. These weights are then truncated at different levels $k$, which breaks the topological symmetry and it is investigated whether this symmetry is restored under coarse graining. While for low-$k$ and high-$k$ cut-off the high and low temperature fixed points are found respectively, there exist intermediate phases showing oscillating behaviour. In \cite{Dittrich:2013uqe} tensor network methods are generalized to non-Abelian groups applied to spin nets for $\mathcal{S}_3$, the permutation group of three elements. To keep these models feasible, the coupling rules are heavily used to optimize the algorithm. The models investigated build upon a holonomy representation of spin foam models \cite{Bahr:2012qj} and their implementation of simplicity constraints. In general they find a non-trivial phase diagram of three phases, a low temperature $\mathcal{S}_3$ ordered phase, a high temperature $\mathcal{S}_3$ disordered phase as well as a $\mathbb{Z}_2$ ordered phase.

As a next step \cite{Dittrich:2013voa}, finite non-Abelian groups are replaced by the quantum group $\text{SU}(2)_\text{k}$ with the deformation parameter $q = \exp(\frac{2 \pi i}{\text{k}+2})$ a root of unity\footnote{
Quantum groups do not allow for a holonomy representation. However, its representation theory is close to the one of $\text{SU}(2)$, such that one defines these models directly from the high temperature expansion of spin net models.
} \cite{biedenharn,yellowbook}. The advantage is that the integer level $\text{k}$ defines a gauge-invariant cut-off $j_{\text{max}} = \frac{\text{k}}{2}$. That way it is possible to study systems with more degrees of freedom by increasing the level $\text{k}$, while the representation theory remains similar. Moreover, one eventually approaches full $\text{SU}(2)$ as $\text{k} \rightarrow \infty$. Moreover, quantum groups are physically motivated from 3D spin foam models, where they describe gravity with a non-vanishing cosmological constant \cite{Turaev:1992hq}. The models studied in \cite{Dittrich:2013voa} are constructed from so-called intertwiner model fixed points \cite{Dittrich:2013aia}, which represent topological field theories. In a nutshell, intertwiner models are ``half'' of a spin net model, with an edge Hilbert space of $V_\rho$ instead of $V_\rho \otimes V_{ \rho ^ \ast}$. These models are interesting since one can directly investigate whether the two copies remain coupled or decouple under coarse graining. Indeed one finds a rich phase structure with potential second order phase transitions.

Eventually, the work \cite{Dittrich:2016tys} investigates spin net models for $\text{SU}(2)_k \times \text{SU}(2)_k$ that mimic the construction of 4D Riemannian spin foam models, namely the Barrett-Crane \cite{Barrett:1997gw} and EPRL / FK model \cite{Engle:2007wy,Freidel:2007py}. For the BC model several attractive fixed points are found, none of which correspond to topological BF theory. While this indicates that simplicity constraints are strongly implemented, no indications for a 2nd order phase transition are observed. This hints towards the fact that the constraints might be too strongly implemented \cite{Alesci:2007tx}. In contrast, the EPRL / FK model shows a highly intricate flow and partially oscillating behaviour, most likely due to exciting only a few representations initially. This is a particularity of the implementation of the simplicity constraints in the Riemannian EPRL / FK model, which relate $\text{Spin}(4)$ representations $(j^+,j^-)$ to an $\text{SU}(2)$ representation $k$ via $j^\pm = \frac{1}{2} \, k \, |1 \pm \gamma|$. Note that $j^\pm$ as well as $k$ must be half integer, such that $\gamma$ must be rational.

These results impressively show the potential of tensor network techniques for studying the renormalization group of spin foam quantum gravity. Moreover, they lead to the development of key optimizations and insights that are crucial for going to higher dimensional gauge systems. This is the subject of the next section.

\subsubsection{Decorated tensor networks for lattice gauge theories and spin foams}

Dimensions larger than $d=2$ and lattice gauge theories pose challenges for tensor network renormalization methods. Since higher dimensional tensors carry more boundary data, the algorithm generically is more costly than its lower dimensional counterpart. For lattice gauge theories, where due to the local gauge symmetry many degrees of freedom are redundant, it is thus imperative to develop an optimal representation if one intends to cast them into a tensor network form. Moreover, these networks are generically more complex than spin systems, since several data are shared among more than two building blocks. One example are spin foams in four dimensions, where a face and the representation it carries are shared among multiple 4-simplices. A possible tensor network representation is to assign a tensor dual to each 4-simplex, yet one must introduce auxiliary tensors \cite{Liu:2013nsa} to ensure the correct identification of shared variables. See \cite{Dittrich:2014mxa} for a more extensive discussion of possible representations.

To improve on these representations, decorated tensor network algorithms are developed and introduced in \cite{Dittrich:2014mxa}. The idea is to shift the perspective away from a pure tensor network representation of the system towards amplitudes with more intricate boundary data. Yet the key ideas are retained to explicitly contract bulk degrees of freedom and to dynamically define effective degrees of freedom using a singular value decomposition. Instead of tensors, represented by vertices and legs, one works with a spin foam inspired representation where amplitudes are assigned to regions, which carry boundary data, e.g. spin network data. While the assignment of amplitudes remains local, the non-local nature of gauge theories requires more complex boundary data and intricate gluing rules that cannot be cast in a simple tensor network form without introducing additional structures. The rest of the algorithm remains essentially the same: amplitudes are glued together by suitably identifying variables among them. Depending on the considered situation, some of the identified variables are not summed over and remain part of the boundary Hilbert space. On this fine amplitude one performs a singular value decomposition to derive an embedding map leading to coarse effective degrees of freedom.

\begin{figure}[h!]
\begin{center}
\includegraphics[width=0.9\textwidth]{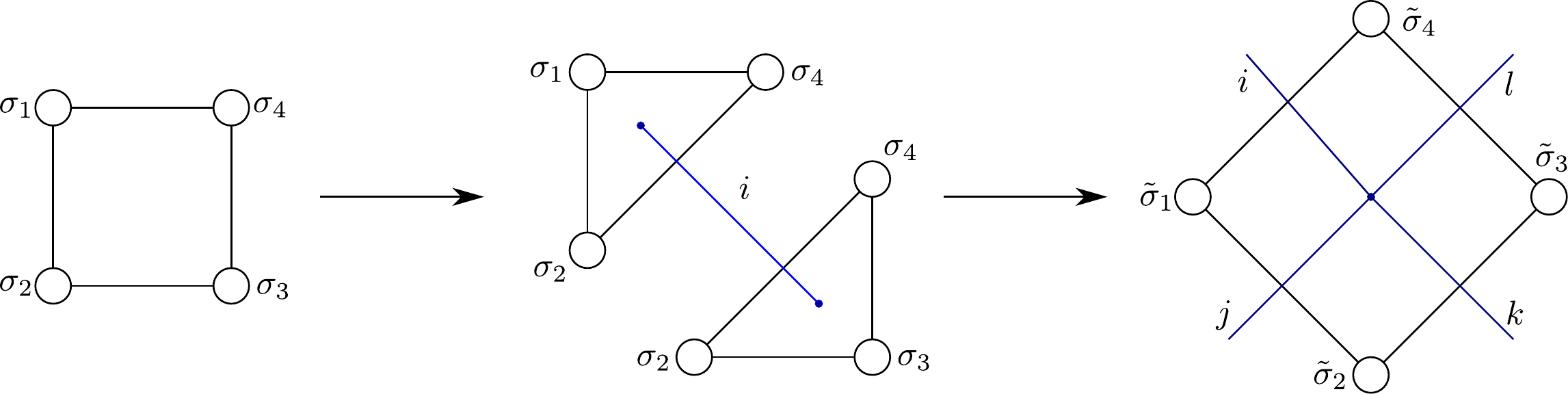}
\caption{\label{fig:Decorated_TNW}
Illustration of decorated tensor networks. Splitting an amplitude, here for four Ising spins, via a singular value decomposition generically gives rise to additional indices. These indices are understood as more general boundary data and are encoded in a tensor network dual to the lattice. This network is thus ``decorated'' by the remaining data.
}
\end{center}
\end{figure}

The original algorithm in \cite{Dittrich:2014mxa} works slightly differently by ``splitting'' the amplitudes explicitly. Let us briefly demonstrate the idea for the usual 2D Ising model, for which a decorated tensor network algorithm exists as well. Consider an amplitude assigned to a square given by four Ising spins $\sigma_i$, $i \in \{1,\dots,4\}$, $\mathcal{A}(\sigma_1, \dots, \sigma_4)$, see fig. \ref{fig:Decorated_TNW}. The idea of the algorithm is alternatingly split the squares into regular triangles, such that four of these form a coarse, rotated square with a single Ising spin to sum over in their center. To do so we split the amplitude in two, separating the dependency on the spins opposite to the cut. For the singular value decomposition, we need to distinguish two sets of variables: the variables that we want to separate are encoded in the two indices of the matrix to decompose, while the shared variables will remain fixed similar to the symmetry preserving algorithm before. Hence, we perform a singular value decomposition for each configuration of shared variables, which is more efficient than a decomposing a big matrix. In a sense, this leads to a doubling of the shared variables, which is necessary for gluing them again in consecutive iterations. In our example, we get:
\begin{equation}
  M_{\sigma_1,\sigma_3}^{(\sigma_2,\sigma_4)} := \sum_{i=1}^2 U^{(\sigma_2,\sigma_4)}_{\sigma_1,i} \; \lambda_i \; (V^\dagger)^{(\sigma_2,\sigma_4)}_{i,\sigma_3} \; .
\end{equation}
By assigning a square root of the singular values $\lambda_i$ to $U$ and $V$, we derive the desired amplitudes assigned to the triangles. Note that each amplitude is not just given by the configuration of three Ising spins, but also by an additional index $i$, assigned to the coarse edge. When combing four triangular amplitudes, the resulting amplitude for the square is given by more general boundary data, four Ising spins and four new indices $\mathcal{A}(\sigma_1, \dots, \sigma_4,i_1,\dots,i_4)$, again see fig. \ref{fig:Decorated_TNW}. Since these indices are shared with neighbouring amplitudes, we represent them by a tensor network on the lattice dual to the squares. Thus, we have a tensor network encoding higher order corrections / more general boundary data ``decorated'' by the original boundary data of the system.

An algorithm is developed for 3D lattice gauge theories and first applied to Abelian $\mathbb{Z}_2$ lattice gauge theory in \cite{Dittrich:2014mxa}. Instead of working with the original lattice, one works with the dual lattice in the strong coupling expansion. Thus the variables are irreducible representations $k_e$ of $\mathbb{Z}_2$ one edge of the dual lattice and Gauss constraints ($\mathbb{Z}_2$ $\delta$-functions) on each face. The Gauss constraints can be explicitly solved to reduce the amount of data saved, and there is a freedom to choose which variables $k_e$ to gauge fix. This choice is adapted to the intended splitting.

Let us briefly sketch the coarse graining algorithm: the idea is to cut cubes in half by cutting along the diagonal of one of its faces. Four of these amplitudes are glued together to form a new (distorted) cube. This coarse graining is continued in the other directions by ``rotating'' the amplitude and iterating the procedure. Compared to the 2D algorithms, the splitting is slightly more complicated. In order to cut the cube along the face, it is necessary to introduce another representation along the intended ``cut''. This variable serves as the variable assigned to the coarse edge, and is introduced by splitting the Gauss constraint on the square face into two assigned to the two triangles. Then, the variables are gauge fixed such that equally distributed the remaining degrees of freedom are equally distributed among the split amplitudes, and some shared by both. As explained above, the shared variables will be kept fixed during the singular value decomposition and label the new amplitudes. Also, again more general boundary data arise in the form of a decorated tensor network due to this splitting, which will be assigned to edges and faces, see \cite{Dittrich:2014mxa} for more details.

In its lowest order approximation, i.e. when truncating all tensor indices, the algorithm reproduces the phase diagram of $\mathbb{Z}_2$ lattice gauge theory with a strong and weak coupling phase, whereas the critical coupling is found within an error of a few percent compared to Monte Carlo simulations \cite{Binder:2001ha}. These results are improved by keeping more degrees of freedom after the singular value decomposition, yet the computational costs grow quickly\footnote{\cite{Dittrich:2014mxa} also proposed an algorithm based on smaller building blocks that generically are more efficient, see also the ``triangular'' algorithm in 2D \cite{Dittrich:2016tys,Steinhaus:2015kxa}.}. As one of the first tensor network algorithms applied to 3D lattice gauge theories it already shows promising qualitative results.

In \cite{Delcamp:2016dqo} this algorithm is generalized to non-Abelian symmetry groups and applied to $\mathcal{S}_3$ lattice gauge theory. While the basic idea and principle of the algorithm remains similar, it is significantly more complicated due to the non-Abelian group. The basic steps, splitting, gluing and choice of variables, are still in place, however in order to define them transformations between the holonomy and spin network representation are necessary. For the details, we refer the reader to the extensive and thorough explanation of the technical details in \cite{Delcamp:2016dqo}. Let us focus instead on the results.

As for the similar work on $\mathcal{S}_3$ spin nets \cite{Dittrich:2013uqe}, analogous simplicity constraints for $\mathcal{S}_3$ are implemented in the holonomy representation \cite{Bahr:2012qj}. The coarse graining flow is studied and three different phases found, as in \cite{Dittrich:2013uqe}, that correspond to a strong coupling $S_3$ phase, a weak coupling $S_3$ phase as well as a weak coupling $S_3 / \mathbb{Z}_3 \simeq \mathbb{Z}_2$ phase. The successful generalisation to non-Abelian groups as well as the derivation of the phase diagram of the theory demonstrate the potential of decorated tensor network techniques. However, this work also revealed a short-coming of using the spin network basis for labeling the boundary Hilbert space.

\begin{figure}[h!]
	\begin{center}
		\includegraphics[width=0.65\textwidth]{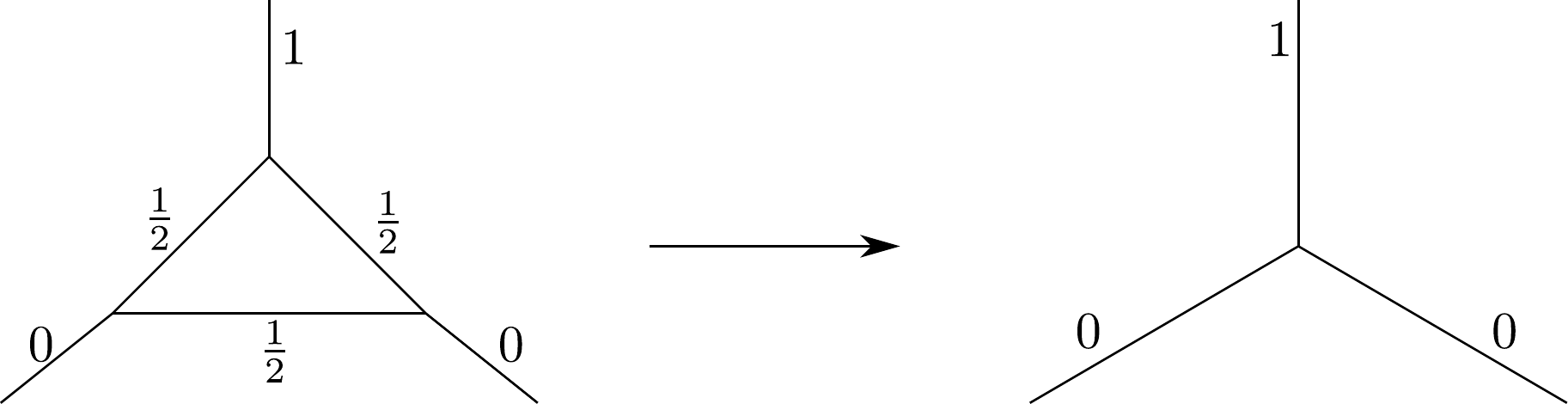}
		\caption{\label{fig:SNW-unstable}
			Gauss constraint violations under coarse graining (for $\text{SU}(2)$): we can define an effective vertex by summing over the $\text{SU}(2)$ representations associated to the inner edges. The configuration on the left is allowed by the coupling rules, thus the effective vertex on the right is allowed. However, this configuration is forbidden by $\text{SU}(2)$ coupling rules.
	}
	\end{center}
\end{figure}

As we discuss in great detail in this review, at some step of the coarse graining process one sums over fine degrees of freedom. In the spin network basis, where one assigns irreducible representations to the links and intertwiners to the nodes, this implies defining an effective vertex /  intertwiner by summing over representations, see fig. \ref{fig:SNW-unstable}. However, usually the coupling rules at the effective vertex are violated such that the vector associated to the vertex is not an intertwiner any more and gauge invariance is broken. This is a well-known shortcoming of the spin network basis under coarse graining \cite{Livine:2013gna,Charles:2016xwc} and it is also expected for lattice gauge theories\footnote{In order to define coarse grained fluxes in the discrete, the fine fluxes must be parallel transported to the same point. If the connection has curvature, the fluxes do not necessarily close any more, which is often referred to as curvature induced torsion.}. This can be overcome by a different representation of the boundary Hilbert space that can accommodate Gauss constraint violations (electric charges) as well as curvature excitations (magnetic fluxes). In 3D this is accomplished by the so-called fusion basis.

Here we will only briefly sketch the main features of the fusion basis, which arises in anyon systems \cite{Koenig_2010,Hu:2015dga}, (2+1)D lattice gauge theories \cite{Delcamp:2016yix} and 3D quantum gravity \cite{Dittrich:2016typ,Dittrich:2018dvs}. The algebraic structures are called Drinfeld Doubles, see \cite{Dittrich:2016typ,Delcamp:2016yix,deWildPropitius:1995hk,Kitaev:1997wr,Bombin:2007qv,Levin:2004mi} for more details. Its main feature is that it diagonalizes a set of commuting operators, so-called Ribbon operators. These Ribbon operators, which contain both a Wilson loop operator as well as a t'Hooft operator, measure both the magnetic (curvature) as well as electric (torsion) excitations. Such excitations are localized on punctures carrying the magnetic and electric excitation. In lattice gauge theory, one can imagine one puncture per plaquette of the lattice. Ribbon operators surrounding a single or a collection of plaquettes then measure excitations associated with the puncture or collection of punctures.

\begin{figure}[h!]
\begin{center}
	\includegraphics[width=0.45\textwidth]{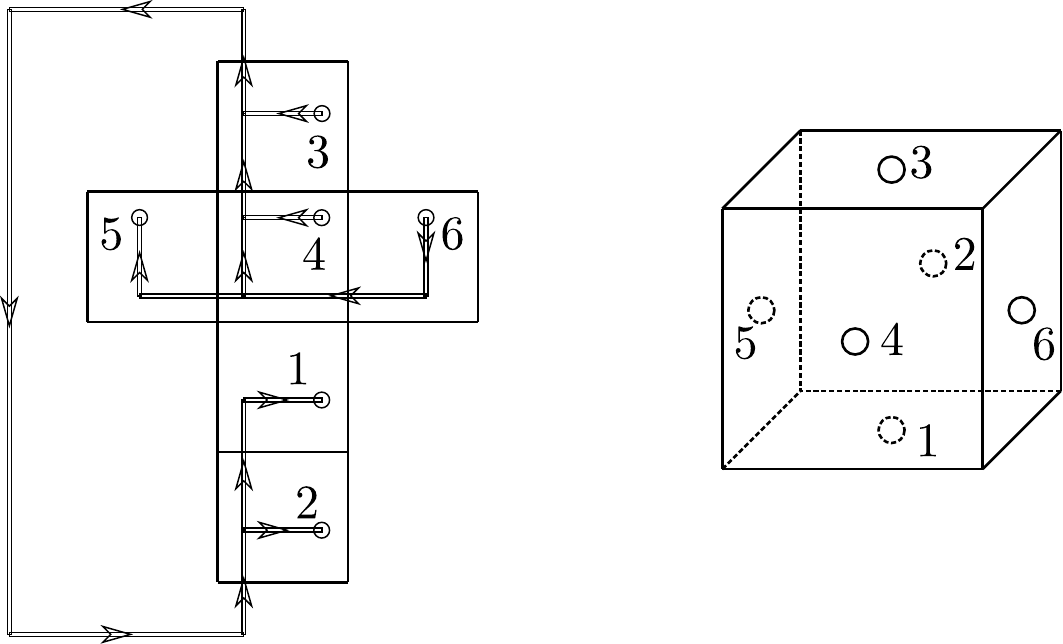}
	\caption{\label{fig:Cube-states}
		An example of a fusion tree for a 3D cube with six punctures. The tree is drawn on the boundary of the 3D cube, here ``unfolded''. The order of the fusion of punctures determines which set Ribbon operators is diagonalized.
	}
\end{center}
\end{figure}

These Ribbon operators commute among each other as long as they do not intersect. Hence, two Ribbon operators that surround a single puncture each commute with each other trivially, and they also commute with the operator surrounding both punctures. A choice of such a set of commuting Ribbon operators is encoded into the fusion basis by the choice of a fusion tree. The plaquettes are the leaves of the tree, and the connectivity of the tree determines which operators / observables are diagonalized by this choice of basis, see fig. \ref{fig:Cube-states}. Moreover, the basis states can be transformed into one another, such that one can translate states to diagonalize the observables one intends to measure. This is a crucial concept that has a notion of coarse graining built into it. Imagine two cubes glued together: in order to derive an effective building block with effective degrees of freedom, one would like fuse the punctures of subdivided faces into one. To do so, the fusion basis must be chosen such that it diagonalizes the Ribbon operator around punctures. This ensures that the expecation values agree in both original and coarse grained case. Therefore, the fusion tree can be used to encode a choice of coarse grained observables, which is crucial for the decorated tensor network algorithm based on the fusion basis.

Such an algorithm is defined for quantum deformed 3D lattice gauge theories on a cubic lattice for the quantum group $\text{SU}(2)_\text{k}$ in \cite{Cunningham:2020uco}. More details about the fusion basis for $\text{SU}(2)_\text{k}$ can be found in \cite{Dittrich:2016typ}. As in previous decorated tensor network algorithms, the basic ingredient is the amplitude associated to a cube. Its boundary Hilbert space is spanned by six-puncture states, for which a fusion tree needs to be specified. Then two cubes are glued together, which results in a cuboid with four course faces, each carrying two punctures. The goal is to compute an embedding map that fuses the two punctures on a subdivided face into one effective puncture. Generically, after gluing the fusion basis of the cuboid is not suited to do so, since it does not diagonalize the Ribbon operator surrounding both punctures. Thus, the basis must be transformed by a series of tree transformations involving $\text{SU}(2)_\text{k}$ recoupling theory, see \cite{Cunningham:2020uco} for details.

\begin{figure}[h!]
	\begin{center}
		\includegraphics[width=0.6\textwidth]{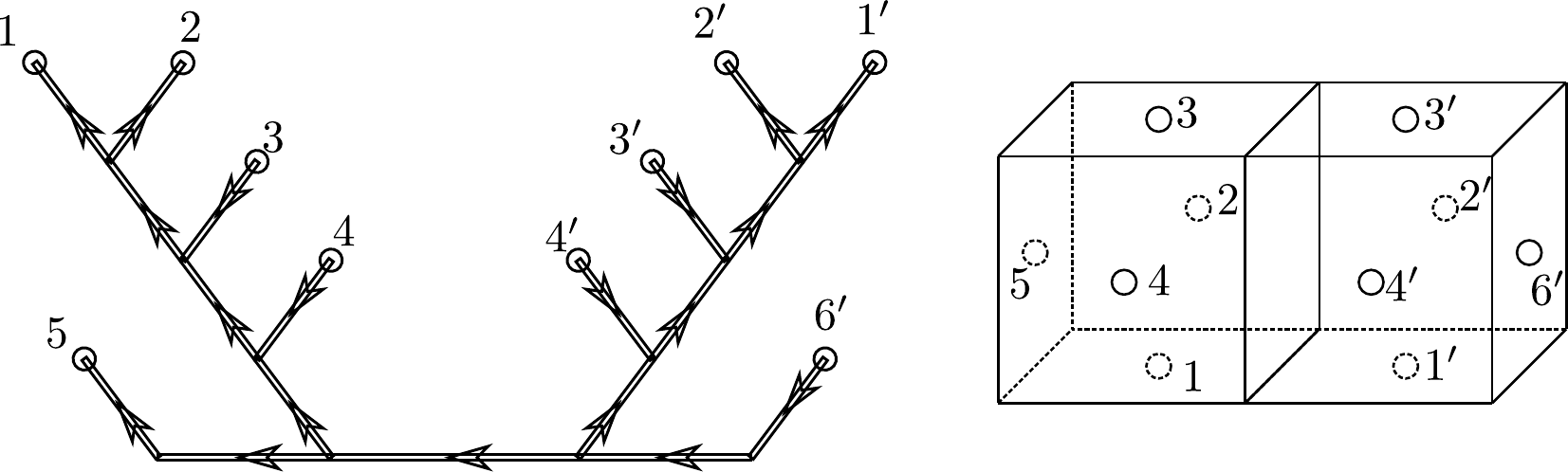}
		\caption{ \label{fig:bare_tree}
			Example for why tree transformations are necessary: after gluing the tree is not suited for coarse graining pairs of punctures $i,i'$. It must be transformed such $i,i'$ are directly fused together. See the glued cubes on the right as reference.
			}
	\end{center}
\end{figure}

Once one has arrived at a fusion basis that directly fuses the punctures as desired, one performs a singular value decomposition that splits the data of two punctures from all other variables (similar to section \ref{sec:TNW}). In order to label the new puncture with the usual data, one keeps fixed the data of fusion tree directly after fusing the punctures. These data label the effective puncture, while the singular value decomposition gives the weight of these data in the effective amplitude.  Also from the perspective of (coarse) observables this choice is viable, since the eigenvalue of a Ribbon operator surrounding two punctures is solely determined by these data. This allows us to use tensor network renormalization to approximately compute expectation values of Ribbon operators that we explain below.

After computing the embedding maps, they are used to coarse grain the punctures in such a way that the partition function is altered as little as possible (see again section \ref{sec:TNW}). Again, only one non-vanishing singular value is kept per puncture label, such that the same boundary Hilbert space is pertained. Once all pairs of punctures are coarse grained, one obtains again an amplitude of six punctures that is associated with a cuboid. To complete one coarse graining iteration, the same procedure is performed in the other spatial directions. The cube is ``rotated'' by reordering the punctures and the same procedure repeated in all directions to arrive at a coarse cube with six punctures.

In \cite{Cunningham:2020uco} this algorithm is applied to 3D lattice gauge theories defined for the quantum group $\text{SU}(2)_\text{k}$. As for quantum group analogue spin foam models \cite{Dittrich:2013voa,Dittrich:2016tys}, the quantum group introduces a gauge-invariant cut-off on the irreducible representations $j_\text{max} = \frac{\text{k}}{2}$. Thus, the boundary Hilbert spaces are finite dimensional and it is possible to study larger ``groups'' by increasing the level $\text{k}$ (and approach $\text{SU}(2)$ in the limit $\text{k} \rightarrow \infty$). The lattice gauge theory is modeled via a Heat kernel action for $\text{SU}(2)_\text{k}$ parametrised by a gauge coupling parameter $g$. Lastly, in the initial state each puncture only carries magnetic excitations as it usually is the case in lattice gauge theory.

Let us summarize a few of the main results: at each level $\text{k}$ there are two phases separated by a phase transition given by a critical coupling $g_c$. For $g < g_c$, the system flows to the weak coupling fixed point $g=0$ and is thus characeterized as the deconfined phase. Conversely for $g > g_c$ it flows to strong coupling $g \rightarrow \infty$, which describes the confined phase. The position of the critical coupling $g_c$ depends on the level $\text{k}$ and decreases apparently linearly for small $\text{k}$. This tentatively suggests that for $\text{SU}(2)$, i.e. the limit $\text{k} \rightarrow \infty$, $g_c \rightarrow 0$ such that only the confining phase exists. Additionally, the fusion basis permits to track the appearance of electric excitations that get excited under coarse graining, even though the initial state had no electric charges. While they do not appear to be vital for the dynamics, e.g. the position of the phase transition is barely affected if electric charges are completely truncated, including electric charges is important for the behaviour of the coarse graining flow as they serve as (non-dynamical) disentangling maps. See \cite{Cunningham:2020uco} for more details.

The final result we would like to mention is the expectation value of observables, here of Ribbon / Wilson loop operators. Since the fusion basis diagonalizes Ribbon operators, it is straightforward to approximately compute the expectation value of coarse Ribbon operators, i.e. Ribbon operators surrounding a larger number of plaquettes. In lattice simulations one usually has to simulate the entire system in order to measure coarse observables. Here, we first coarse grain the amplitude to arrive at an effective amplitude for the coarse cube, for which we measure the coarse Ribbon operator around the coarse plaquette. Thus, we first coarse grain / integrate out the fine degrees of freedom and account for them (with some truncations) in the effective amplitude, for which we then calculate the expectation values of the operators. Using this method, we derive different scaling behaviours of the expectation value with the enclosed area of the plaquette, in particular we recover the area law of the Wilson loop in the confined phase.

\subsection{Restricted, semi-classical path integrals} \label{sec:restricted}

Despite the tremendous progress in developing tensor network methods for spin foam models and lattice gauge theories, applying them directly to spin foam models of 4D quantum gravity (either Riemannian or Lorentzian) is still out of reach, in particular for a continuous symmetry group. An attempt to make the 4D Riemannian spin foam models accessible is to study simpler models that represent a subset of the full gravitational path integral. These simplifications include restricting the degrees of freedom to specific intertwiners and representations as well as using asymptotic expansions of spin foam amplitudes valid only for large representations. Let us explain these assumptions in more detail.

Intertwiners, which determine the shape of dual 3D building blocks, can be expressed in terms of Perelomov coherent states / Livine-Speziale intertwiners \cite{perelomov,Livine:2007vk}: to each face of the intertwiner one assigns an $\text{SU}(2)$ coherent state $|j,\vec{n}\rangle$, where $j$ labels the irreducible representation and $\vec{n}$ is a vector on $S^2$. This vector is a maximum weight state diagonalizing the angular momentum operator $J_{\vec{n}} = \vec{n} \cdot \vec{J}$ in $\vec{n}$ direction. Given these states, the coherent interwiner is given by:
\begin{equation}
  |\iota \rangle := \int_{\text{SU}(2)} dg \, g \, \triangleright \, \bigotimes_{i=1}^N |j_i,\vec{n}_i \rangle \; .
\end{equation}
Each coherent state $\sim |j_i,\vec{n}_i\rangle$ represents a face with area $\sqrt{j_i(j_i+1)}$ peaked on a normal vector pointing in direction $\vec{n}_i$. The tensor product of these coherent states represents a 3D quantum building block sharply peaked on a classical geometry (if it exists) with areas and outward pointing normal encoded in the labels $j_i$ and $\vec{n}_i$. The group integration (with Haar measure $dg$) defines an intertwiner. Note that the spin foam partition function itself can also be expressed in terms of coherent states and an integral over the labels of coherent states, e.g. see the review \cite{Perez:2012wv} for more details.

These coherent states play an important role in deriving semi-classical expressions for spin foam (vertex) amplitudes. When computing the vertex amplitude as a contraction of coherent intertwiners, these can be rewritten as several group integrations of contracted $\text{SU}(2)$ coherent states. The latter part is then exponentiated and the group integration performed via a stationary phase approximation:
\begin{equation}
  \mathcal{A}_v = \int_{\text{SU}(2)^E} \prod_{e} dg_e \prod_{f \supset e} \langle j_{ab}, -\vec{n}_{ba} | g_b^{-1} g_a | j_{ab}, \vec{n}_{ab} \rangle
  =: \int_{\text{SU}(2)^E} \prod_{e} dg_e e^{\sum_{f \supset e} 2 j_{ab} \ln \langle -\vec{n}_{ba} | g_b^{-1} g_a | \vec{n}_{ab} \rangle} \; .
\end{equation}
Since the stationary phase approximation is only valid when the argument in the exponential is highly oscillating, all representation $j_f$ must be large. Hence this expansion is often called the large-$j$ limit. For single vertex amplitudes it is shown that the ``action'' in the exponential evaluated on stationary and critical points is given by the Regge action of the building block dual to the vertex \cite{Conrady:2008mk,Barrett:2009gg}.

Given these familiar results from spin foam literature, the idea is to restrict the spin foam partition function of the EPRL / FK model to specific coherent intertwiners (and representations) and use only the amplitudes derived in the large-$j$ limit. That way the system depends on significantly fewer variables and the spin foam amplitudes, in particular the vertex amplitude, can be expressed in terms of closed formulas of the representations. Additionally, in the large-$j$ limit the sum over representations can be approximated by an integral. The motivation is to employ numerical integration techniques, e.g. the Cuba package \cite{Hahn:2004fe}. From now on, we only consider and discuss models defined on a 2-complex with hypercubic combinatorics which makes iterating the coarse graining steps straightforward.

\begin{figure}[h!]
\begin{center}
\includegraphics[width=0.75\textwidth]{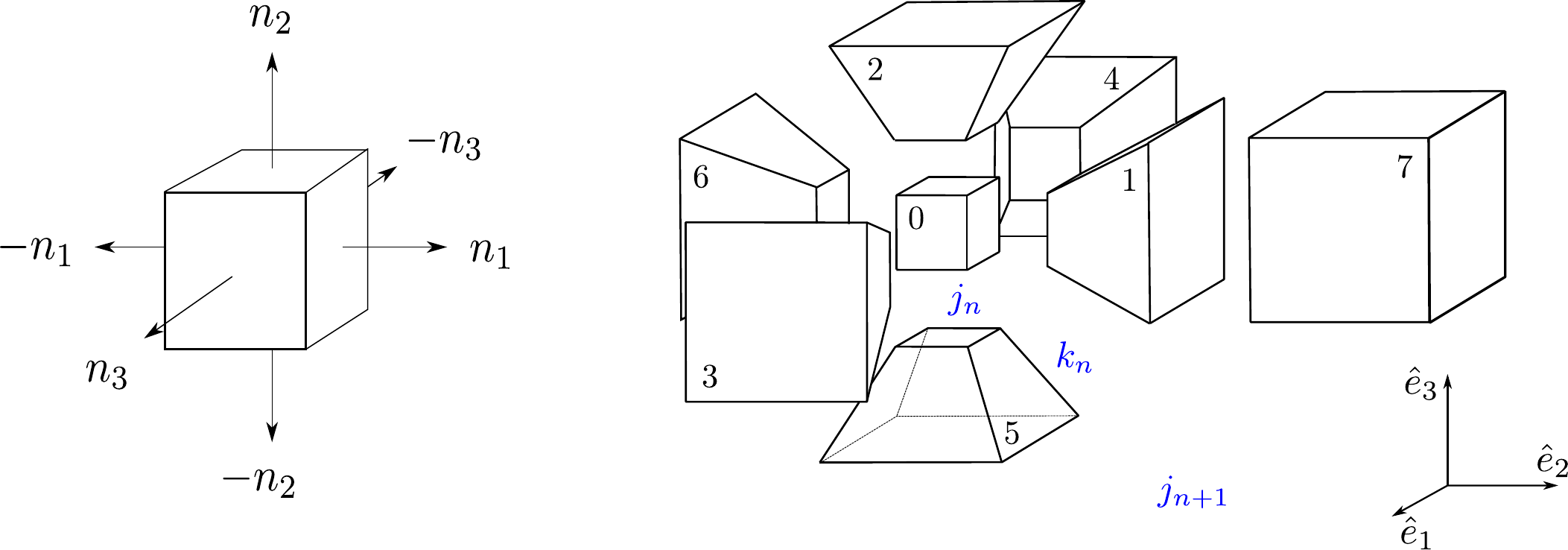}
\caption{\label{fig:restricted_ints}
Left: cuboid intertwiners intertwiners are peaked on the classical discrete geometry of a 3D cuboid. Right: the construction of the frusta vertex amplitude from the contraction of intertwiners.
}
\end{center}
\end{figure}

So far, two models of restricted spin foams are defined that are also studied under coarse graining. The first one are so-called quantum cuboids \cite{Bahr:2015gxa}, where the intertwiners are sharply peaked on a classical cuboid geometry. Opposite faces of the intertwiner carry the same representation and their normals are anti-parallel. Moreover, the outward pointing normal of a face is orthogonal to all normals of the adjacent faces, see the left part of fig. \ref{fig:restricted_ints}. Indeed these are severe restrictions, in particular the requirement that opposite faces in each intertwiner carry the same representation translate through the entire lattice. The asymptotic expansion of the vertex amplitudes depends again on the Regge action, which generically vanishes for cuboid configurations. For larger complexes this implies that the flat cuboid building blocks are glued in a flat way. Thus, this model describes a superposition of flat discrete space-times of different distribution of sizes across its building blocks. While this is by no means a realistic model of quantum gravity, it captures an Abelian subgroup of diffeomorphisms corresponding to shifts of entire hypersurfaces in the lattice\footnote{
Given a flat space-time decomposed into flat hypercuboid lattices, it should not matter how the 4-volume is split among the building blocks.
}.
Notably, this spin foam model is not invariant under these transformations \cite{Bahr:2015gxa}. Due to its simplicity it does not have any free parameters, thus an additional parameter $\alpha$ is introduced in the face amplitude of the model, $(2 j_f)^\alpha$, which can be understood as a modification of the path integral measure. This exponent simply emphasizes small or large representations / face areas in the partition function, and is motivated by a discussion in the community on the right choice of face amplitude \cite{Bianchi:2010fj}.

A physically more interesting model is based on so-called frusta \cite{Bahr:2017eyi}. A frustum is a higher-dimensional analogue of a trapezoid. In 4D, it consists of two cubes at its top and bottom, potentially of different size, which are connected by six 3D frusta, see the right of fig. \ref{fig:restricted_ints}. That way, a hyperfrustum can describe the evolution from one spatial cube to a larger / smaller spatial cube. Thus the idea is to restrict the intertwiners to be cube / frusta shaped in order to study the expanding / contracting cubulated 3D spatial slices. The frustum intertwiner is then given by three representations, $j_i$ and $j_f$ correspond to the initial and final area of the initial and final cubes respectively and $k$ gives the side-face area, which also determines the ``opening angle'' $\phi$ of the frustum\footnote{
The representation $j_i$, $j_f$ and $k$ must satisfy a relation that is spelled out in \cite{Bahr:2017eyi}.
}. Crucially, in contrast to the cuboid model, the Regge action associated to a hyperfrustum in the asymptotic expansion no longer vanishes and one obtains the familiar cosine formula \cite{Bahr:2017eyi}:
\begin{equation}
  \mathcal{A}_v \sim \cos(\frac{S_{\text{R}}}{G} + \varphi) + \cos(\frac{\gamma S_{\text{R}}}{G} + \frac{\Lambda}{G} \mathcal{V}) \; ,
\end{equation}
where $S_{\text{R}}$ denotes the Regge action of the hyperfrustum, $\mathcal{V}$ its volume, $G$ Netwons' constant, $\Lambda$ the cosmological constant and $\gamma$ the Barbero-Immirzi parameter.

Thus, this model captures several important generalizations compared to the cuboid model. As signified by the non-vanishing Regge action, frusta configurations allow for curvature to appear. Moreover, more parameters play a role in the dynamics: Newton's constant $G$ enters here as providing an explicit scale to the representations / areas on the boundary that serve as initial and final states. The cosmological constant $\Lambda$ is added in \cite{Bahr:2018ewi} (analogous to \cite{Han:2011aa}) by deforming the vertex amplitude. The parameter $\alpha$ remains as in the cuboid model\footnote{
The Immirzi parameter $\gamma$ is explicitly kept fixed. Due to the particularities of the Riemannian EPRL / FK model, $\gamma$ is necessarily a rational number, which significantly impacts the amount of allowed representations. Hence, the models for slightly different $\gamma$ are substantially different.
}.

After this basic introduction of these models, let us discuss the simplified coarse graining scheme and results.

\subsubsection{Coarse graining setup and results}

While in spirit the coarse graining setup is similar to the general method outlined in section \ref{sec:embedding}, there are several noteworthy difference and assumptions being made. Firstly, the embedding maps are chosen on geometric grounds instead of determining from the dynamics / the amplitudes. The intuitive idea is, e.g. in case of the hypercuboids, that a coarse hypercuboid arises as a superposition of fine hypercuboids consistent with the coarse hypercuboid geometry. Secondly, the coarse graining flow is computed for one fixed coarse boundary state, not the whole coarse boundary Hilbert space. Thus, in the path integral context, this coarse graining flow is performed for a fixed transition. The third and final assumption includes a projection back onto the original form of the amplitude, such that the flow is formulated as a flow in parameter space of the theory. This projection is defined by comparing expectation values of observables in the coarse and fine calculation. In a sense, the logic of the consistent boundary formulation is inverted: the RG flow assigns a family of amplitudes to different lattices such that expectation values of observables agree on all lattices. Instead, we derive the RG flow by identifying theories / parameters across lattices for which the expectation values of some (sensitive) observables agree. In case of the cuboids, which are just given by the parameter $\alpha$, this would read:
\begin{equation}
  \langle \mathcal{O} \rangle^{\alpha'}_{b'} \approx \langle \mathcal{O} \rangle^{\alpha}_{b} \; ,
\end{equation}
which defines the flow $\alpha' \rightarrow \alpha$ from fine $b'$ to coarse $b$. Note that all these three assumptions are strong simplifications that need to be lifted to verify their validity.

Let us first discuss the setup for coarse graining in the quantum cuboid model reported in \cite{Bahr:2016hwc}. Consider two hypercuboids glued together along a common 3D cuboid. The total geometry of both hypercuboids is fixed in the coarse boundary state, i.e. the total area of each coarse face is fixed, yet the distribution of 4-volume among the two cuboids fluctuates. Obviously, the expectation value of the volume of a single hypercuboid is always exactly half of the total volume, yet its variance depends sensitively on the parameter $\alpha$. Thus, for fixed coarse boundary state, one studies the variance of a single coarse in the coarse and in the fine case, where each of the hypercuboids is subdivided into 16. The geometric embedding map is prescribed such that the fine areas sum up to the coarse area. This setup and observable is particularly interesting since it is closely connected to the Abelian subgroup of diffeomorphisms that can be represented in the quantum cuboid model.

With this setup, the variance of the 4-volume is computed in both the coarse and fine case for various $\alpha$ / $\alpha'$ respectively. In both cases the observable is monotonously decreasing and both curves intersect once in the value $\alpha^\ast$. This particular value of $\alpha$ defines a fixed point of the renormalization group flow $\alpha' \rightarrow \alpha$, which is repulsive, i.e. $\alpha > \alpha'$ for $\alpha' > \alpha^\ast$ and $\alpha < \alpha'$ for $\alpha' < \alpha'$. Thus, the fixed point also separates two phases, which are dominated by different configurations. For $\alpha < \alpha^\ast$, small representations $j$ are preferred, such that subdivided faces contain one large area and several small ones. In contrast for $\alpha > \alpha^\ast $, the configuration dominates in which a face is equally subdivided since then all spins are as large as possible. Remarkably, the value $\alpha^\ast$ is close to the one at which diffeomorphism symmetry is almost restored \cite{Bahr:2015gxa}. This result together with the repulsive behaviour at the phase transition indicate that this transition might be of second order, and that on it the subgroup of diffeomorphisms might be restored.

The same calculation is repeated for different coarse boundary states in \cite{Bahr:2017klw} and results in the same qualitative behaviour, yet the position of the fixed point changes. Thus, we do not include the exact value. This result sheds a light on the possible interpretation of these results. Since the coarse boundary state is kept fixed, this coarse graining derives a family of amplitudes on a family of lattices for this specific transition. Therefore it contains the information whether and for which parameters the regulator / the lattice can be removed and the results are consistent (within the given approximations and truncations). Note however that this is a weaker condition than the coarse graining flow defined in section \ref{sec:embedding}, which refers to all transitions / boundary states. In a sense, the fixed state becomes part of the observable for which the coarse graining flow is defined. That way it provides first insights of coarse graining flow in a truncated theory space.

A similar analysis of the coarse graining flow is performed for frusta spin foams in \cite{Bahr:2018gwf} with a slighly different setup. Here the boundary is made up of two parts, an initial and a final 3D spatial cubulations each prescribed by a single representation $j_i$ and $j_f$, which are chosen to be equal. Again, the goal is to compute expectation values of observables in a fine and a coarse setting and define a renormalization group flow in parameter space $(\alpha,G,\Lambda)$ such that these observables agree. In \cite{Bahr:2018gwf} a few different setups are examined, here we just discuss the main result of the RG flow in three dimensional parameter space.

Due to the high symmetry of frusta configurations, the lattice is prescribed by spatial and temporal subdivisions. The former fixes the fineness of the spatial cubulation while the latter determines the number of time steps. The coarse lattice then has two spatial subdivisions, i.e. $4^3$ spatial cubes, and one intermediate time step. The fine lattice has one more spatial and temporal subdivision, i.e. $8^3$ cubes and two intermediate time steps. The boundary states in both settings are straightforwardly related by requiring that the total 3D spatial volume encoded in initial / final state agrees in both settings. Moreover, the total ``height'' / ``time'' is fixed in both settings as well as each time step is chosen to be equal. That way, only the intermediate spatial volume is integrated over, while the side panels are fixed, which greatly reduces the numerical cost.

In order to derive a renormalization group flow in a parameter space with three parameters, one must consider at least three observables. In \cite{Bahr:2018gwf}, the 3D spatial volume of the intermediate slice, its variance as well as the total 4D volume are considered. The expectation values for all these observables are computed in a range of all parameters $G$, $\Lambda$ and $\alpha$ and compared for both settings. Then a coarse graining flow is derived by matching theories with the smallest relative error of observables\footnote{Moreover, one is only comparing theories that are ``close'' in parameter space. This is justified since one is mainly interested in fixed points of the coarse graining flow, given the truncations introduced in the model.}. Under this premise, indications for a fixed point around $\alpha^\ast \approx 0.677$, $G^\ast \approx 0.037$ and $\Lambda^\ast \approx 0.08$ are found. While the exact numerical values are less relevant and most likely subject to change for different boundary states, qualitatively the numerics indicate that this fixed point has one repulsive and two attractive directions. As for the cuboid case, the repulsive direction appears to be (mostly) related to the parameter $\alpha$, while $G$ and $\Lambda$ seem to be the attractive directions. In standard lore, this would imply that both $G$ and $\Lambda$ are irrelevant couplings and fixed by the RG trajectory.

At first sight, this result appears to be at odds with results in asymptotic safety \cite{Reuter:2019byg}, where both $G$ and $\Lambda$ correspond to free parameters / relevant directions. However note that this setting here is significantly different: due to considering only a specific transition, a scale is introduced into the system, which is not changed by the coarse graining flow. Thus, in contrast to asymptotic safety where one derives theories at different scales, this coarse graining flow teaches us whether and for which parameters the regulator / lattice might be removed. In this sense, the fixed point gives the correct discretisation independent amplitude (given the introduced approximations) for this specific transition. So $G^\ast$ and $\Lambda^\ast$ mark the correct parameters for one specific transition and are thus irrelevant in this flow, yet they might correspond to relevant directions when different scales are related.

\subsubsection{Numerical methods}

A short comment on the numerical methods is in order. In the semi-classical, restricted spin foam models, Monte-Carlo and numerical integration techniques are used \cite{Hahn:2004fe}. This works particularly well for the quantum cuboid model, where the action vanishes and the amplitudes do not show an oscillating behaviour. Nevertheless, also for the frusta models, which feature oscillating amplitudes, can be explored with these methods with slower convergence. In general, convergence slows down for higher dimensional integrals and larger discretisations, such that this method appears to be feasible for systems with a few building blocks and symmetries that reduce the amount of degrees of freedom.

\subsection{Semi-classical continuum limit}

Before concluding this review, we would like to briefly discuss the semi-classical continuum limit approach \cite{Han:2017xwo,Han:2018fmu}, since it aims at defining a flow across a family of triangulations and might at first sight be similar to the restricted path integral method.

This approach discussed in the papers \cite{Han:2017xwo,Han:2018fmu} aims at defining a semi-classical continuum limit for spin foam models in the following sense. As mentioned before, it is a well-established result that one obtains (area) Regge calculus in the asymptotic expansion of spin foam vertex amplitues \cite{Barrett:1998gs,Conrady:2008mk,Barrett:2009gg,Kaminski:2017eew,Liu:2018gfc}. Often this is called a semi-classical limit, by scaling all representations $j \rightarrow \lambda j$ by a parameter $\lambda$. In \cite{Han:2018fmu} a Gaussian weight is introduced into this semi-classical limit that suppresses non-length-Regge like geometries, i.e. geometries prescribed by areas that do not correspond to triangulation given by edge lengths. The parameter for this Gaussian weight is $\delta$, where $\delta \rightarrow 0$ removes this weight. These systems are studied in the regime $\lambda \gg \delta^{-1} \gg 1$, such that the semi-classical formula is valid and higher curvature corrections (in the deficit angle) are suppressed.

Essentially the idea is to define a continuum limit as in Regge calculus: for a sequence of triangulations $\mathcal{K}_N$ of the same manifold continuum general relativity is restored if all lengths and deficit angles converge to zero as $N \rightarrow \infty$. In \cite{Han:2018fmu} it is argued that this is achieved for particular scaling relations for $\lambda$, $\delta$ and $\mu$, where $\mu$ rescales the Planck length. Thus, one defines a flow across triangulations in this parameter space, where in the limit $N \rightarrow \infty$, both areas and deficit angles converge to zero. Invoking the continuum limit of Regge calculus, it is argued that general relativity is obtained in this limit.

The existence of such a regime, where one obtains general relativity as the continuum limit of Regge calculus, would be intriguing and it is suggestive to think it should exist, given the close relation of spin foam models and Regge calculus. However, there are several points that must be considered before this can be confirmed: firstly, the assumptions and modifications made that must be carefully cross-checked. Most strikingly, a term that suppresses non-Regge like geometries is not present in spin foam models and one might argue that such a role ought to be already implemented in the simplicty constraints. Secondly, the conditions under which the formulas are valid are highly specific and it must be validated whether these are satisfied in generic situations. Finally, the defined flow of parameters is not dynamical, in the sense that it is not derived by relating dynamics across different triangulations. Thus, it is not clear whether this continuum limit gives well-defined continuum dynamics.

\section{Outlook: Towards renormalization in 4D} \label{sec:directions}

In this article we provide a detailed review of coarse graining in spin foams at the conceptual and practical level. Attentive readers notice that these methods have not been applied yet to the full 4D theory, e.g. the EPRL / FK in the Riemannian or Lorentzian setting, and it is currently out of reach. In this outlook, we would like to discuss the open issues and questions that need to be addressed.

A first point, which is relevant for all calculations performed in spin foam quantum gravity, is the computability of spin foam amplitudes, more precisely the vertex amplitude. As the amplitude associated to a 4D building block, it is the centerpiece of the theory and the most intricate to compute. Analytical formulas are known for the asymptotic expansion of the amplitude \cite{Barrett:1998gs,Conrady:2008mk,Barrett:2009gg,Kaminski:2017eew,Liu:2018gfc}, where the boundary data is given by coherent states peaked on classical discrete geometries. However, these results are not valid for small representations, the quantum regime of the theory. To compute the amplitude in this regime requires numerical techniques, e.g. by explicitly contracting intertwiners to obtain the vertex amplitude. Significant progress was made in recent years for the Lorentzian EPRL model in \cite{Dona:2017dvf,Dona:2018nev,Dona:2019dkf} using the results form \cite{Speziale:2016axj}. Nevertheless, these calculations require significant numerical resources, which makes it difficult to explore systems with multiple vertex amplitudes. Two ideas might be helpful in exploring larger 2-complexes: Firstly, storing computed vertex amplitudes, e.g. for an orthonromal basis of intertwiners, in an open-data database such as the ``Encyclopedia of Quantum Geometries''\footnote{\url{https://zenodo.org/communities/enqugeo/?page=1&size=20}} would make them accessible to interested researchers and avoid computing the same amplitudes multiple times. The second idea relies on the fact that the asymptotic formula well approximates the vertex amplitude for fairly small representations, $j \sim 10$ for a 4-simplex in the Riemannian EPRL model \cite{Bayle:2016doe}. Exploiting this fact could lead to an efficient hybrid algorithm, similar to the idea used in loop quantum cosmology \cite{Diener:2013uka}, that only uses the costly to compute quantum amplitude in case the asymptotic formula is not accurate.

An alternative route towards studying spin foams with multiple simplices lies in defining simplified models. In section \ref{sec:restricted} we review one example for such models, namely restricted spin foam models. Instead of exploring the full spin foam path integral, only a subset of configurations is explored using the asymptotic formula. Thus the number of degress of freedom is drastically reduced and the issue of exactly computing the vertex amplitude is circumvented, which makes it possible to renormalize these models. Clearly, as next steps these restrictions need to be lifted in order to explore more of the dynamics of the theory. This could either be by allowing more configurations in the path integral, see e.g. \cite{Assanioussi:2020fml}, or by going beyond the asymptotic formula and including the full vertex amplitude. Recently, another simplified model has been constructed in a similar direction \cite{Asante:2020qpa}. Again, the asymptotic formula of the vertex amplitude is invoked to define a simplified vertex amplitude. Special emphasis is given to an implementation of simplicity constraints akin to spin foam models as weak conditions on 3D dihedral angles, which might give new insights into spin foam dynamics for large 2-complexes. Note that this model does not restrict the allowed configurations in contrast to the restricted models discussed in this review.

The most holistic approach to coarse graining spin foam models, tensor network renormalization discussed in detail in section \ref{sec:TNW}, faces two main challenges when going to 4D. One is the increased complexity of the amplitude which results both in larger memory cost as well as computational time. Related to this is the second issue, how to define a tensor network algorithm for systems with infinitely many configurations or continuous variables like the 4D spin foam models defined for Lie groups. A solution to the former challenge might lie in defining a representation of the model suited for renormalization, similar to the fusion basis in 3D \cite{Dittrich:2016typ,Cunningham:2020uco}. Alternative formulations of 4D models are investigated e.g. in \cite{Delcamp:2016lux,Dittrich:2017nmq}. Using observables might again serve as a guiding principle to find such representations. The second issue might be tackled in a similar direction, where a rewriting of the model might lead to an efficient tensor network description. One such example is \cite{Delcamp:2020hzo}, where the renormalization group flow of $\phi^4$ scalar field theory is accessible for tensor network methods by performing a simple transformation.

Another important research direction, on which renormalization and coarse graining can shed a new light, is matter coupling in spin foam quantum gravity. Since spin foams are a purely gravitational theory, matter degrees of freedom must be added in order to adequately describe the universe. Different ways to couple matter to spin foams exist in the literature \cite{Mikovic:2001xi,Oriti:2002bn,Speziale:2007mt,Smolin:2007rx,Han:2011as,Bianchi:2010bn}, yet the intriguing dynamics of the coupled matter gravity system are hardly explored. Applying a coarse graining scheme to the combined system allows us to renormalize matter and gravitational degrees of freedom at the same time, uncovering the phase diagram of the system. This idea is realized for a simplified toy model in \cite{Steinhaus:2015kxa}. Without a question, a system consisting of both spin foam and matter degrees of freedom is more difficult to study than the former alone. Nevertheless, adding matter to simplified models might be accessible and lead towards intriguing new features and insights, e.g. it would be interesting to see how the matter sector influences the quantum gravitational theory as in \cite{Dona:2013qba}.

Beyond the methods discussed in the review, ideas from other fields and approaches to quantum gravity might help us advance coarse graining in spin foam models to 4D. These might be novel numerical techniques, like deep learning, or well-established ones like Monte Carlo methods, which might be efficiently applicable in certain settings.

\section*{Conflict of Interest Statement}
%All financial, commercial or other relationships that might be perceived by the academic community as representing a potential conflict of interest must be disclosed. If no such relationship exists, authors will be asked to confirm the following statement:

The authors declare that the research was conducted in the absence of any commercial or financial relationships that could be construed as a potential conflict of interest.

%\section*{Author Contributions}

%The Author Contributions section is mandatory for all articles, including articles by sole authors. If an appropriate statement is not provided on submission, a standard one will be inserted during the production process. The Author Contributions statement must describe the contributions of individual authors referred to by their initials and, in doing so, all authors agree to be accountable for the content of the work. Please see  \href{http://home.frontiersin.org/about/author-guidelines#AuthorandContributors}{here} for full authorship criteria.

\section*{Funding}
SSt  is  funded  by  the  Deutsche  Forschungsgemeinschaft  (DFG, German  Research  Foundation)  -  Projektnummer  /  project  number  422809950.

\section*{Acknowledgments}
The author would like to thank both anonymous referees for their thorough and constructive feedback that helped to improve the accessibility of this review.

%\section*{Supplemental Data}
% \href{http://home.frontiersin.org/about/author-guidelines#SupplementaryMaterial}{Supplementary Material} should be uploaded separately on submission, if there are Supplementary Figures, please include the caption in the same file as the figure. LaTeX Supplementary Material templates can be found in the Frontiers LaTeX folder.

%\section*{Data Availability Statement}
%The datasets [GENERATED/ANALYZED] for this study can be found in the [NAME OF REPOSITORY] [LINK].
% Please see the availability of data guidelines for more information, at https://www.frontiersin.org/about/author-guidelines#AvailabilityofData

%\bibliographystyle{frontiersinSCNS_ENG_HUMS} % for Science, Engineering and Humanities and Social Sciences articles, for Humanities and Social Sciences articles please include page numbers in the in-text citations
%\bibliographystyle{frontiersinHLTH&FPHY} % for Health, Physics and Mathematics articles
%\bibliographystyle{utphys.bst}
\bibliography{test.bib}

%merlin.mbs apsrev4-1.bst 2010-07-25 4.21a (PWD, AO, DPC) hacked
%Control: key (0)
%Control: author (0) dotless jnrlst
%Control: editor formatted (1) identically to author
%Control: production of article title (0) allowed
%Control: page (1) range
%Control: year (0) verbatim
%Control: production of eprint (0) enabled
\begin{thebibliography}{132}%
\makeatletter
\providecommand \@ifxundefined [1]{%
 \@ifx{#1\undefined}
}%
\providecommand \@ifnum [1]{%
 \ifnum #1\expandafter \@firstoftwo
 \else \expandafter \@secondoftwo
 \fi
}%
\providecommand \@ifx [1]{%
 \ifx #1\expandafter \@firstoftwo
 \else \expandafter \@secondoftwo
 \fi
}%
\providecommand \natexlab [1]{#1}%
\providecommand \enquote  [1]{``#1''}%
\providecommand \bibnamefont  [1]{#1}%
\providecommand \bibfnamefont [1]{#1}%
\providecommand \citenamefont [1]{#1}%
\providecommand \href@noop [0]{\@secondoftwo}%
\providecommand \href [0]{\begingroup \@sanitize@url \@href}%
\providecommand \@href[1]{\@@startlink{#1}\@@href}%
\providecommand \@@href[1]{\endgroup#1\@@endlink}%
\providecommand \@sanitize@url [0]{\catcode `\\12\catcode `\$12\catcode
  `\&12\catcode `\#12\catcode `\^12\catcode `\_12\catcode `\%12\relax}%
\providecommand \@@startlink[1]{}%
\providecommand \@@endlink[0]{}%
\providecommand \url  [0]{\begingroup\@sanitize@url \@url }%
\providecommand \@url [1]{\endgroup\@href {#1}{\urlprefix }}%
\providecommand \urlprefix  [0]{URL }%
\providecommand \Eprint [0]{\href }%
\providecommand \doibase [0]{http://dx.doi.org/}%
\providecommand \selectlanguage [0]{\@gobble}%
\providecommand \bibinfo  [0]{\@secondoftwo}%
\providecommand \bibfield  [0]{\@secondoftwo}%
\providecommand \translation [1]{[#1]}%
\providecommand \BibitemOpen [0]{}%
\providecommand \bibitemStop [0]{}%
\providecommand \bibitemNoStop [0]{.\EOS\space}%
\providecommand \EOS [0]{\spacefactor3000\relax}%
\providecommand \BibitemShut  [1]{\csname bibitem#1\endcsname}%
\let\auto@bib@innerbib\@empty
%</preamble>
\bibitem [{\citenamefont {Perez}(2013)}]{Perez:2012wv}%
  \BibitemOpen
  \bibfield  {author} {\bibinfo {author} {\bibfnamefont {Alejandro}\
  \bibnamefont {Perez}},\ }\bibfield  {title} {\enquote {\bibinfo {title} {{The
  Spin Foam Approach to Quantum Gravity}},}\ }\href {\doibase
  10.12942/lrr-2013-3} {\bibfield  {journal} {\bibinfo  {journal} {Living Rev.\
  Rel.}\ }\textbf {\bibinfo {volume} {16}},\ \bibinfo {pages} {3} (\bibinfo
  {year} {2013})},\ \Eprint {http://arxiv.org/abs/1205.2019} {arXiv:1205.2019
  [gr-qc]} \BibitemShut {NoStop}%
\bibitem [{\citenamefont {Rovelli}(2004)}]{carlobook}%
  \BibitemOpen
  \bibfield  {author} {\bibinfo {author} {\bibfnamefont {C.}~\bibnamefont
  {Rovelli}},\ }\href@noop {} {\emph {\bibinfo {title} {Quantum gravity}}}\
  (\bibinfo  {publisher} {Cambridge University Press},\ \bibinfo {address}
  {Cambridge},\ \bibinfo {year} {2004})\BibitemShut {NoStop}%
\bibitem [{\citenamefont {Thiemann}(2008)}]{thomasbook}%
  \BibitemOpen
  \bibfield  {author} {\bibinfo {author} {\bibfnamefont {T.}~\bibnamefont
  {Thiemann}},\ }\href@noop {} {\emph {\bibinfo {title} {Modern {C}anonical
  {Q}uantum {G}eneral {R}elativity}}}\ (\bibinfo  {publisher} {Cambridge
  University Press},\ \bibinfo {address} {Cambridge},\ \bibinfo {year}
  {2008})\BibitemShut {NoStop}%
\bibitem [{\citenamefont {Plebanski}(1977)}]{Plebanski:1977zz}%
  \BibitemOpen
  \bibfield  {author} {\bibinfo {author} {\bibfnamefont {Jerzy~F.}\
  \bibnamefont {Plebanski}},\ }\bibfield  {title} {\enquote {\bibinfo {title}
  {{On the separation of Einsteinian substructures}},}\ }\href {\doibase
  10.1063/1.523215} {\bibfield  {journal} {\bibinfo  {journal} {J.\ Math.\
  Phys.}\ }\textbf {\bibinfo {volume} {18}},\ \bibinfo {pages} {2511--2520}
  (\bibinfo {year} {1977})}\BibitemShut {NoStop}%
\bibitem [{\citenamefont {Baez}(2000)}]{Baez:1999sr}%
  \BibitemOpen
  \bibfield  {author} {\bibinfo {author} {\bibfnamefont {J.~C.}\ \bibnamefont
  {Baez}},\ }\bibfield  {title} {\enquote {\bibinfo {title} {{An Introduction
  to spin foam models of quantum gravity and BF theory}},}\ }\href {\doibase
  10.1007/3-540-46552-9\_2} {\bibfield  {journal} {\bibinfo  {journal} {Lect.\
  Notes Phys.}\ }\textbf {\bibinfo {volume} {543}},\ \bibinfo {pages} {25--94}
  (\bibinfo {year} {2000})},\ \Eprint {http://arxiv.org/abs/gr-qc/9905087}
  {arXiv:gr-qc/9905087} \BibitemShut {NoStop}%
\bibitem [{\citenamefont {Ponzano}\ and\ \citenamefont {Regge}(1968)}]{PR}%
  \BibitemOpen
  \bibfield  {author} {\bibinfo {author} {\bibfnamefont {G.}~\bibnamefont
  {Ponzano}}\ and\ \bibinfo {author} {\bibfnamefont {T.}~\bibnamefont
  {Regge}},\ }\href@noop {} {\emph {\bibinfo {title} {Semiclassical limit of
  {Racah} coefficients, in: Spectroscopy and group theoretical methods in
  physics}}},\ edited by\ \bibinfo {editor} {\bibfnamefont {F.}~\bibnamefont
  {Bloch}}\ (\bibinfo  {publisher} {North Holland Publ. Co.},\ \bibinfo
  {address} {Amsterdam},\ \bibinfo {year} {1968})\ pp.\ \bibinfo {pages}
  {1--58}\BibitemShut {NoStop}%
\bibitem [{\citenamefont {Freidel}\ and\ \citenamefont
  {Louapre}(2004)}]{Freidel:2004vi}%
  \BibitemOpen
  \bibfield  {author} {\bibinfo {author} {\bibfnamefont {Laurent}\ \bibnamefont
  {Freidel}}\ and\ \bibinfo {author} {\bibfnamefont {David}\ \bibnamefont
  {Louapre}},\ }\bibfield  {title} {\enquote {\bibinfo {title} {{Ponzano-Regge
  model revisited I: Gauge fixing, observables and interacting spinning
  particles}},}\ }\href {\doibase 10.1088/0264-9381/21/24/002} {\bibfield
  {journal} {\bibinfo  {journal} {Class.\ Quant.\ Grav.}\ }\textbf {\bibinfo
  {volume} {21}},\ \bibinfo {pages} {5685--5726} (\bibinfo {year} {2004})},\
  \Eprint {http://arxiv.org/abs/hep-th/0401076} {arXiv:hep-th/0401076}
  \BibitemShut {NoStop}%
\bibitem [{\citenamefont {Barrett}\ and\ \citenamefont
  {Naish-Guzman}(2009)}]{Barrett:2008wh}%
  \BibitemOpen
  \bibfield  {author} {\bibinfo {author} {\bibfnamefont {John~W.}\ \bibnamefont
  {Barrett}}\ and\ \bibinfo {author} {\bibfnamefont {Ileana}\ \bibnamefont
  {Naish-Guzman}},\ }\bibfield  {title} {\enquote {\bibinfo {title} {{The
  Ponzano-Regge model}},}\ }\href {\doibase 10.1088/0264-9381/26/15/155014}
  {\bibfield  {journal} {\bibinfo  {journal} {Class.\ Quant.\ Grav.}\ }\textbf
  {\bibinfo {volume} {26}},\ \bibinfo {pages} {155014} (\bibinfo {year}
  {2009})},\ \Eprint {http://arxiv.org/abs/0803.3319} {arXiv:0803.3319 [gr-qc]}
  \BibitemShut {NoStop}%
\bibitem [{\citenamefont {Holst}(1996)}]{Holst:1995pc}%
  \BibitemOpen
  \bibfield  {author} {\bibinfo {author} {\bibfnamefont {Soren}\ \bibnamefont
  {Holst}},\ }\bibfield  {title} {\enquote {\bibinfo {title} {{Barbero's
  Hamiltonian derived from a generalized Hilbert-Palatini action}},}\ }\href
  {\doibase 10.1103/PhysRevD.53.5966} {\bibfield  {journal} {\bibinfo
  {journal} {Phys.\ Rev.\ D}\ }\textbf {\bibinfo {volume} {53}},\ \bibinfo
  {pages} {5966--5969} (\bibinfo {year} {1996})},\ \Eprint
  {http://arxiv.org/abs/gr-qc/9511026} {arXiv:gr-qc/9511026} \BibitemShut
  {NoStop}%
\bibitem [{\citenamefont {Barrett}\ and\ \citenamefont
  {Crane}(1998)}]{Barrett:1997gw}%
  \BibitemOpen
  \bibfield  {author} {\bibinfo {author} {\bibfnamefont {John~W.}\ \bibnamefont
  {Barrett}}\ and\ \bibinfo {author} {\bibfnamefont {Louis}\ \bibnamefont
  {Crane}},\ }\bibfield  {title} {\enquote {\bibinfo {title} {{Relativistic
  spin networks and quantum gravity}},}\ }\href {\doibase 10.1063/1.532254}
  {\bibfield  {journal} {\bibinfo  {journal} {J.\ Math.\ Phys.}\ }\textbf
  {\bibinfo {volume} {39}},\ \bibinfo {pages} {3296--3302} (\bibinfo {year}
  {1998})},\ \Eprint {http://arxiv.org/abs/gr-qc/9709028} {arXiv:gr-qc/9709028}
  \BibitemShut {NoStop}%
\bibitem [{\citenamefont {Barrett}\ and\ \citenamefont
  {Crane}(2000)}]{Barrett:1999qw}%
  \BibitemOpen
  \bibfield  {author} {\bibinfo {author} {\bibfnamefont {John~W.}\ \bibnamefont
  {Barrett}}\ and\ \bibinfo {author} {\bibfnamefont {Louis}\ \bibnamefont
  {Crane}},\ }\bibfield  {title} {\enquote {\bibinfo {title} {{A Lorentzian
  signature model for quantum general relativity}},}\ }\href {\doibase
  10.1088/0264-9381/17/16/302} {\bibfield  {journal} {\bibinfo  {journal}
  {Class.\ Quant.\ Grav.}\ }\textbf {\bibinfo {volume} {17}},\ \bibinfo {pages}
  {3101--3118} (\bibinfo {year} {2000})},\ \Eprint
  {http://arxiv.org/abs/gr-qc/9904025} {arXiv:gr-qc/9904025} \BibitemShut
  {NoStop}%
\bibitem [{\citenamefont {Engle}\ \emph
  {et~al.}(2008{\natexlab{a}})\citenamefont {Engle}, \citenamefont {Pereira},\
  and\ \citenamefont {Rovelli}}]{Engle:2007qf}%
  \BibitemOpen
  \bibfield  {author} {\bibinfo {author} {\bibfnamefont {Jonathan}\
  \bibnamefont {Engle}}, \bibinfo {author} {\bibfnamefont {Roberto}\
  \bibnamefont {Pereira}}, \ and\ \bibinfo {author} {\bibfnamefont {Carlo}\
  \bibnamefont {Rovelli}},\ }\bibfield  {title} {\enquote {\bibinfo {title}
  {{Flipped spinfoam vertex and loop gravity}},}\ }\href {\doibase
  10.1016/j.nuclphysb.2008.02.002} {\bibfield  {journal} {\bibinfo  {journal}
  {Nucl.\ Phys.\ B}\ }\textbf {\bibinfo {volume} {798}},\ \bibinfo {pages}
  {251--290} (\bibinfo {year} {2008}{\natexlab{a}})},\ \Eprint
  {http://arxiv.org/abs/0708.1236} {arXiv:0708.1236 [gr-qc]} \BibitemShut
  {NoStop}%
\bibitem [{\citenamefont {Engle}\ \emph
  {et~al.}(2008{\natexlab{b}})\citenamefont {Engle}, \citenamefont {Livine},
  \citenamefont {Pereira},\ and\ \citenamefont {Rovelli}}]{Engle:2007wy}%
  \BibitemOpen
  \bibfield  {author} {\bibinfo {author} {\bibfnamefont {Jonathan}\
  \bibnamefont {Engle}}, \bibinfo {author} {\bibfnamefont {Etera}\ \bibnamefont
  {Livine}}, \bibinfo {author} {\bibfnamefont {Roberto}\ \bibnamefont
  {Pereira}}, \ and\ \bibinfo {author} {\bibfnamefont {Carlo}\ \bibnamefont
  {Rovelli}},\ }\bibfield  {title} {\enquote {\bibinfo {title} {{LQG vertex
  with finite Immirzi parameter}},}\ }\href {\doibase
  10.1016/j.nuclphysb.2008.02.018} {\bibfield  {journal} {\bibinfo  {journal}
  {Nucl.\ Phys.\ B}\ }\textbf {\bibinfo {volume} {799}},\ \bibinfo {pages}
  {136--149} (\bibinfo {year} {2008}{\natexlab{b}})},\ \Eprint
  {http://arxiv.org/abs/0711.0146} {arXiv:0711.0146 [gr-qc]} \BibitemShut
  {NoStop}%
\bibitem [{\citenamefont {Freidel}\ and\ \citenamefont
  {Krasnov}(2008)}]{Freidel:2007py}%
  \BibitemOpen
  \bibfield  {author} {\bibinfo {author} {\bibfnamefont {Laurent}\ \bibnamefont
  {Freidel}}\ and\ \bibinfo {author} {\bibfnamefont {Kirill}\ \bibnamefont
  {Krasnov}},\ }\bibfield  {title} {\enquote {\bibinfo {title} {{A New Spin
  Foam Model for 4d Gravity}},}\ }\href {\doibase
  10.1088/0264-9381/25/12/125018} {\bibfield  {journal} {\bibinfo  {journal}
  {Class.\ Quant.\ Grav.}\ }\textbf {\bibinfo {volume} {25}},\ \bibinfo {pages}
  {125018} (\bibinfo {year} {2008})},\ \Eprint {http://arxiv.org/abs/0708.1595}
  {arXiv:0708.1595 [gr-qc]} \BibitemShut {NoStop}%
\bibitem [{\citenamefont {Alesci}\ and\ \citenamefont
  {Rovelli}(2007)}]{Alesci:2007tx}%
  \BibitemOpen
  \bibfield  {author} {\bibinfo {author} {\bibfnamefont {Emanuele}\
  \bibnamefont {Alesci}}\ and\ \bibinfo {author} {\bibfnamefont {Carlo}\
  \bibnamefont {Rovelli}},\ }\bibfield  {title} {\enquote {\bibinfo {title}
  {{The Complete LQG propagator. I. Difficulties with the Barrett-Crane
  vertex}},}\ }\href {\doibase 10.1103/PhysRevD.76.104012} {\bibfield
  {journal} {\bibinfo  {journal} {Phys.\ Rev.\ D}\ }\textbf {\bibinfo {volume}
  {76}},\ \bibinfo {pages} {104012} (\bibinfo {year} {2007})},\ \Eprint
  {http://arxiv.org/abs/0708.0883} {arXiv:0708.0883 [gr-qc]} \BibitemShut
  {NoStop}%
\bibitem [{\citenamefont {Alesci}\ and\ \citenamefont
  {Rovelli}(2008)}]{Alesci:2007tg}%
  \BibitemOpen
  \bibfield  {author} {\bibinfo {author} {\bibfnamefont {Emanuele}\
  \bibnamefont {Alesci}}\ and\ \bibinfo {author} {\bibfnamefont {Carlo}\
  \bibnamefont {Rovelli}},\ }\bibfield  {title} {\enquote {\bibinfo {title}
  {{The Complete LQG propagator. II. Asymptotic behavior of the vertex}},}\
  }\href {\doibase 10.1103/PhysRevD.77.044024} {\bibfield  {journal} {\bibinfo
  {journal} {Phys.\ Rev.\ D}\ }\textbf {\bibinfo {volume} {77}},\ \bibinfo
  {pages} {044024} (\bibinfo {year} {2008})},\ \Eprint
  {http://arxiv.org/abs/0711.1284} {arXiv:0711.1284 [gr-qc]} \BibitemShut
  {NoStop}%
\bibitem [{\citenamefont {Baez}\ and\ \citenamefont
  {Barrett}(1999)}]{Baez:1999tk}%
  \BibitemOpen
  \bibfield  {author} {\bibinfo {author} {\bibfnamefont {John~C.}\ \bibnamefont
  {Baez}}\ and\ \bibinfo {author} {\bibfnamefont {John~W.}\ \bibnamefont
  {Barrett}},\ }\bibfield  {title} {\enquote {\bibinfo {title} {{The Quantum
  tetrahedron in three-dimensions and four-dimensions}},}\ }\href {\doibase
  10.4310/ATMP.1999.v3.n4.a3} {\bibfield  {journal} {\bibinfo  {journal} {Adv.
  Theor. Math. Phys.}\ }\textbf {\bibinfo {volume} {3}},\ \bibinfo {pages}
  {815--850} (\bibinfo {year} {1999})},\ \Eprint
  {http://arxiv.org/abs/gr-qc/9903060} {arXiv:gr-qc/9903060} \BibitemShut
  {NoStop}%
\bibitem [{\citenamefont {Livine}\ and\ \citenamefont
  {Speziale}(2007)}]{Livine:2007vk}%
  \BibitemOpen
  \bibfield  {author} {\bibinfo {author} {\bibfnamefont {Etera~R.}\
  \bibnamefont {Livine}}\ and\ \bibinfo {author} {\bibfnamefont {Simone}\
  \bibnamefont {Speziale}},\ }\bibfield  {title} {\enquote {\bibinfo {title}
  {{A New spinfoam vertex for quantum gravity}},}\ }\href {\doibase
  10.1103/PhysRevD.76.084028} {\bibfield  {journal} {\bibinfo  {journal}
  {Phys.\ Rev.\ D}\ }\textbf {\bibinfo {volume} {76}},\ \bibinfo {pages}
  {084028} (\bibinfo {year} {2007})},\ \Eprint {http://arxiv.org/abs/0705.0674}
  {arXiv:0705.0674 [gr-qc]} \BibitemShut {NoStop}%
\bibitem [{\citenamefont {Minkowski}(1989)}]{Minkowski}%
  \BibitemOpen
  \bibfield  {author} {\bibinfo {author} {\bibfnamefont {Hermann}\ \bibnamefont
  {Minkowski}},\ }\href@noop {} {\emph {\bibinfo {title} {Allgemeine Lehrsätze
  über die konvexen Polyeder}}},\ \bibinfo {series} {Ausgewählte Arbeiten zur
  Zahlentheorie und zur Geometrie. Teubner-Archiv zur Mathematik}, Vol.\
  \bibinfo {volume} {vol. 12}\ (\bibinfo  {publisher} {Springer},\ \bibinfo
  {address} {Vienna},\ \bibinfo {year} {1989})\BibitemShut {NoStop}%
\bibitem [{\citenamefont {Bianchi}\ \emph {et~al.}(2011)\citenamefont
  {Bianchi}, \citenamefont {Dona},\ and\ \citenamefont
  {Speziale}}]{Bianchi:2010gc}%
  \BibitemOpen
  \bibfield  {author} {\bibinfo {author} {\bibfnamefont {Eugenio}\ \bibnamefont
  {Bianchi}}, \bibinfo {author} {\bibfnamefont {Pietro}\ \bibnamefont {Dona}},
  \ and\ \bibinfo {author} {\bibfnamefont {Simone}\ \bibnamefont {Speziale}},\
  }\bibfield  {title} {\enquote {\bibinfo {title} {{Polyhedra in loop quantum
  gravity}},}\ }\href {\doibase 10.1103/PhysRevD.83.044035} {\bibfield
  {journal} {\bibinfo  {journal} {Phys. Rev. D}\ }\textbf {\bibinfo {volume}
  {83}},\ \bibinfo {pages} {044035} (\bibinfo {year} {2011})},\ \Eprint
  {http://arxiv.org/abs/1009.3402} {arXiv:1009.3402 [gr-qc]} \BibitemShut
  {NoStop}%
\bibitem [{\citenamefont {Barrett}\ and\ \citenamefont
  {Williams}(1999)}]{Barrett:1998gs}%
  \BibitemOpen
  \bibfield  {author} {\bibinfo {author} {\bibfnamefont {John~W.}\ \bibnamefont
  {Barrett}}\ and\ \bibinfo {author} {\bibfnamefont {Ruth~M.}\ \bibnamefont
  {Williams}},\ }\bibfield  {title} {\enquote {\bibinfo {title} {{The
  Asymptotics of an amplitude for the four simplex}},}\ }\href {\doibase
  10.4310/ATMP.1999.v3.n2.a1} {\bibfield  {journal} {\bibinfo  {journal} {Adv.\
  Theor.\ Math.\ Phys.}\ }\textbf {\bibinfo {volume} {3}},\ \bibinfo {pages}
  {209--215} (\bibinfo {year} {1999})},\ \Eprint
  {http://arxiv.org/abs/gr-qc/9809032} {arXiv:gr-qc/9809032} \BibitemShut
  {NoStop}%
\bibitem [{\citenamefont {Conrady}\ and\ \citenamefont
  {Freidel}(2008)}]{Conrady:2008mk}%
  \BibitemOpen
  \bibfield  {author} {\bibinfo {author} {\bibfnamefont {Florian}\ \bibnamefont
  {Conrady}}\ and\ \bibinfo {author} {\bibfnamefont {Laurent}\ \bibnamefont
  {Freidel}},\ }\bibfield  {title} {\enquote {\bibinfo {title} {{On the
  semiclassical limit of 4d spin foam models}},}\ }\href {\doibase
  10.1103/PhysRevD.78.104023} {\bibfield  {journal} {\bibinfo  {journal}
  {Phys.\ Rev.\ D}\ }\textbf {\bibinfo {volume} {78}},\ \bibinfo {pages}
  {104023} (\bibinfo {year} {2008})},\ \Eprint {http://arxiv.org/abs/0809.2280}
  {arXiv:0809.2280 [gr-qc]} \BibitemShut {NoStop}%
\bibitem [{\citenamefont {Barrett}\ \emph {et~al.}(2009)\citenamefont
  {Barrett}, \citenamefont {Dowdall}, \citenamefont {Fairbairn}, \citenamefont
  {Gomes},\ and\ \citenamefont {Hellmann}}]{Barrett:2009gg}%
  \BibitemOpen
  \bibfield  {author} {\bibinfo {author} {\bibfnamefont {John~W.}\ \bibnamefont
  {Barrett}}, \bibinfo {author} {\bibfnamefont {R.J.}\ \bibnamefont {Dowdall}},
  \bibinfo {author} {\bibfnamefont {Winston~J.}\ \bibnamefont {Fairbairn}},
  \bibinfo {author} {\bibfnamefont {Henrique}\ \bibnamefont {Gomes}}, \ and\
  \bibinfo {author} {\bibfnamefont {Frank}\ \bibnamefont {Hellmann}},\
  }\bibfield  {title} {\enquote {\bibinfo {title} {{Asymptotic analysis of the
  EPRL four-simplex amplitude}},}\ }\href {\doibase 10.1063/1.3244218}
  {\bibfield  {journal} {\bibinfo  {journal} {J.\ Math.\ Phys.}\ }\textbf
  {\bibinfo {volume} {50}},\ \bibinfo {pages} {112504} (\bibinfo {year}
  {2009})},\ \Eprint {http://arxiv.org/abs/0902.1170} {arXiv:0902.1170 [gr-qc]}
  \BibitemShut {NoStop}%
\bibitem [{\citenamefont {Kaminski}\ \emph {et~al.}(2018)\citenamefont
  {Kaminski}, \citenamefont {Kisielowski},\ and\ \citenamefont
  {Sahlmann}}]{Kaminski:2017eew}%
  \BibitemOpen
  \bibfield  {author} {\bibinfo {author} {\bibfnamefont {Wojciech}\
  \bibnamefont {Kaminski}}, \bibinfo {author} {\bibfnamefont {Marcin}\
  \bibnamefont {Kisielowski}}, \ and\ \bibinfo {author} {\bibfnamefont {Hanno}\
  \bibnamefont {Sahlmann}},\ }\bibfield  {title} {\enquote {\bibinfo {title}
  {{Asymptotic analysis of the EPRL model with timelike tetrahedra}},}\ }\href
  {\doibase 10.1088/1361-6382/aac6a4} {\bibfield  {journal} {\bibinfo
  {journal} {Class.\ Quant.\ Grav.}\ }\textbf {\bibinfo {volume} {35}},\
  \bibinfo {pages} {135012} (\bibinfo {year} {2018})},\ \Eprint
  {http://arxiv.org/abs/1705.02862} {arXiv:1705.02862 [gr-qc]} \BibitemShut
  {NoStop}%
\bibitem [{\citenamefont {Liu}\ and\ \citenamefont {Han}(2019)}]{Liu:2018gfc}%
  \BibitemOpen
  \bibfield  {author} {\bibinfo {author} {\bibfnamefont {Hongguang}\
  \bibnamefont {Liu}}\ and\ \bibinfo {author} {\bibfnamefont {Muxin}\
  \bibnamefont {Han}},\ }\bibfield  {title} {\enquote {\bibinfo {title}
  {{Asymptotic analysis of spin foam amplitude with timelike triangles}},}\
  }\href {\doibase 10.1103/PhysRevD.99.084040} {\bibfield  {journal} {\bibinfo
  {journal} {Phys.\ Rev.\ D}\ }\textbf {\bibinfo {volume} {99}},\ \bibinfo
  {pages} {084040} (\bibinfo {year} {2019})},\ \Eprint
  {http://arxiv.org/abs/1810.09042} {arXiv:1810.09042 [gr-qc]} \BibitemShut
  {NoStop}%
\bibitem [{\citenamefont {Regge}(1961)}]{Regge:1961px}%
  \BibitemOpen
  \bibfield  {author} {\bibinfo {author} {\bibfnamefont {T.}~\bibnamefont
  {Regge}},\ }\bibfield  {title} {\enquote {\bibinfo {title} {{GENERAL
  RELATIVITY WITHOUT COORDINATES}},}\ }\href {\doibase 10.1007/BF02733251}
  {\bibfield  {journal} {\bibinfo  {journal} {Nuovo Cim.}\ }\textbf {\bibinfo
  {volume} {19}},\ \bibinfo {pages} {558--571} (\bibinfo {year}
  {1961})}\BibitemShut {NoStop}%
\bibitem [{\citenamefont {Barrett}\ \emph {et~al.}(2018)\citenamefont
  {Barrett}, \citenamefont {Oriti},\ and\ \citenamefont
  {Williams}}]{Barrett:2018ybl}%
  \BibitemOpen
  \bibfield  {author} {\bibinfo {author} {\bibfnamefont {John~W.}\ \bibnamefont
  {Barrett}}, \bibinfo {author} {\bibfnamefont {Daniele}\ \bibnamefont
  {Oriti}}, \ and\ \bibinfo {author} {\bibfnamefont {Ruth~M.}\ \bibnamefont
  {Williams}},\ }\enquote {\bibinfo {title} {{Tullio Regge's legacy: Regge
  calculus and discrete gravity}},}\ \ (\bibinfo {year} {2018})\ \Eprint
  {http://arxiv.org/abs/1812.06193} {arXiv:1812.06193 [gr-qc]} \BibitemShut
  {NoStop}%
\bibitem [{\citenamefont {Donà}\ \emph {et~al.}(2018)\citenamefont {Donà},
  \citenamefont {Fanizza}, \citenamefont {Sarno},\ and\ \citenamefont
  {Speziale}}]{Dona:2017dvf}%
  \BibitemOpen
  \bibfield  {author} {\bibinfo {author} {\bibfnamefont {Pietro}\ \bibnamefont
  {Donà}}, \bibinfo {author} {\bibfnamefont {Marco}\ \bibnamefont {Fanizza}},
  \bibinfo {author} {\bibfnamefont {Giorgio}\ \bibnamefont {Sarno}}, \ and\
  \bibinfo {author} {\bibfnamefont {Simone}\ \bibnamefont {Speziale}},\
  }\bibfield  {title} {\enquote {\bibinfo {title} {{SU(2) graph invariants,
  Regge actions and polytopes}},}\ }\href {\doibase 10.1088/1361-6382/aaa53a}
  {\bibfield  {journal} {\bibinfo  {journal} {Class.\ Quant.\ Grav.}\ }\textbf
  {\bibinfo {volume} {35}},\ \bibinfo {pages} {045011} (\bibinfo {year}
  {2018})},\ \Eprint {http://arxiv.org/abs/1708.01727} {arXiv:1708.01727
  [gr-qc]} \BibitemShut {NoStop}%
\bibitem [{\citenamefont {Donà}\ \emph {et~al.}(2019)\citenamefont {Donà},
  \citenamefont {Fanizza}, \citenamefont {Sarno},\ and\ \citenamefont
  {Speziale}}]{Dona:2019dkf}%
  \BibitemOpen
  \bibfield  {author} {\bibinfo {author} {\bibfnamefont {Pietro}\ \bibnamefont
  {Donà}}, \bibinfo {author} {\bibfnamefont {Marco}\ \bibnamefont {Fanizza}},
  \bibinfo {author} {\bibfnamefont {Giorgio}\ \bibnamefont {Sarno}}, \ and\
  \bibinfo {author} {\bibfnamefont {Simone}\ \bibnamefont {Speziale}},\
  }\bibfield  {title} {\enquote {\bibinfo {title} {{Numerical study of the
  Lorentzian Engle-Pereira-Rovelli-Livine spin foam amplitude}},}\ }\href
  {\doibase 10.1103/PhysRevD.100.106003} {\bibfield  {journal} {\bibinfo
  {journal} {Phys.\ Rev.\ D}\ }\textbf {\bibinfo {volume} {100}},\ \bibinfo
  {pages} {106003} (\bibinfo {year} {2019})},\ \Eprint
  {http://arxiv.org/abs/1903.12624} {arXiv:1903.12624 [gr-qc]} \BibitemShut
  {NoStop}%
\bibitem [{\citenamefont {Dittrich}\ and\ \citenamefont
  {Steinhaus}(2014)}]{Dittrich:2013xwa}%
  \BibitemOpen
  \bibfield  {author} {\bibinfo {author} {\bibfnamefont {Bianca}\ \bibnamefont
  {Dittrich}}\ and\ \bibinfo {author} {\bibfnamefont {Sebastian}\ \bibnamefont
  {Steinhaus}},\ }\bibfield  {title} {\enquote {\bibinfo {title} {{Time
  evolution as refining, coarse graining and entangling}},}\ }\href {\doibase
  10.1088/1367-2630/16/12/123041} {\bibfield  {journal} {\bibinfo  {journal}
  {New J.\ Phys.}\ }\textbf {\bibinfo {volume} {16}},\ \bibinfo {pages}
  {123041} (\bibinfo {year} {2014})},\ \Eprint {http://arxiv.org/abs/1311.7565}
  {arXiv:1311.7565 [gr-qc]} \BibitemShut {NoStop}%
\bibitem [{\citenamefont {Dittrich}(2017{\natexlab{a}})}]{Dittrich:2014ala}%
  \BibitemOpen
  \bibfield  {author} {\bibinfo {author} {\bibfnamefont {Bianca}\ \bibnamefont
  {Dittrich}},\ }\enquote {\bibinfo {title} {{The continuum limit of loop
  quantum gravity - a framework for solving the theory}},}\ in\ \href {\doibase
  10.1142/9789813220003\_0006} {\emph {\bibinfo {booktitle} {{Loop Quantum
  Gravity}: {The First 30 Years}}}}\ (\bibinfo {year} {2017})\ pp.\ \bibinfo
  {pages} {153--179},\ \Eprint {http://arxiv.org/abs/1409.1450}
  {arXiv:1409.1450 [gr-qc]} \BibitemShut {NoStop}%
\bibitem [{\citenamefont {Dirac}(1964)}]{Dirac}%
  \BibitemOpen
  \bibfield  {author} {\bibinfo {author} {\bibfnamefont {P.A.}\ \bibnamefont
  {Dirac}},\ }\href@noop {} {\emph {\bibinfo {title} {Lectures on Quantum
  Mechanics}}}\ (\bibinfo  {publisher} {Yeshiva University Press},\ \bibinfo
  {year} {1964})\BibitemShut {NoStop}%
\bibitem [{\citenamefont {Henneaux}\ and\ \citenamefont
  {Teitelboim}(1992)}]{Henneaux:1992ig}%
  \BibitemOpen
  \bibfield  {author} {\bibinfo {author} {\bibfnamefont {M.}~\bibnamefont
  {Henneaux}}\ and\ \bibinfo {author} {\bibfnamefont {C.}~\bibnamefont
  {Teitelboim}},\ }\href@noop {} {\emph {\bibinfo {title} {{Quantization of
  gauge systems}}}}\ (\bibinfo  {publisher} {Princeton University Press},\
  \bibinfo {year} {1992})\BibitemShut {NoStop}%
\bibitem [{\citenamefont {Dittrich}(2008)}]{Dittrich:2008pw}%
  \BibitemOpen
  \bibfield  {author} {\bibinfo {author} {\bibfnamefont {Bianca}\ \bibnamefont
  {Dittrich}},\ }\bibfield  {title} {\enquote {\bibinfo {title}
  {{Diffeomorphism symmetry in quantum gravity models}},}\ }\href {\doibase
  10.1166/asl.2009.1022} {\bibfield  {journal} {\bibinfo  {journal} {Adv.\
  Sci.\ Lett.}\ }\textbf {\bibinfo {volume} {2}},\ \bibinfo {pages} {151}
  (\bibinfo {year} {2008})},\ \Eprint {http://arxiv.org/abs/0810.3594}
  {arXiv:0810.3594 [gr-qc]} \BibitemShut {NoStop}%
\bibitem [{\citenamefont {Bahr}\ and\ \citenamefont
  {Dittrich}(2009{\natexlab{a}})}]{Bahr:2009ku}%
  \BibitemOpen
  \bibfield  {author} {\bibinfo {author} {\bibfnamefont {Benjamin}\
  \bibnamefont {Bahr}}\ and\ \bibinfo {author} {\bibfnamefont {Bianca}\
  \bibnamefont {Dittrich}},\ }\bibfield  {title} {\enquote {\bibinfo {title}
  {{(Broken) Gauge Symmetries and Constraints in Regge Calculus}},}\ }\href
  {\doibase 10.1088/0264-9381/26/22/225011} {\bibfield  {journal} {\bibinfo
  {journal} {Class.\ Quant.\ Grav.}\ }\textbf {\bibinfo {volume} {26}},\
  \bibinfo {pages} {225011} (\bibinfo {year} {2009}{\natexlab{a}})},\ \Eprint
  {http://arxiv.org/abs/0905.1670} {arXiv:0905.1670 [gr-qc]} \BibitemShut
  {NoStop}%
\bibitem [{\citenamefont {Bahr}\ and\ \citenamefont
  {Steinhaus}(2016{\natexlab{a}})}]{Bahr:2015gxa}%
  \BibitemOpen
  \bibfield  {author} {\bibinfo {author} {\bibfnamefont {Benjamin}\
  \bibnamefont {Bahr}}\ and\ \bibinfo {author} {\bibfnamefont {Sebastian}\
  \bibnamefont {Steinhaus}},\ }\bibfield  {title} {\enquote {\bibinfo {title}
  {{Investigation of the Spinfoam Path integral with Quantum Cuboid
  Intertwiners}},}\ }\href {\doibase 10.1103/PhysRevD.93.104029} {\bibfield
  {journal} {\bibinfo  {journal} {Phys.\ Rev.\ D}\ }\textbf {\bibinfo {volume}
  {93}},\ \bibinfo {pages} {104029} (\bibinfo {year} {2016}{\natexlab{a}})},\
  \Eprint {http://arxiv.org/abs/1508.07961} {arXiv:1508.07961 [gr-qc]}
  \BibitemShut {NoStop}%
\bibitem [{\citenamefont {Bahr}\ \emph {et~al.}(2011)\citenamefont {Bahr},
  \citenamefont {Dittrich},\ and\ \citenamefont {Steinhaus}}]{Bahr:2011uj}%
  \BibitemOpen
  \bibfield  {author} {\bibinfo {author} {\bibfnamefont {Benjamin}\
  \bibnamefont {Bahr}}, \bibinfo {author} {\bibfnamefont {Bianca}\ \bibnamefont
  {Dittrich}}, \ and\ \bibinfo {author} {\bibfnamefont {Sebastian}\
  \bibnamefont {Steinhaus}},\ }\bibfield  {title} {\enquote {\bibinfo {title}
  {{Perfect discretization of reparametrization invariant path integrals}},}\
  }\href {\doibase 10.1103/PhysRevD.83.105026} {\bibfield  {journal} {\bibinfo
  {journal} {Phys.\ Rev.\ D}\ }\textbf {\bibinfo {volume} {83}},\ \bibinfo
  {pages} {105026} (\bibinfo {year} {2011})},\ \Eprint
  {http://arxiv.org/abs/1101.4775} {arXiv:1101.4775 [gr-qc]} \BibitemShut
  {NoStop}%
\bibitem [{\citenamefont {Bahr}\ and\ \citenamefont
  {Dittrich}(2009{\natexlab{b}})}]{Bahr:2009qc}%
  \BibitemOpen
  \bibfield  {author} {\bibinfo {author} {\bibfnamefont {Benjamin}\
  \bibnamefont {Bahr}}\ and\ \bibinfo {author} {\bibfnamefont {Bianca}\
  \bibnamefont {Dittrich}},\ }\bibfield  {title} {\enquote {\bibinfo {title}
  {{Improved and Perfect Actions in Discrete Gravity}},}\ }\href {\doibase
  10.1103/PhysRevD.80.124030} {\bibfield  {journal} {\bibinfo  {journal}
  {Phys.\ Rev.\ D}\ }\textbf {\bibinfo {volume} {80}},\ \bibinfo {pages}
  {124030} (\bibinfo {year} {2009}{\natexlab{b}})},\ \Eprint
  {http://arxiv.org/abs/0907.4323} {arXiv:0907.4323 [gr-qc]} \BibitemShut
  {NoStop}%
\bibitem [{\citenamefont {Dittrich}(2011)}]{Dittrich:2012qb}%
  \BibitemOpen
  \bibfield  {author} {\bibinfo {author} {\bibfnamefont {Bianca}\ \bibnamefont
  {Dittrich}},\ }\bibfield  {title} {\enquote {\bibinfo {title} {{How to
  construct diffeomorphism symmetry on the lattice}},}\ }\href {\doibase
  10.22323/1.140.0012} {\bibfield  {journal} {\bibinfo  {journal} {PoS}\
  }\textbf {\bibinfo {volume} {QGQGS2011}},\ \bibinfo {pages} {012} (\bibinfo
  {year} {2011})},\ \Eprint {http://arxiv.org/abs/1201.3840} {arXiv:1201.3840
  [gr-qc]} \BibitemShut {NoStop}%
\bibitem [{\citenamefont {Dittrich}(2012)}]{Dittrich:2012jq}%
  \BibitemOpen
  \bibfield  {author} {\bibinfo {author} {\bibfnamefont {Bianca}\ \bibnamefont
  {Dittrich}},\ }\bibfield  {title} {\enquote {\bibinfo {title} {{From the
  discrete to the continuous: Towards a cylindrically consistent dynamics}},}\
  }\href {\doibase 10.1088/1367-2630/14/12/123004} {\bibfield  {journal}
  {\bibinfo  {journal} {New J.\ Phys.}\ }\textbf {\bibinfo {volume} {14}},\
  \bibinfo {pages} {123004} (\bibinfo {year} {2012})},\ \Eprint
  {http://arxiv.org/abs/1205.6127} {arXiv:1205.6127 [gr-qc]} \BibitemShut
  {NoStop}%
\bibitem [{\citenamefont {Rovelli}\ and\ \citenamefont
  {Smerlak}(2012)}]{Rovelli:2010qx}%
  \BibitemOpen
  \bibfield  {author} {\bibinfo {author} {\bibfnamefont {Carlo}\ \bibnamefont
  {Rovelli}}\ and\ \bibinfo {author} {\bibfnamefont {Matteo}\ \bibnamefont
  {Smerlak}},\ }\bibfield  {title} {\enquote {\bibinfo {title} {{In quantum
  gravity, summing is refining}},}\ }\href {\doibase
  10.1088/0264-9381/29/5/055004} {\bibfield  {journal} {\bibinfo  {journal}
  {Class.\ Quant.\ Grav.}\ }\textbf {\bibinfo {volume} {29}},\ \bibinfo {pages}
  {055004} (\bibinfo {year} {2012})},\ \Eprint {http://arxiv.org/abs/1010.5437}
  {arXiv:1010.5437 [gr-qc]} \BibitemShut {NoStop}%
\bibitem [{\citenamefont {Freidel}(2005)}]{Freidel:2005qe}%
  \BibitemOpen
  \bibfield  {author} {\bibinfo {author} {\bibfnamefont {Laurent}\ \bibnamefont
  {Freidel}},\ }\bibfield  {title} {\enquote {\bibinfo {title} {{Group field
  theory: An Overview}},}\ }\href {\doibase 10.1007/s10773-005-8894-1}
  {\bibfield  {journal} {\bibinfo  {journal} {Int.\ J.\ Theor.\ Phys.}\
  }\textbf {\bibinfo {volume} {44}},\ \bibinfo {pages} {1769--1783} (\bibinfo
  {year} {2005})},\ \Eprint {http://arxiv.org/abs/hep-th/0505016}
  {arXiv:hep-th/0505016} \BibitemShut {NoStop}%
\bibitem [{\citenamefont {Oriti}(2016)}]{Oriti:2013aqa}%
  \BibitemOpen
  \bibfield  {author} {\bibinfo {author} {\bibfnamefont {Daniele}\ \bibnamefont
  {Oriti}},\ }\bibfield  {title} {\enquote {\bibinfo {title} {{Group field
  theory as the 2nd quantization of Loop Quantum Gravity}},}\ }\href {\doibase
  10.1088/0264-9381/33/8/085005} {\bibfield  {journal} {\bibinfo  {journal}
  {Class.\ Quant.\ Grav.}\ }\textbf {\bibinfo {volume} {33}},\ \bibinfo {pages}
  {085005} (\bibinfo {year} {2016})},\ \Eprint {http://arxiv.org/abs/1310.7786}
  {arXiv:1310.7786 [gr-qc]} \BibitemShut {NoStop}%
\bibitem [{\citenamefont {Carrozza}(2016)}]{Carrozza:2016vsq}%
  \BibitemOpen
  \bibfield  {author} {\bibinfo {author} {\bibfnamefont {Sylvain}\ \bibnamefont
  {Carrozza}},\ }\bibfield  {title} {\enquote {\bibinfo {title} {{Flowing in
  Group Field Theory Space: a Review}},}\ }\href {\doibase
  10.3842/SIGMA.2016.070} {\bibfield  {journal} {\bibinfo  {journal} {SIGMA}\
  }\textbf {\bibinfo {volume} {12}},\ \bibinfo {pages} {070} (\bibinfo {year}
  {2016})},\ \Eprint {http://arxiv.org/abs/1603.01902} {arXiv:1603.01902
  [gr-qc]} \BibitemShut {NoStop}%
\bibitem [{\citenamefont {Dona}\ and\ \citenamefont
  {Sarno}(2018)}]{Dona:2018nev}%
  \BibitemOpen
  \bibfield  {author} {\bibinfo {author} {\bibfnamefont {Pietro}\ \bibnamefont
  {Dona}}\ and\ \bibinfo {author} {\bibfnamefont {Giorgio}\ \bibnamefont
  {Sarno}},\ }\bibfield  {title} {\enquote {\bibinfo {title} {{Numerical
  methods for EPRL spin foam transition amplitudes and Lorentzian recoupling
  theory}},}\ }\href {\doibase 10.1007/s10714-018-2452-7} {\bibfield  {journal}
  {\bibinfo  {journal} {Gen.\ Rel.\ Grav.}\ }\textbf {\bibinfo {volume} {50}},\
  \bibinfo {pages} {127} (\bibinfo {year} {2018})},\ \Eprint
  {http://arxiv.org/abs/1807.03066} {arXiv:1807.03066 [gr-qc]} \BibitemShut
  {NoStop}%
\bibitem [{\citenamefont {Levin}\ and\ \citenamefont {Nave}(2007)}]{Levin}%
  \BibitemOpen
  \bibfield  {author} {\bibinfo {author} {\bibfnamefont {M.}~\bibnamefont
  {Levin}}\ and\ \bibinfo {author} {\bibfnamefont {C.~P.}\ \bibnamefont
  {Nave}},\ }\bibfield  {title} {\enquote {\bibinfo {title} {Tensor
  renormalization group approach to 2d classical lattice models},}\ }\href
  {\doibase 10.1103/PhysRevLett.99.120601} {\bibfield  {journal} {\bibinfo
  {journal} {Phys. Rev. Lett.}\ }\textbf {\bibinfo {volume} {99}},\ \bibinfo
  {pages} {120601} (\bibinfo {year} {2007})},\ \Eprint
  {http://arxiv.org/abs/cond-mat/0611687} {arXiv:cond-mat/0611687 [cond-mat]}
  \BibitemShut {NoStop}%
\bibitem [{\citenamefont {Gu}\ and\ \citenamefont {Wen}(2009)}]{GuWen}%
  \BibitemOpen
  \bibfield  {author} {\bibinfo {author} {\bibfnamefont {Z.-C.}\ \bibnamefont
  {Gu}}\ and\ \bibinfo {author} {\bibfnamefont {X.-G.}\ \bibnamefont {Wen}},\
  }\bibfield  {title} {\enquote {\bibinfo {title}
  {Tensor-{E}ntanglement-{F}iltering {R}enormalization {A}pproach and
  {S}ymmetry {P}rotected {T}opological {O}rder},}\ }\href {\doibase
  10.1103/PhysRevB.80.155131} {\bibfield  {journal} {\bibinfo  {journal} {Phys.
  Rev. B}\ }\textbf {\bibinfo {volume} {80}},\ \bibinfo {pages} {155131}
  (\bibinfo {year} {2009})},\ \Eprint {http://arxiv.org/abs/0903.1069}
  {arXiv:0903.1069 [cond-mat.str-el]} \BibitemShut {NoStop}%
\bibitem [{\citenamefont {Evenbly}\ and\ \citenamefont
  {Vidal}(2015)}]{vidal-TNR}%
  \BibitemOpen
  \bibfield  {author} {\bibinfo {author} {\bibfnamefont {G.}~\bibnamefont
  {Evenbly}}\ and\ \bibinfo {author} {\bibfnamefont {G.}~\bibnamefont
  {Vidal}},\ }\bibfield  {title} {\enquote {\bibinfo {title} {Tensor network
  renormalization},}\ }\href {\doibase 10.1103/PhysRevLett.115.180405}
  {\bibfield  {journal} {\bibinfo  {journal} {Phys. Rev. Lett.}\ }\textbf
  {\bibinfo {volume} {115}},\ \bibinfo {pages} {180405} (\bibinfo {year}
  {2015})}\BibitemShut {NoStop}%
\bibitem [{\citenamefont {Cunningham}\ \emph {et~al.}(2020)\citenamefont
  {Cunningham}, \citenamefont {Dittrich},\ and\ \citenamefont
  {Steinhaus}}]{Cunningham:2020uco}%
  \BibitemOpen
  \bibfield  {author} {\bibinfo {author} {\bibfnamefont {William~J.}\
  \bibnamefont {Cunningham}}, \bibinfo {author} {\bibfnamefont {Bianca}\
  \bibnamefont {Dittrich}}, \ and\ \bibinfo {author} {\bibfnamefont
  {Sebastian}\ \bibnamefont {Steinhaus}},\ }\bibfield  {title} {\enquote
  {\bibinfo {title} {{Tensor network renormalization with fusion charges:
  applications to 3d lattice gauge theory}},}\ }\href@noop {} {\  (\bibinfo
  {year} {2020})},\ \Eprint {http://arxiv.org/abs/2002.10472} {arXiv:2002.10472
  [hep-th]} \BibitemShut {NoStop}%
\bibitem [{\citenamefont {Rocek}\ and\ \citenamefont
  {Williams}(1981)}]{Rocek:1982fr}%
  \BibitemOpen
  \bibfield  {author} {\bibinfo {author} {\bibfnamefont {M.}~\bibnamefont
  {Rocek}}\ and\ \bibinfo {author} {\bibfnamefont {Ruth~M.}\ \bibnamefont
  {Williams}},\ }\bibfield  {title} {\enquote {\bibinfo {title} {{QUANTUM REGGE
  CALCULUS}},}\ }\href {\doibase 10.1016/0370-2693(81)90848-0} {\bibfield
  {journal} {\bibinfo  {journal} {Phys.\ Lett.\ B}\ }\textbf {\bibinfo {volume}
  {104}},\ \bibinfo {pages} {31} (\bibinfo {year} {1981})}\BibitemShut
  {NoStop}%
\bibitem [{\citenamefont {Pachner}(1986)}]{pachner}%
  \BibitemOpen
  \bibfield  {author} {\bibinfo {author} {\bibfnamefont {U.}~\bibnamefont
  {Pachner}},\ }\bibfield  {title} {\enquote {\bibinfo {title}
  {Konstruktionsmethoden und das kombinatorische {H}om{\"o}omorphieproblem
  f{\"u}r {T}riangulationen kompakter semilinearer {M}annigfaltigkeiten},}\
  }\href {\doibase 10.1007/BF02941601} {\bibfield  {journal} {\bibinfo
  {journal} {Abh. Math. Sem. Univ. Hamburg}\ }\textbf {\bibinfo {volume}
  {57}},\ \bibinfo {pages} {69} (\bibinfo {year} {1986})}\BibitemShut {NoStop}%
\bibitem [{\citenamefont {Pachner}(1991)}]{pachner1}%
  \BibitemOpen
  \bibfield  {author} {\bibinfo {author} {\bibfnamefont {U.}~\bibnamefont
  {Pachner}},\ }\bibfield  {title} {\enquote {\bibinfo {title} {P. {L}.
  {H}omeomorphic {M}anifolds are {E}quivalent by {E}lementary {S}hellings},}\
  }\href {\doibase 10.1016/S0195-6698(13)80080-7} {\bibfield  {journal}
  {\bibinfo  {journal} {Europ. J. Combinatorics}\ }\textbf {\bibinfo {volume}
  {12}},\ \bibinfo {pages} {129--145} (\bibinfo {year} {1991})}\BibitemShut
  {NoStop}%
\bibitem [{\citenamefont {Turaev}\ and\ \citenamefont
  {Viro}(1992)}]{Turaev:1992hq}%
  \BibitemOpen
  \bibfield  {author} {\bibinfo {author} {\bibfnamefont {V.G.}\ \bibnamefont
  {Turaev}}\ and\ \bibinfo {author} {\bibfnamefont {O.Y.}\ \bibnamefont
  {Viro}},\ }\bibfield  {title} {\enquote {\bibinfo {title} {{State sum
  invariants of 3 manifolds and quantum 6j symbols}},}\ }\href {\doibase
  10.1016/0040-9383(92)90015-A} {\bibfield  {journal} {\bibinfo  {journal}
  {Topology}\ }\textbf {\bibinfo {volume} {31}},\ \bibinfo {pages} {865--902}
  (\bibinfo {year} {1992})}\BibitemShut {NoStop}%
\bibitem [{\citenamefont {Dittrich}\ and\ \citenamefont
  {Steinhaus}(2012)}]{Dittrich:2011vz}%
  \BibitemOpen
  \bibfield  {author} {\bibinfo {author} {\bibfnamefont {Bianca}\ \bibnamefont
  {Dittrich}}\ and\ \bibinfo {author} {\bibfnamefont {Sebastian}\ \bibnamefont
  {Steinhaus}},\ }\bibfield  {title} {\enquote {\bibinfo {title} {{Path
  integral measure and triangulation independence in discrete gravity}},}\
  }\href {\doibase 10.1103/PhysRevD.85.044032} {\bibfield  {journal} {\bibinfo
  {journal} {Phys.\ Rev.\ D}\ }\textbf {\bibinfo {volume} {85}},\ \bibinfo
  {pages} {044032} (\bibinfo {year} {2012})},\ \Eprint
  {http://arxiv.org/abs/1110.6866} {arXiv:1110.6866 [gr-qc]} \BibitemShut
  {NoStop}%
\bibitem [{\citenamefont {Dittrich}\ \emph
  {et~al.}(2014{\natexlab{a}})\citenamefont {Dittrich}, \citenamefont
  {Kami\'nski},\ and\ \citenamefont {Steinhaus}}]{Dittrich:2014rha}%
  \BibitemOpen
  \bibfield  {author} {\bibinfo {author} {\bibfnamefont {Bianca}\ \bibnamefont
  {Dittrich}}, \bibinfo {author} {\bibfnamefont {Wojciech}\ \bibnamefont
  {Kami\'nski}}, \ and\ \bibinfo {author} {\bibfnamefont {Sebastian}\
  \bibnamefont {Steinhaus}},\ }\bibfield  {title} {\enquote {\bibinfo {title}
  {{Discretization independence implies non-locality in 4D discrete quantum
  gravity}},}\ }\href {\doibase 10.1088/0264-9381/31/24/245009} {\bibfield
  {journal} {\bibinfo  {journal} {Class.\ Quant.\ Grav.}\ }\textbf {\bibinfo
  {volume} {31}},\ \bibinfo {pages} {245009} (\bibinfo {year}
  {2014}{\natexlab{a}})},\ \Eprint {http://arxiv.org/abs/1404.5288}
  {arXiv:1404.5288 [gr-qc]} \BibitemShut {NoStop}%
\bibitem [{\citenamefont {Banburski}\ \emph {et~al.}(2015)\citenamefont
  {Banburski}, \citenamefont {Chen}, \citenamefont {Freidel},\ and\
  \citenamefont {Hnybida}}]{Banburski:2014cwa}%
  \BibitemOpen
  \bibfield  {author} {\bibinfo {author} {\bibfnamefont {Andrzej}\ \bibnamefont
  {Banburski}}, \bibinfo {author} {\bibfnamefont {Lin-Qing}\ \bibnamefont
  {Chen}}, \bibinfo {author} {\bibfnamefont {Laurent}\ \bibnamefont {Freidel}},
  \ and\ \bibinfo {author} {\bibfnamefont {Jeff}\ \bibnamefont {Hnybida}},\
  }\bibfield  {title} {\enquote {\bibinfo {title} {{Pachner moves in a 4d
  Riemannian holomorphic Spin Foam model}},}\ }\href {\doibase
  10.1103/PhysRevD.92.124014} {\bibfield  {journal} {\bibinfo  {journal}
  {Phys.\ Rev.\ D}\ }\textbf {\bibinfo {volume} {92}},\ \bibinfo {pages}
  {124014} (\bibinfo {year} {2015})},\ \Eprint {http://arxiv.org/abs/1412.8247}
  {arXiv:1412.8247 [gr-qc]} \BibitemShut {NoStop}%
\bibitem [{\citenamefont {Dupuis}\ and\ \citenamefont
  {Livine}(2011)}]{Dupuis:2011fz}%
  \BibitemOpen
  \bibfield  {author} {\bibinfo {author} {\bibfnamefont {Maite}\ \bibnamefont
  {Dupuis}}\ and\ \bibinfo {author} {\bibfnamefont {Etera~R.}\ \bibnamefont
  {Livine}},\ }\bibfield  {title} {\enquote {\bibinfo {title} {{Holomorphic
  Simplicity Constraints for 4d Spinfoam Models}},}\ }\href {\doibase
  10.1088/0264-9381/28/21/215022} {\bibfield  {journal} {\bibinfo  {journal}
  {Class.\ Quant.\ Grav.}\ }\textbf {\bibinfo {volume} {28}},\ \bibinfo {pages}
  {215022} (\bibinfo {year} {2011})},\ \Eprint {http://arxiv.org/abs/1104.3683}
  {arXiv:1104.3683 [gr-qc]} \BibitemShut {NoStop}%
\bibitem [{\citenamefont {Onsager}(1944)}]{PhysRev.65.117}%
  \BibitemOpen
  \bibfield  {author} {\bibinfo {author} {\bibfnamefont {Lars}\ \bibnamefont
  {Onsager}},\ }\bibfield  {title} {\enquote {\bibinfo {title} {Crystal
  statistics. i. a two-dimensional model with an order-disorder transition},}\
  }\href {\doibase 10.1103/PhysRev.65.117} {\bibfield  {journal} {\bibinfo
  {journal} {Phys. Rev.}\ }\textbf {\bibinfo {volume} {65}},\ \bibinfo {pages}
  {117--149} (\bibinfo {year} {1944})}\BibitemShut {NoStop}%
\bibitem [{\citenamefont {Efrati}\ \emph {et~al.}(2014)\citenamefont {Efrati},
  \citenamefont {Wang}, \citenamefont {Kolan},\ and\ \citenamefont
  {Kadanoff}}]{RevModPhys.86.647}%
  \BibitemOpen
  \bibfield  {author} {\bibinfo {author} {\bibfnamefont {Efi}\ \bibnamefont
  {Efrati}}, \bibinfo {author} {\bibfnamefont {Zhe}\ \bibnamefont {Wang}},
  \bibinfo {author} {\bibfnamefont {Amy}\ \bibnamefont {Kolan}}, \ and\
  \bibinfo {author} {\bibfnamefont {Leo~P.}\ \bibnamefont {Kadanoff}},\
  }\bibfield  {title} {\enquote {\bibinfo {title} {Real-space renormalization
  in statistical mechanics},}\ }\href {\doibase 10.1103/RevModPhys.86.647}
  {\bibfield  {journal} {\bibinfo  {journal} {Rev. Mod. Phys.}\ }\textbf
  {\bibinfo {volume} {86}},\ \bibinfo {pages} {647--667} (\bibinfo {year}
  {2014})},\ \Eprint {http://arxiv.org/abs/1301.6323} {arXiv:1301.6323
  [cond-mat.stat-mech]} \BibitemShut {NoStop}%
\bibitem [{\citenamefont {Ashtekar}\ and\ \citenamefont
  {Isham}(1992)}]{ashtekar-lewan1}%
  \BibitemOpen
  \bibfield  {author} {\bibinfo {author} {\bibfnamefont {Abhay}\ \bibnamefont
  {Ashtekar}}\ and\ \bibinfo {author} {\bibfnamefont {C.J.}\ \bibnamefont
  {Isham}},\ }\bibfield  {title} {\enquote {\bibinfo {title} {{Representations
  of the holonomy algebras of gravity and nonAbelian gauge theories}},}\ }\href
  {\doibase 10.1088/0264-9381/9/6/004} {\bibfield  {journal} {\bibinfo
  {journal} {Class.Quant.Grav.}\ }\textbf {\bibinfo {volume} {9}},\ \bibinfo
  {pages} {1433--1468} (\bibinfo {year} {1992})},\ \Eprint
  {http://arxiv.org/abs/hep-th/9202053} {arXiv:hep-th/9202053 [hep-th]}
  \BibitemShut {NoStop}%
%%CITATION = HEP-TH/9202053;%%
\bibitem [{\citenamefont {Ashtekar}\ and\ \citenamefont
  {Lewandowski}(1995)}]{ashtekar-lewan2}%
  \BibitemOpen
  \bibfield  {author} {\bibinfo {author} {\bibfnamefont {Abhay}\ \bibnamefont
  {Ashtekar}}\ and\ \bibinfo {author} {\bibfnamefont {Jerzy}\ \bibnamefont
  {Lewandowski}},\ }\bibfield  {title} {\enquote {\bibinfo {title} {{Projective
  techniques and functional integration for gauge theories}},}\ }\href
  {\doibase 10.1063/1.531037} {\bibfield  {journal} {\bibinfo  {journal}
  {J.Math.Phys.}\ }\textbf {\bibinfo {volume} {36}},\ \bibinfo {pages}
  {2170--2191} (\bibinfo {year} {1995})},\ \Eprint
  {http://arxiv.org/abs/gr-qc/9411046} {arXiv:gr-qc/9411046 [gr-qc]}
  \BibitemShut {NoStop}%
%%CITATION = GR-QC/9411046;%%
\bibitem [{\citenamefont {Dittrich}\ and\ \citenamefont
  {Geiller}(2015{\natexlab{a}})}]{Dittrich:2014wpa}%
  \BibitemOpen
  \bibfield  {author} {\bibinfo {author} {\bibfnamefont {Bianca}\ \bibnamefont
  {Dittrich}}\ and\ \bibinfo {author} {\bibfnamefont {Marc}\ \bibnamefont
  {Geiller}},\ }\bibfield  {title} {\enquote {\bibinfo {title} {{A new vacuum
  for Loop Quantum Gravity}},}\ }\href {\doibase
  10.1088/0264-9381/32/11/112001} {\bibfield  {journal} {\bibinfo  {journal}
  {Class.\ Quant.\ Grav.}\ }\textbf {\bibinfo {volume} {32}},\ \bibinfo {pages}
  {112001} (\bibinfo {year} {2015}{\natexlab{a}})},\ \Eprint
  {http://arxiv.org/abs/1401.6441} {arXiv:1401.6441 [gr-qc]} \BibitemShut
  {NoStop}%
\bibitem [{\citenamefont {Dittrich}\ and\ \citenamefont
  {Geiller}(2015{\natexlab{b}})}]{Dittrich:2014wda}%
  \BibitemOpen
  \bibfield  {author} {\bibinfo {author} {\bibfnamefont {Bianca}\ \bibnamefont
  {Dittrich}}\ and\ \bibinfo {author} {\bibfnamefont {Marc}\ \bibnamefont
  {Geiller}},\ }\bibfield  {title} {\enquote {\bibinfo {title} {{Flux
  formulation of loop quantum gravity: Classical framework}},}\ }\href
  {\doibase 10.1088/0264-9381/32/13/135016} {\bibfield  {journal} {\bibinfo
  {journal} {Class.\ Quant.\ Grav.}\ }\textbf {\bibinfo {volume} {32}},\
  \bibinfo {pages} {135016} (\bibinfo {year} {2015}{\natexlab{b}})},\ \Eprint
  {http://arxiv.org/abs/1412.3752} {arXiv:1412.3752 [gr-qc]} \BibitemShut
  {NoStop}%
\bibitem [{\citenamefont {Bahr}\ \emph {et~al.}(2015)\citenamefont {Bahr},
  \citenamefont {Dittrich},\ and\ \citenamefont {Geiller}}]{Bahr:2015bra}%
  \BibitemOpen
  \bibfield  {author} {\bibinfo {author} {\bibfnamefont {Benjamin}\
  \bibnamefont {Bahr}}, \bibinfo {author} {\bibfnamefont {Bianca}\ \bibnamefont
  {Dittrich}}, \ and\ \bibinfo {author} {\bibfnamefont {Marc}\ \bibnamefont
  {Geiller}},\ }\bibfield  {title} {\enquote {\bibinfo {title} {{A new
  realization of quantum geometry}},}\ }\href@noop {} {\  (\bibinfo {year}
  {2015})},\ \Eprint {http://arxiv.org/abs/1506.08571} {arXiv:1506.08571
  [gr-qc]} \BibitemShut {NoStop}%
\bibitem [{\citenamefont {Anderson}(2012)}]{time-anderson2}%
  \BibitemOpen
  \bibfield  {author} {\bibinfo {author} {\bibfnamefont {Edward}\ \bibnamefont
  {Anderson}},\ }\bibfield  {title} {\enquote {\bibinfo {title} {{Problem of
  Time in Quantum Gravity}},}\ }\href {\doibase 10.1002/andp.201200147}
  {\bibfield  {journal} {\bibinfo  {journal} {Annalen Phys.}\ }\textbf
  {\bibinfo {volume} {524}},\ \bibinfo {pages} {757--786} (\bibinfo {year}
  {2012})},\ \Eprint {http://arxiv.org/abs/1206.2403} {arXiv:1206.2403 [gr-qc]}
  \BibitemShut {NoStop}%
%%CITATION = ARXIV:1206.2403;%%
\bibitem [{\citenamefont {Dittrich}\ and\ \citenamefont
  {Hoehn}(2013)}]{Dittrich:2013jaa}%
  \BibitemOpen
  \bibfield  {author} {\bibinfo {author} {\bibfnamefont {Bianca}\ \bibnamefont
  {Dittrich}}\ and\ \bibinfo {author} {\bibfnamefont {Philipp~A}\ \bibnamefont
  {Hoehn}},\ }\bibfield  {title} {\enquote {\bibinfo {title} {{Constraint
  analysis for variational discrete systems}},}\ }\href {\doibase
  10.1063/1.4818895} {\bibfield  {journal} {\bibinfo  {journal} {J.\ Math.\
  Phys.}\ }\textbf {\bibinfo {volume} {54}},\ \bibinfo {pages} {093505}
  (\bibinfo {year} {2013})},\ \Eprint {http://arxiv.org/abs/1303.4294}
  {arXiv:1303.4294 [math-ph]} \BibitemShut {NoStop}%
\bibitem [{\citenamefont {Höhn}(2014{\natexlab{a}})}]{Hoehn:2014fka}%
  \BibitemOpen
  \bibfield  {author} {\bibinfo {author} {\bibfnamefont {Philipp~A.}\
  \bibnamefont {Höhn}},\ }\bibfield  {title} {\enquote {\bibinfo {title}
  {{Quantization of systems with temporally varying discretization I: Evolving
  Hilbert spaces}},}\ }\href {\doibase 10.1063/1.4890558} {\bibfield  {journal}
  {\bibinfo  {journal} {J.\ Math.\ Phys.}\ }\textbf {\bibinfo {volume} {55}},\
  \bibinfo {pages} {083508} (\bibinfo {year} {2014}{\natexlab{a}})},\ \Eprint
  {http://arxiv.org/abs/1401.6062} {arXiv:1401.6062 [gr-qc]} \BibitemShut
  {NoStop}%
\bibitem [{\citenamefont {Höhn}(2014{\natexlab{b}})}]{Hoehn:2014wwa}%
  \BibitemOpen
  \bibfield  {author} {\bibinfo {author} {\bibfnamefont {Philipp~A}\
  \bibnamefont {Höhn}},\ }\bibfield  {title} {\enquote {\bibinfo {title}
  {{Quantization of systems with temporally varying discretization II: Local
  evolution moves}},}\ }\href {\doibase 10.1063/1.4898764} {\bibfield
  {journal} {\bibinfo  {journal} {J.\ Math.\ Phys.}\ }\textbf {\bibinfo
  {volume} {55}},\ \bibinfo {pages} {103507} (\bibinfo {year}
  {2014}{\natexlab{b}})},\ \Eprint {http://arxiv.org/abs/1401.7731}
  {arXiv:1401.7731 [gr-qc]} \BibitemShut {NoStop}%
\bibitem [{\citenamefont {Thiemann}\ and\ \citenamefont
  {Zipfel}(2014)}]{Thiemann:2013lka}%
  \BibitemOpen
  \bibfield  {author} {\bibinfo {author} {\bibfnamefont {Thomas}\ \bibnamefont
  {Thiemann}}\ and\ \bibinfo {author} {\bibfnamefont {Antonia}\ \bibnamefont
  {Zipfel}},\ }\bibfield  {title} {\enquote {\bibinfo {title} {{Linking
  covariant and canonical LQG II: Spin foam projector}},}\ }\href {\doibase
  10.1088/0264-9381/31/12/125008} {\bibfield  {journal} {\bibinfo  {journal}
  {Class.\ Quant.\ Grav.}\ }\textbf {\bibinfo {volume} {31}},\ \bibinfo {pages}
  {125008} (\bibinfo {year} {2014})},\ \Eprint {http://arxiv.org/abs/1307.5885}
  {arXiv:1307.5885 [gr-qc]} \BibitemShut {NoStop}%
\bibitem [{\citenamefont {Loll}(2020)}]{Loll:2019rdj}%
  \BibitemOpen
  \bibfield  {author} {\bibinfo {author} {\bibfnamefont {R.}~\bibnamefont
  {Loll}},\ }\bibfield  {title} {\enquote {\bibinfo {title} {{Quantum Gravity
  from Causal Dynamical Triangulations: A Review}},}\ }\href {\doibase
  10.1088/1361-6382/ab57c7} {\bibfield  {journal} {\bibinfo  {journal} {Class.
  Quant. Grav.}\ }\textbf {\bibinfo {volume} {37}},\ \bibinfo {pages} {013002}
  (\bibinfo {year} {2020})},\ \Eprint {http://arxiv.org/abs/1905.08669}
  {arXiv:1905.08669 [hep-th]} \BibitemShut {NoStop}%
\bibitem [{\citenamefont {Ambjorn}\ \emph {et~al.}(2004)\citenamefont
  {Ambjorn}, \citenamefont {Jurkiewicz},\ and\ \citenamefont
  {Loll}}]{Ambjorn:2004qm}%
  \BibitemOpen
  \bibfield  {author} {\bibinfo {author} {\bibfnamefont {J.}~\bibnamefont
  {Ambjorn}}, \bibinfo {author} {\bibfnamefont {J.}~\bibnamefont {Jurkiewicz}},
  \ and\ \bibinfo {author} {\bibfnamefont {R.}~\bibnamefont {Loll}},\
  }\bibfield  {title} {\enquote {\bibinfo {title} {{Emergence of a 4-D world
  from causal quantum gravity}},}\ }\href {\doibase
  10.1103/PhysRevLett.93.131301} {\bibfield  {journal} {\bibinfo  {journal}
  {Phys. Rev. Lett.}\ }\textbf {\bibinfo {volume} {93}},\ \bibinfo {pages}
  {131301} (\bibinfo {year} {2004})},\ \Eprint
  {http://arxiv.org/abs/hep-th/0404156} {arXiv:hep-th/0404156} \BibitemShut
  {NoStop}%
\bibitem [{\citenamefont {Bahr}(2017)}]{Bahr:2014qza}%
  \BibitemOpen
  \bibfield  {author} {\bibinfo {author} {\bibfnamefont {Benjamin}\
  \bibnamefont {Bahr}},\ }\bibfield  {title} {\enquote {\bibinfo {title} {{On
  background-independent renormalization of spin foam models}},}\ }\href
  {\doibase 10.1088/1361-6382/aa5e13} {\bibfield  {journal} {\bibinfo
  {journal} {Class.\ Quant.\ Grav.}\ }\textbf {\bibinfo {volume} {34}},\
  \bibinfo {pages} {075001} (\bibinfo {year} {2017})},\ \Eprint
  {http://arxiv.org/abs/1407.7746} {arXiv:1407.7746 [gr-qc]} \BibitemShut
  {NoStop}%
\bibitem [{\citenamefont {Orús}(2014)}]{Orus_2014}%
  \BibitemOpen
  \bibfield  {author} {\bibinfo {author} {\bibfnamefont {Román}\ \bibnamefont
  {Orús}},\ }\bibfield  {title} {\enquote {\bibinfo {title} {A practical
  introduction to tensor networks: Matrix product states and projected
  entangled pair states},}\ }\href {\doibase 10.1016/j.aop.2014.06.013}
  {\bibfield  {journal} {\bibinfo  {journal} {Annals of Physics}\ }\textbf
  {\bibinfo {volume} {349}},\ \bibinfo {pages} {117–158} (\bibinfo {year}
  {2014})}\BibitemShut {NoStop}%
\bibitem [{\citenamefont {Vidal}(2008)}]{mera}%
  \BibitemOpen
  \bibfield  {author} {\bibinfo {author} {\bibfnamefont {G.}~\bibnamefont
  {Vidal}},\ }\bibfield  {title} {\enquote {\bibinfo {title} {{Class of Quantum
  Many-Body States That Can Be Efficiently Simulated}},}\ }\href {\doibase
  10.1103/PhysRevLett.101.110501} {\bibfield  {journal} {\bibinfo  {journal}
  {Phys.Rev.Lett.}\ }\textbf {\bibinfo {volume} {101}},\ \bibinfo {pages}
  {110501} (\bibinfo {year} {2008})}\BibitemShut {NoStop}%
%%CITATION = PRLTA,101,110501;%%
\bibitem [{\citenamefont {Hauru}\ \emph {et~al.}(2018)\citenamefont {Hauru},
  \citenamefont {Delcamp},\ and\ \citenamefont {Mizera}}]{Hauru:2017jbf}%
  \BibitemOpen
  \bibfield  {author} {\bibinfo {author} {\bibfnamefont {Markus}\ \bibnamefont
  {Hauru}}, \bibinfo {author} {\bibfnamefont {Clement}\ \bibnamefont
  {Delcamp}}, \ and\ \bibinfo {author} {\bibfnamefont {Sebastian}\ \bibnamefont
  {Mizera}},\ }\bibfield  {title} {\enquote {\bibinfo {title} {{Renormalization
  of tensor networks using graph independent local truncations}},}\ }\href
  {\doibase 10.1103/PhysRevB.97.045111} {\bibfield  {journal} {\bibinfo
  {journal} {Phys.\ Rev.\ B}\ }\textbf {\bibinfo {volume} {97}},\ \bibinfo
  {pages} {045111} (\bibinfo {year} {2018})},\ \Eprint
  {http://arxiv.org/abs/1709.07460} {arXiv:1709.07460 [cond-mat.str-el]}
  \BibitemShut {NoStop}%
\bibitem [{\citenamefont {Ferris}\ and\ \citenamefont
  {Vidal}(2012)}]{Ferris_2012}%
  \BibitemOpen
  \bibfield  {author} {\bibinfo {author} {\bibfnamefont {Andrew~J.}\
  \bibnamefont {Ferris}}\ and\ \bibinfo {author} {\bibfnamefont {Guifre}\
  \bibnamefont {Vidal}},\ }\bibfield  {title} {\enquote {\bibinfo {title}
  {Perfect sampling with unitary tensor networks},}\ }\href {\doibase
  10.1103/physrevb.85.165146} {\bibfield  {journal} {\bibinfo  {journal}
  {Physical Review B}\ }\textbf {\bibinfo {volume} {85}} (\bibinfo {year}
  {2012}),\ 10.1103/physrevb.85.165146}\BibitemShut {NoStop}%
\bibitem [{\citenamefont {Ferris}(2015)}]{ferris2015unbiased}%
  \BibitemOpen
  \bibfield  {author} {\bibinfo {author} {\bibfnamefont {Andrew~J.}\
  \bibnamefont {Ferris}},\ }\href@noop {} {\enquote {\bibinfo {title} {Unbiased
  monte carlo for the age of tensor networks},}\ } (\bibinfo {year} {2015}),\
  \Eprint {http://arxiv.org/abs/1507.00767} {arXiv:1507.00767
  [cond-mat.stat-mech]} \BibitemShut {NoStop}%
\bibitem [{\citenamefont {Bahr}\ \emph
  {et~al.}(2013{\natexlab{a}})\citenamefont {Bahr}, \citenamefont {Dittrich},\
  and\ \citenamefont {Ryan}}]{Bahr:2011yc}%
  \BibitemOpen
  \bibfield  {author} {\bibinfo {author} {\bibfnamefont {Benjamin}\
  \bibnamefont {Bahr}}, \bibinfo {author} {\bibfnamefont {Bianca}\ \bibnamefont
  {Dittrich}}, \ and\ \bibinfo {author} {\bibfnamefont {James~P.}\ \bibnamefont
  {Ryan}},\ }\bibfield  {title} {\enquote {\bibinfo {title} {{Spin foam models
  with finite groups}},}\ }\href {\doibase 10.1155/2013/549824} {\bibfield
  {journal} {\bibinfo  {journal} {J.\ Grav.}\ }\textbf {\bibinfo {volume}
  {2013}},\ \bibinfo {pages} {549824} (\bibinfo {year} {2013}{\natexlab{a}})},\
  \Eprint {http://arxiv.org/abs/1103.6264} {arXiv:1103.6264 [gr-qc]}
  \BibitemShut {NoStop}%
\bibitem [{\citenamefont {Hall}(2015)}]{liegroups}%
  \BibitemOpen
  \bibfield  {author} {\bibinfo {author} {\bibfnamefont {Brian~C.}\
  \bibnamefont {Hall}},\ }\href@noop {} {\emph {\bibinfo {title} {Lie Groups,
  Lie Algebras, and Representations: An Elementary Introduction}}},\ \bibinfo
  {edition} {2nd}\ ed.,\ Vol.\ \bibinfo {volume} {Graduate Texts in
  Mathematics, 222 (2nd ed.)}\ (\bibinfo  {publisher} {Springer},\ \bibinfo
  {year} {2015})\BibitemShut {NoStop}%
\bibitem [{\citenamefont {Kogut}(1979)}]{Kogut:1979wt}%
  \BibitemOpen
  \bibfield  {author} {\bibinfo {author} {\bibfnamefont {John~B.}\ \bibnamefont
  {Kogut}},\ }\bibfield  {title} {\enquote {\bibinfo {title} {{An Introduction
  to Lattice Gauge Theory and Spin Systems}},}\ }\href {\doibase
  10.1103/RevModPhys.51.659} {\bibfield  {journal} {\bibinfo  {journal} {Rev.\
  Mod.\ Phys.}\ }\textbf {\bibinfo {volume} {51}},\ \bibinfo {pages} {659}
  (\bibinfo {year} {1979})}\BibitemShut {NoStop}%
\bibitem [{\citenamefont {Dittrich}\ \emph
  {et~al.}(2014{\natexlab{b}})\citenamefont {Dittrich}, \citenamefont
  {Martin-Benito},\ and\ \citenamefont {Steinhaus}}]{Dittrich:2013voa}%
  \BibitemOpen
  \bibfield  {author} {\bibinfo {author} {\bibfnamefont {Bianca}\ \bibnamefont
  {Dittrich}}, \bibinfo {author} {\bibfnamefont {Mercedes}\ \bibnamefont
  {Martin-Benito}}, \ and\ \bibinfo {author} {\bibfnamefont {Sebastian}\
  \bibnamefont {Steinhaus}},\ }\bibfield  {title} {\enquote {\bibinfo {title}
  {{Quantum group spin nets: refinement limit and relation to spin foams}},}\
  }\href {\doibase 10.1103/PhysRevD.90.024058} {\bibfield  {journal} {\bibinfo
  {journal} {Phys.\ Rev.\ D}\ }\textbf {\bibinfo {volume} {90}},\ \bibinfo
  {pages} {024058} (\bibinfo {year} {2014}{\natexlab{b}})},\ \Eprint
  {http://arxiv.org/abs/1312.0905} {arXiv:1312.0905 [gr-qc]} \BibitemShut
  {NoStop}%
\bibitem [{\citenamefont {Dittrich}\ \emph {et~al.}(2013)\citenamefont
  {Dittrich}, \citenamefont {Martín-Benito},\ and\ \citenamefont
  {Schnetter}}]{Dittrich:2013uqe}%
  \BibitemOpen
  \bibfield  {author} {\bibinfo {author} {\bibfnamefont {Bianca}\ \bibnamefont
  {Dittrich}}, \bibinfo {author} {\bibfnamefont {Mercedes}\ \bibnamefont
  {Martín-Benito}}, \ and\ \bibinfo {author} {\bibfnamefont {Erik}\
  \bibnamefont {Schnetter}},\ }\bibfield  {title} {\enquote {\bibinfo {title}
  {{Coarse graining of spin net models: dynamics of intertwiners}},}\ }\href
  {\doibase 10.1088/1367-2630/15/10/103004} {\bibfield  {journal} {\bibinfo
  {journal} {New J.\ Phys.}\ }\textbf {\bibinfo {volume} {15}},\ \bibinfo
  {pages} {103004} (\bibinfo {year} {2013})},\ \Eprint
  {http://arxiv.org/abs/1306.2987} {arXiv:1306.2987 [gr-qc]} \BibitemShut
  {NoStop}%
\bibitem [{\citenamefont {Dittrich}\ \emph
  {et~al.}(2016{\natexlab{a}})\citenamefont {Dittrich}, \citenamefont
  {Schnetter}, \citenamefont {Seth},\ and\ \citenamefont
  {Steinhaus}}]{Dittrich:2016tys}%
  \BibitemOpen
  \bibfield  {author} {\bibinfo {author} {\bibfnamefont {Bianca}\ \bibnamefont
  {Dittrich}}, \bibinfo {author} {\bibfnamefont {Erik}\ \bibnamefont
  {Schnetter}}, \bibinfo {author} {\bibfnamefont {Cameron~J.}\ \bibnamefont
  {Seth}}, \ and\ \bibinfo {author} {\bibfnamefont {Sebastian}\ \bibnamefont
  {Steinhaus}},\ }\bibfield  {title} {\enquote {\bibinfo {title} {{Coarse
  graining flow of spin foam intertwiners}},}\ }\href {\doibase
  10.1103/PhysRevD.94.124050} {\bibfield  {journal} {\bibinfo  {journal}
  {Phys.\ Rev.\ D}\ }\textbf {\bibinfo {volume} {94}},\ \bibinfo {pages}
  {124050} (\bibinfo {year} {2016}{\natexlab{a}})},\ \Eprint
  {http://arxiv.org/abs/1609.02429} {arXiv:1609.02429 [gr-qc]} \BibitemShut
  {NoStop}%
\bibitem [{\citenamefont {Dittrich}\ \emph {et~al.}(2012)\citenamefont
  {Dittrich}, \citenamefont {Eckert},\ and\ \citenamefont
  {Martin-Benito}}]{Dittrich:2011zh}%
  \BibitemOpen
  \bibfield  {author} {\bibinfo {author} {\bibfnamefont {Bianca}\ \bibnamefont
  {Dittrich}}, \bibinfo {author} {\bibfnamefont {Frank~C.}\ \bibnamefont
  {Eckert}}, \ and\ \bibinfo {author} {\bibfnamefont {Mercedes}\ \bibnamefont
  {Martin-Benito}},\ }\bibfield  {title} {\enquote {\bibinfo {title} {{Coarse
  graining methods for spin net and spin foam models}},}\ }\href {\doibase
  10.1088/1367-2630/14/3/035008} {\bibfield  {journal} {\bibinfo  {journal}
  {New J.\ Phys.}\ }\textbf {\bibinfo {volume} {14}},\ \bibinfo {pages}
  {035008} (\bibinfo {year} {2012})},\ \Eprint {http://arxiv.org/abs/1109.4927}
  {arXiv:1109.4927 [gr-qc]} \BibitemShut {NoStop}%
\bibitem [{\citenamefont {Dittrich}\ and\ \citenamefont
  {Eckert}(2012)}]{Dittrich:2011av}%
  \BibitemOpen
  \bibfield  {author} {\bibinfo {author} {\bibfnamefont {Bianca}\ \bibnamefont
  {Dittrich}}\ and\ \bibinfo {author} {\bibfnamefont {Frank~C.}\ \bibnamefont
  {Eckert}},\ }\bibfield  {title} {\enquote {\bibinfo {title} {{Towards
  computational insights into the large-scale structure of spin foams}},}\
  }\href {\doibase 10.1088/1742-6596/360/1/012004} {\bibfield  {journal}
  {\bibinfo  {journal} {J.\ Phys.\ Conf.\ Ser.}\ }\textbf {\bibinfo {volume}
  {360}},\ \bibinfo {pages} {012004} (\bibinfo {year} {2012})},\ \Eprint
  {http://arxiv.org/abs/1111.0967} {arXiv:1111.0967 [gr-qc]} \BibitemShut
  {NoStop}%
\bibitem [{\citenamefont {Bahr}\ \emph
  {et~al.}(2013{\natexlab{b}})\citenamefont {Bahr}, \citenamefont {Dittrich},
  \citenamefont {Hellmann},\ and\ \citenamefont {Kaminski}}]{Bahr:2012qj}%
  \BibitemOpen
  \bibfield  {author} {\bibinfo {author} {\bibfnamefont {Benjamin}\
  \bibnamefont {Bahr}}, \bibinfo {author} {\bibfnamefont {Bianca}\ \bibnamefont
  {Dittrich}}, \bibinfo {author} {\bibfnamefont {Frank}\ \bibnamefont
  {Hellmann}}, \ and\ \bibinfo {author} {\bibfnamefont {Wojciech}\ \bibnamefont
  {Kaminski}},\ }\bibfield  {title} {\enquote {\bibinfo {title} {{Holonomy Spin
  Foam Models: Definition and Coarse Graining}},}\ }\href {\doibase
  10.1103/PhysRevD.87.044048} {\bibfield  {journal} {\bibinfo  {journal}
  {Phys.\ Rev.\ D}\ }\textbf {\bibinfo {volume} {87}},\ \bibinfo {pages}
  {044048} (\bibinfo {year} {2013}{\natexlab{b}})},\ \Eprint
  {http://arxiv.org/abs/1208.3388} {arXiv:1208.3388 [gr-qc]} \BibitemShut
  {NoStop}%
\bibitem [{\citenamefont {Biedenharn}\ and\ \citenamefont
  {Lohe}(1995)}]{biedenharn}%
  \BibitemOpen
  \bibfield  {author} {\bibinfo {author} {\bibfnamefont {L.~C.}\ \bibnamefont
  {Biedenharn}}\ and\ \bibinfo {author} {\bibfnamefont {M.~A.}\ \bibnamefont
  {Lohe}},\ }\href@noop {} {\emph {\bibinfo {title} {{Quantum Group Symmetries
  and q-Tensor Algebras}}}}\ (\bibinfo  {publisher} {World Scientific},\
  \bibinfo {address} {Singapore},\ \bibinfo {year} {1995})\BibitemShut
  {NoStop}%
\bibitem [{\citenamefont {Carter}\ \emph {et~al.}(1995)\citenamefont {Carter},
  \citenamefont {Flath},\ and\ \citenamefont {Saito}}]{yellowbook}%
  \BibitemOpen
  \bibfield  {author} {\bibinfo {author} {\bibfnamefont {J.~S.}\ \bibnamefont
  {Carter}}, \bibinfo {author} {\bibfnamefont {D.~E.}\ \bibnamefont {Flath}}, \
  and\ \bibinfo {author} {\bibfnamefont {M.}~\bibnamefont {Saito}},\
  }\href@noop {} {\emph {\bibinfo {title} {{The Classical and Quantum
  $6j$--symbols}}}}\ (\bibinfo  {publisher} {Princeton University Press},\
  \bibinfo {address} {Princeton},\ \bibinfo {year} {1995})\BibitemShut
  {NoStop}%
\bibitem [{\citenamefont {Dittrich}\ and\ \citenamefont
  {Kaminski}(2013)}]{Dittrich:2013aia}%
  \BibitemOpen
  \bibfield  {author} {\bibinfo {author} {\bibfnamefont {Bianca}\ \bibnamefont
  {Dittrich}}\ and\ \bibinfo {author} {\bibfnamefont {Wojciech}\ \bibnamefont
  {Kaminski}},\ }\bibfield  {title} {\enquote {\bibinfo {title} {{Topological
  lattice field theories from intertwiner dynamics}},}\ }\href@noop {} {\
  (\bibinfo {year} {2013})},\ \Eprint {http://arxiv.org/abs/1311.1798}
  {arXiv:1311.1798 [gr-qc]} \BibitemShut {NoStop}%
\bibitem [{\citenamefont {Liu}\ \emph {et~al.}(2013)\citenamefont {Liu},
  \citenamefont {Meurice}, \citenamefont {Qin}, \citenamefont {Unmuth-Yockey},
  \citenamefont {Xiang}, \citenamefont {Xie}, \citenamefont {Yu},\ and\
  \citenamefont {Zou}}]{Liu:2013nsa}%
  \BibitemOpen
  \bibfield  {author} {\bibinfo {author} {\bibfnamefont {Yuzhi}\ \bibnamefont
  {Liu}}, \bibinfo {author} {\bibfnamefont {Y.}~\bibnamefont {Meurice}},
  \bibinfo {author} {\bibfnamefont {M.P.}\ \bibnamefont {Qin}}, \bibinfo
  {author} {\bibfnamefont {J.}~\bibnamefont {Unmuth-Yockey}}, \bibinfo {author}
  {\bibfnamefont {T.}~\bibnamefont {Xiang}}, \bibinfo {author} {\bibfnamefont
  {Z.Y.}\ \bibnamefont {Xie}}, \bibinfo {author} {\bibfnamefont {J.F.}\
  \bibnamefont {Yu}}, \ and\ \bibinfo {author} {\bibfnamefont {Haiyuan}\
  \bibnamefont {Zou}},\ }\bibfield  {title} {\enquote {\bibinfo {title} {{Exact
  Blocking Formulas for Spin and Gauge Models}},}\ }\href {\doibase
  10.1103/PhysRevD.88.056005} {\bibfield  {journal} {\bibinfo  {journal}
  {Phys.\ Rev.\ D}\ }\textbf {\bibinfo {volume} {88}},\ \bibinfo {pages}
  {056005} (\bibinfo {year} {2013})},\ \Eprint {http://arxiv.org/abs/1307.6543}
  {arXiv:1307.6543 [hep-lat]} \BibitemShut {NoStop}%
\bibitem [{\citenamefont {Dittrich}\ \emph
  {et~al.}(2016{\natexlab{b}})\citenamefont {Dittrich}, \citenamefont
  {Mizera},\ and\ \citenamefont {Steinhaus}}]{Dittrich:2014mxa}%
  \BibitemOpen
  \bibfield  {author} {\bibinfo {author} {\bibfnamefont {Bianca}\ \bibnamefont
  {Dittrich}}, \bibinfo {author} {\bibfnamefont {Sebastian}\ \bibnamefont
  {Mizera}}, \ and\ \bibinfo {author} {\bibfnamefont {Sebastian}\ \bibnamefont
  {Steinhaus}},\ }\bibfield  {title} {\enquote {\bibinfo {title} {{Decorated
  tensor network renormalization for lattice gauge theories and spin foam
  models}},}\ }\href {\doibase 10.1088/1367-2630/18/5/053009} {\bibfield
  {journal} {\bibinfo  {journal} {New J.\ Phys.}\ }\textbf {\bibinfo {volume}
  {18}},\ \bibinfo {pages} {053009} (\bibinfo {year} {2016}{\natexlab{b}})},\
  \Eprint {http://arxiv.org/abs/1409.2407} {arXiv:1409.2407 [gr-qc]}
  \BibitemShut {NoStop}%
\bibitem [{\citenamefont {Binder}\ and\ \citenamefont
  {Luijten}(2001)}]{Binder:2001ha}%
  \BibitemOpen
  \bibfield  {author} {\bibinfo {author} {\bibfnamefont {K.}~\bibnamefont
  {Binder}}\ and\ \bibinfo {author} {\bibfnamefont {E.}~\bibnamefont
  {Luijten}},\ }\bibfield  {title} {\enquote {\bibinfo {title} {{Monte Carlo
  tests of renormalization group predictions for critical phenomena in Ising
  models}},}\ }\href {\doibase 10.1016/S0370-1573(00)00127-7} {\bibfield
  {journal} {\bibinfo  {journal} {Phys.\ Rept.}\ }\textbf {\bibinfo {volume}
  {344}},\ \bibinfo {pages} {179--253} (\bibinfo {year} {2001})}\BibitemShut
  {NoStop}%
\bibitem [{\citenamefont {Steinhaus}(2015)}]{Steinhaus:2015kxa}%
  \BibitemOpen
  \bibfield  {author} {\bibinfo {author} {\bibfnamefont {Sebastian}\
  \bibnamefont {Steinhaus}},\ }\bibfield  {title} {\enquote {\bibinfo {title}
  {{Coupled intertwiner dynamics: A toy model for coupling matter to spin foam
  models}},}\ }\href {\doibase 10.1103/PhysRevD.92.064007} {\bibfield
  {journal} {\bibinfo  {journal} {Phys.\ Rev.\ D}\ }\textbf {\bibinfo {volume}
  {92}},\ \bibinfo {pages} {064007} (\bibinfo {year} {2015})},\ \Eprint
  {http://arxiv.org/abs/1506.04749} {arXiv:1506.04749 [gr-qc]} \BibitemShut
  {NoStop}%
\bibitem [{\citenamefont {Delcamp}\ and\ \citenamefont
  {Dittrich}(2017{\natexlab{a}})}]{Delcamp:2016dqo}%
  \BibitemOpen
  \bibfield  {author} {\bibinfo {author} {\bibfnamefont {Clement}\ \bibnamefont
  {Delcamp}}\ and\ \bibinfo {author} {\bibfnamefont {Bianca}\ \bibnamefont
  {Dittrich}},\ }\bibfield  {title} {\enquote {\bibinfo {title} {{Towards a
  phase diagram for spin foams}},}\ }\href {\doibase 10.1088/1361-6382/aa8f24}
  {\bibfield  {journal} {\bibinfo  {journal} {Class.\ Quant.\ Grav.}\ }\textbf
  {\bibinfo {volume} {34}},\ \bibinfo {pages} {225006} (\bibinfo {year}
  {2017}{\natexlab{a}})},\ \Eprint {http://arxiv.org/abs/1612.04506}
  {arXiv:1612.04506 [gr-qc]} \BibitemShut {NoStop}%
\bibitem [{\citenamefont {Livine}(2014)}]{Livine:2013gna}%
  \BibitemOpen
  \bibfield  {author} {\bibinfo {author} {\bibfnamefont {Etera~R.}\
  \bibnamefont {Livine}},\ }\bibfield  {title} {\enquote {\bibinfo {title}
  {{Deformation Operators of Spin Networks and Coarse-Graining}},}\ }\href
  {\doibase 10.1088/0264-9381/31/7/075004} {\bibfield  {journal} {\bibinfo
  {journal} {Class.\ Quant.\ Grav.}\ }\textbf {\bibinfo {volume} {31}},\
  \bibinfo {pages} {075004} (\bibinfo {year} {2014})},\ \Eprint
  {http://arxiv.org/abs/1310.3362} {arXiv:1310.3362 [gr-qc]} \BibitemShut
  {NoStop}%
\bibitem [{\citenamefont {Charles}\ and\ \citenamefont
  {Livine}(2016)}]{Charles:2016xwc}%
  \BibitemOpen
  \bibfield  {author} {\bibinfo {author} {\bibfnamefont {Christoph}\
  \bibnamefont {Charles}}\ and\ \bibinfo {author} {\bibfnamefont {Etera~R.}\
  \bibnamefont {Livine}},\ }\bibfield  {title} {\enquote {\bibinfo {title}
  {{The Fock Space of Loopy Spin Networks for Quantum Gravity}},}\ }\href
  {\doibase 10.1007/s10714-016-2107-5} {\bibfield  {journal} {\bibinfo
  {journal} {Gen.\ Rel.\ Grav.}\ }\textbf {\bibinfo {volume} {48}},\ \bibinfo
  {pages} {113} (\bibinfo {year} {2016})},\ \Eprint
  {http://arxiv.org/abs/1603.01117} {arXiv:1603.01117 [gr-qc]} \BibitemShut
  {NoStop}%
\bibitem [{\citenamefont {Koenig}\ \emph {et~al.}(2010)\citenamefont {Koenig},
  \citenamefont {Kuperberg},\ and\ \citenamefont {Reichardt}}]{Koenig_2010}%
  \BibitemOpen
  \bibfield  {author} {\bibinfo {author} {\bibfnamefont {Robert}\ \bibnamefont
  {Koenig}}, \bibinfo {author} {\bibfnamefont {Greg}\ \bibnamefont
  {Kuperberg}}, \ and\ \bibinfo {author} {\bibfnamefont {Ben~W.}\ \bibnamefont
  {Reichardt}},\ }\bibfield  {title} {\enquote {\bibinfo {title} {Quantum
  computation with turaev–viro codes},}\ }\href {\doibase
  10.1016/j.aop.2010.08.001} {\bibfield  {journal} {\bibinfo  {journal} {Annals
  of Physics}\ }\textbf {\bibinfo {volume} {325}},\ \bibinfo {pages}
  {2707–2749} (\bibinfo {year} {2010})}\BibitemShut {NoStop}%
\bibitem [{\citenamefont {Hu}\ \emph {et~al.}(2018)\citenamefont {Hu},
  \citenamefont {Geer},\ and\ \citenamefont {Wu}}]{Hu:2015dga}%
  \BibitemOpen
  \bibfield  {author} {\bibinfo {author} {\bibfnamefont {Yuting}\ \bibnamefont
  {Hu}}, \bibinfo {author} {\bibfnamefont {Nathan}\ \bibnamefont {Geer}}, \
  and\ \bibinfo {author} {\bibfnamefont {Yong-Shi}\ \bibnamefont {Wu}},\
  }\bibfield  {title} {\enquote {\bibinfo {title} {{Full dyon excitation
  spectrum in extended Levin-Wen models}},}\ }\href {\doibase
  10.1103/PhysRevB.97.195154} {\bibfield  {journal} {\bibinfo  {journal}
  {Phys.\ Rev.\ B}\ }\textbf {\bibinfo {volume} {97}},\ \bibinfo {pages}
  {195154} (\bibinfo {year} {2018})},\ \Eprint
  {http://arxiv.org/abs/1502.03433} {arXiv:1502.03433 [cond-mat.str-el]}
  \BibitemShut {NoStop}%
\bibitem [{\citenamefont {Delcamp}\ \emph {et~al.}(2017)\citenamefont
  {Delcamp}, \citenamefont {Dittrich},\ and\ \citenamefont
  {Riello}}]{Delcamp:2016yix}%
  \BibitemOpen
  \bibfield  {author} {\bibinfo {author} {\bibfnamefont {Clement}\ \bibnamefont
  {Delcamp}}, \bibinfo {author} {\bibfnamefont {Bianca}\ \bibnamefont
  {Dittrich}}, \ and\ \bibinfo {author} {\bibfnamefont {Aldo}\ \bibnamefont
  {Riello}},\ }\bibfield  {title} {\enquote {\bibinfo {title} {{Fusion basis
  for lattice gauge theory and loop quantum gravity}},}\ }\href {\doibase
  10.1007/JHEP02(2017)061} {\bibfield  {journal} {\bibinfo  {journal} {JHEP}\
  }\textbf {\bibinfo {volume} {02}},\ \bibinfo {pages} {061} (\bibinfo {year}
  {2017})},\ \Eprint {http://arxiv.org/abs/1607.08881} {arXiv:1607.08881
  [hep-th]} \BibitemShut {NoStop}%
\bibitem [{\citenamefont {Dittrich}\ and\ \citenamefont
  {Geiller}(2017)}]{Dittrich:2016typ}%
  \BibitemOpen
  \bibfield  {author} {\bibinfo {author} {\bibfnamefont {Bianca}\ \bibnamefont
  {Dittrich}}\ and\ \bibinfo {author} {\bibfnamefont {Marc}\ \bibnamefont
  {Geiller}},\ }\bibfield  {title} {\enquote {\bibinfo {title} {{Quantum
  gravity kinematics from extended TQFTs}},}\ }\href {\doibase
  10.1088/1367-2630/aa54e2} {\bibfield  {journal} {\bibinfo  {journal} {New J.\
  Phys.}\ }\textbf {\bibinfo {volume} {19}},\ \bibinfo {pages} {013003}
  (\bibinfo {year} {2017})},\ \Eprint {http://arxiv.org/abs/1604.05195}
  {arXiv:1604.05195 [hep-th]} \BibitemShut {NoStop}%
\bibitem [{\citenamefont {Dittrich}(2018)}]{Dittrich:2018dvs}%
  \BibitemOpen
  \bibfield  {author} {\bibinfo {author} {\bibfnamefont {Bianca}\ \bibnamefont
  {Dittrich}},\ }\bibfield  {title} {\enquote {\bibinfo {title} {{Cosmological
  constant from condensation of defect excitations}},}\ }\href {\doibase
  10.3390/universe4070081} {\bibfield  {journal} {\bibinfo  {journal}
  {Universe}\ }\textbf {\bibinfo {volume} {4}},\ \bibinfo {pages} {81}
  (\bibinfo {year} {2018})},\ \Eprint {http://arxiv.org/abs/1802.09439}
  {arXiv:1802.09439 [gr-qc]} \BibitemShut {NoStop}%
\bibitem [{\citenamefont {de~Wild~Propitius}\ and\ \citenamefont
  {Bais}(1995)}]{deWildPropitius:1995hk}%
  \BibitemOpen
  \bibfield  {author} {\bibinfo {author} {\bibfnamefont {Mark}\ \bibnamefont
  {de~Wild~Propitius}}\ and\ \bibinfo {author} {\bibfnamefont {F.Alexander}\
  \bibnamefont {Bais}},\ }\bibfield  {title} {\enquote {\bibinfo {title}
  {{Discrete gauge theories}},}\ }in\ \href@noop {} {\emph {\bibinfo
  {booktitle} {{Particles and fields. Proceedings, CAP-CRM Summer School,
  Banff, Canada, August 16-24, 1994}}}}\ (\bibinfo {year} {1995})\ pp.\
  \bibinfo {pages} {353--439},\ \Eprint {http://arxiv.org/abs/hep-th/9511201}
  {arXiv:hep-th/9511201} \BibitemShut {NoStop}%
\bibitem [{\citenamefont {Kitaev}(2003)}]{Kitaev:1997wr}%
  \BibitemOpen
  \bibfield  {author} {\bibinfo {author} {\bibfnamefont {A.Yu.}\ \bibnamefont
  {Kitaev}},\ }\bibfield  {title} {\enquote {\bibinfo {title} {{Fault tolerant
  quantum computation by anyons}},}\ }\href {\doibase
  10.1016/S0003-4916(02)00018-0} {\bibfield  {journal} {\bibinfo  {journal}
  {Annals Phys.}\ }\textbf {\bibinfo {volume} {303}},\ \bibinfo {pages} {2--30}
  (\bibinfo {year} {2003})},\ \Eprint {http://arxiv.org/abs/quant-ph/9707021}
  {arXiv:quant-ph/9707021} \BibitemShut {NoStop}%
\bibitem [{\citenamefont {Bombin}\ and\ \citenamefont
  {Martin-Delgado}(2008)}]{Bombin:2007qv}%
  \BibitemOpen
  \bibfield  {author} {\bibinfo {author} {\bibfnamefont {H.}~\bibnamefont
  {Bombin}}\ and\ \bibinfo {author} {\bibfnamefont {M.A.}\ \bibnamefont
  {Martin-Delgado}},\ }\bibfield  {title} {\enquote {\bibinfo {title} {{A
  Family of Non-Abelian Kitaev Models on a Lattice: Topological Confinement and
  Condensation}},}\ }\href {\doibase 10.1103/PhysRevB.78.115421} {\bibfield
  {journal} {\bibinfo  {journal} {Phys.\ Rev.\ B}\ }\textbf {\bibinfo {volume}
  {78}},\ \bibinfo {pages} {115421} (\bibinfo {year} {2008})},\ \Eprint
  {http://arxiv.org/abs/0712.0190} {arXiv:0712.0190 [cond-mat.str-el]}
  \BibitemShut {NoStop}%
\bibitem [{\citenamefont {Levin}\ and\ \citenamefont
  {Wen}(2005)}]{Levin:2004mi}%
  \BibitemOpen
  \bibfield  {author} {\bibinfo {author} {\bibfnamefont {Michael~A.}\
  \bibnamefont {Levin}}\ and\ \bibinfo {author} {\bibfnamefont {Xiao-Gang}\
  \bibnamefont {Wen}},\ }\bibfield  {title} {\enquote {\bibinfo {title}
  {{String net condensation: A Physical mechanism for topological phases}},}\
  }\href {\doibase 10.1103/PhysRevB.71.045110} {\bibfield  {journal} {\bibinfo
  {journal} {Phys.\ Rev.\ B}\ }\textbf {\bibinfo {volume} {71}},\ \bibinfo
  {pages} {045110} (\bibinfo {year} {2005})},\ \Eprint
  {http://arxiv.org/abs/cond-mat/0404617} {arXiv:cond-mat/0404617} \BibitemShut
  {NoStop}%
\bibitem [{\citenamefont {Perelomov}(1986)}]{perelomov}%
  \BibitemOpen
  \bibfield  {author} {\bibinfo {author} {\bibfnamefont {A.}~\bibnamefont
  {Perelomov}},\ }\href@noop {} {\emph {\bibinfo {title} {Generalized coherent
  states and their applications}}}\ (\bibinfo  {publisher} {Springer},\
  \bibinfo {address} {Berlin},\ \bibinfo {year} {1986})\BibitemShut {NoStop}%
\bibitem [{\citenamefont {Hahn}(2005)}]{Hahn:2004fe}%
  \BibitemOpen
  \bibfield  {author} {\bibinfo {author} {\bibfnamefont {T.}~\bibnamefont
  {Hahn}},\ }\bibfield  {title} {\enquote {\bibinfo {title} {{CUBA: A Library
  for multidimensional numerical integration}},}\ }\href {\doibase
  10.1016/j.cpc.2005.01.010} {\bibfield  {journal} {\bibinfo  {journal}
  {Comput.\ Phys.\ Commun.}\ }\textbf {\bibinfo {volume} {168}},\ \bibinfo
  {pages} {78--95} (\bibinfo {year} {2005})},\ \Eprint
  {http://arxiv.org/abs/hep-ph/0404043} {arXiv:hep-ph/0404043} \BibitemShut
  {NoStop}%
\bibitem [{\citenamefont {Bianchi}\ \emph {et~al.}(2010)\citenamefont
  {Bianchi}, \citenamefont {Regoli},\ and\ \citenamefont
  {Rovelli}}]{Bianchi:2010fj}%
  \BibitemOpen
  \bibfield  {author} {\bibinfo {author} {\bibfnamefont {Eugenio}\ \bibnamefont
  {Bianchi}}, \bibinfo {author} {\bibfnamefont {Daniele}\ \bibnamefont
  {Regoli}}, \ and\ \bibinfo {author} {\bibfnamefont {Carlo}\ \bibnamefont
  {Rovelli}},\ }\bibfield  {title} {\enquote {\bibinfo {title} {{Face amplitude
  of spinfoam quantum gravity}},}\ }\href {\doibase
  10.1088/0264-9381/27/18/185009} {\bibfield  {journal} {\bibinfo  {journal}
  {Class.\ Quant.\ Grav.}\ }\textbf {\bibinfo {volume} {27}},\ \bibinfo {pages}
  {185009} (\bibinfo {year} {2010})},\ \Eprint {http://arxiv.org/abs/1005.0764}
  {arXiv:1005.0764 [gr-qc]} \BibitemShut {NoStop}%
\bibitem [{\citenamefont {Bahr}\ \emph {et~al.}(2017)\citenamefont {Bahr},
  \citenamefont {Kl\"oser},\ and\ \citenamefont {Rabuffo}}]{Bahr:2017eyi}%
  \BibitemOpen
  \bibfield  {author} {\bibinfo {author} {\bibfnamefont {Benjamin}\
  \bibnamefont {Bahr}}, \bibinfo {author} {\bibfnamefont {Sebastian}\
  \bibnamefont {Kl\"oser}}, \ and\ \bibinfo {author} {\bibfnamefont {Giovanni}\
  \bibnamefont {Rabuffo}},\ }\bibfield  {title} {\enquote {\bibinfo {title}
  {{Towards a Cosmological subsector of Spin Foam Quantum Gravity}},}\ }\href
  {\doibase 10.1103/PhysRevD.96.086009} {\bibfield  {journal} {\bibinfo
  {journal} {Phys.\ Rev.\ D}\ }\textbf {\bibinfo {volume} {96}},\ \bibinfo
  {pages} {086009} (\bibinfo {year} {2017})},\ \Eprint
  {http://arxiv.org/abs/1704.03691} {arXiv:1704.03691 [gr-qc]} \BibitemShut
  {NoStop}%
\bibitem [{\citenamefont {Bahr}\ and\ \citenamefont
  {Rabuffo}(2018)}]{Bahr:2018ewi}%
  \BibitemOpen
  \bibfield  {author} {\bibinfo {author} {\bibfnamefont {Benjamin}\
  \bibnamefont {Bahr}}\ and\ \bibinfo {author} {\bibfnamefont {Giovanni}\
  \bibnamefont {Rabuffo}},\ }\bibfield  {title} {\enquote {\bibinfo {title}
  {{Deformation of the Engle-Livine-Pereira-Rovelli spin foam model by a
  cosmological constant}},}\ }\href {\doibase 10.1103/PhysRevD.97.086010}
  {\bibfield  {journal} {\bibinfo  {journal} {Phys.\ Rev.\ D}\ }\textbf
  {\bibinfo {volume} {97}},\ \bibinfo {pages} {086010} (\bibinfo {year}
  {2018})},\ \Eprint {http://arxiv.org/abs/1803.01838} {arXiv:1803.01838
  [gr-qc]} \BibitemShut {NoStop}%
\bibitem [{\citenamefont {Han}(2011)}]{Han:2011aa}%
  \BibitemOpen
  \bibfield  {author} {\bibinfo {author} {\bibfnamefont {Muxin}\ \bibnamefont
  {Han}},\ }\bibfield  {title} {\enquote {\bibinfo {title} {{Cosmological
  Constant in LQG Vertex Amplitude}},}\ }\href {\doibase
  10.1103/PhysRevD.84.064010} {\bibfield  {journal} {\bibinfo  {journal}
  {Phys.\ Rev.\ D}\ }\textbf {\bibinfo {volume} {84}},\ \bibinfo {pages}
  {064010} (\bibinfo {year} {2011})},\ \Eprint {http://arxiv.org/abs/1105.2212}
  {arXiv:1105.2212 [gr-qc]} \BibitemShut {NoStop}%
\bibitem [{\citenamefont {Bahr}\ and\ \citenamefont
  {Steinhaus}(2016{\natexlab{b}})}]{Bahr:2016hwc}%
  \BibitemOpen
  \bibfield  {author} {\bibinfo {author} {\bibfnamefont {Benjamin}\
  \bibnamefont {Bahr}}\ and\ \bibinfo {author} {\bibfnamefont {Sebastian}\
  \bibnamefont {Steinhaus}},\ }\bibfield  {title} {\enquote {\bibinfo {title}
  {{Numerical evidence for a phase transition in 4d spin foam quantum
  gravity}},}\ }\href {\doibase 10.1103/PhysRevLett.117.141302} {\bibfield
  {journal} {\bibinfo  {journal} {Phys.\ Rev.\ Lett.}\ }\textbf {\bibinfo
  {volume} {117}},\ \bibinfo {pages} {141302} (\bibinfo {year}
  {2016}{\natexlab{b}})},\ \Eprint {http://arxiv.org/abs/1605.07649}
  {arXiv:1605.07649 [gr-qc]} \BibitemShut {NoStop}%
\bibitem [{\citenamefont {Bahr}\ and\ \citenamefont
  {Steinhaus}(2017)}]{Bahr:2017klw}%
  \BibitemOpen
  \bibfield  {author} {\bibinfo {author} {\bibfnamefont {Benjamin}\
  \bibnamefont {Bahr}}\ and\ \bibinfo {author} {\bibfnamefont {Sebastian}\
  \bibnamefont {Steinhaus}},\ }\bibfield  {title} {\enquote {\bibinfo {title}
  {{Hypercuboidal renormalization in spin foam quantum gravity}},}\ }\href
  {\doibase 10.1103/PhysRevD.95.126006} {\bibfield  {journal} {\bibinfo
  {journal} {Phys.\ Rev.\ D}\ }\textbf {\bibinfo {volume} {95}},\ \bibinfo
  {pages} {126006} (\bibinfo {year} {2017})},\ \Eprint
  {http://arxiv.org/abs/1701.02311} {arXiv:1701.02311 [gr-qc]} \BibitemShut
  {NoStop}%
\bibitem [{\citenamefont {Bahr}\ \emph {et~al.}(2018)\citenamefont {Bahr},
  \citenamefont {Rabuffo},\ and\ \citenamefont {Steinhaus}}]{Bahr:2018gwf}%
  \BibitemOpen
  \bibfield  {author} {\bibinfo {author} {\bibfnamefont {Benjamin}\
  \bibnamefont {Bahr}}, \bibinfo {author} {\bibfnamefont {Giovanni}\
  \bibnamefont {Rabuffo}}, \ and\ \bibinfo {author} {\bibfnamefont {Sebastian}\
  \bibnamefont {Steinhaus}},\ }\bibfield  {title} {\enquote {\bibinfo {title}
  {{Renormalization of symmetry restricted spin foam models with curvature in
  the asymptotic regime}},}\ }\href {\doibase 10.1103/PhysRevD.98.106026}
  {\bibfield  {journal} {\bibinfo  {journal} {Phys.\ Rev.\ D}\ }\textbf
  {\bibinfo {volume} {98}},\ \bibinfo {pages} {106026} (\bibinfo {year}
  {2018})},\ \Eprint {http://arxiv.org/abs/1804.00023} {arXiv:1804.00023
  [gr-qc]} \BibitemShut {NoStop}%
\bibitem [{\citenamefont {Reuter}\ and\ \citenamefont
  {Saueressig}(2019)}]{Reuter:2019byg}%
  \BibitemOpen
  \bibfield  {author} {\bibinfo {author} {\bibfnamefont {Martin}\ \bibnamefont
  {Reuter}}\ and\ \bibinfo {author} {\bibfnamefont {Frank}\ \bibnamefont
  {Saueressig}},\ }\href@noop {} {\emph {\bibinfo {title} {{Quantum Gravity and
  the Functional Renormalization Group}: {The Road towards Asymptotic
  Safety}}}}\ (\bibinfo  {publisher} {Cambridge University Press},\ \bibinfo
  {year} {2019})\BibitemShut {NoStop}%
\bibitem [{\citenamefont {Han}(2017)}]{Han:2017xwo}%
  \BibitemOpen
  \bibfield  {author} {\bibinfo {author} {\bibfnamefont {Muxin}\ \bibnamefont
  {Han}},\ }\bibfield  {title} {\enquote {\bibinfo {title} {{Einstein Equation
  from Covariant Loop Quantum Gravity in Semiclassical Continuum Limit}},}\
  }\href {\doibase 10.1103/PhysRevD.96.024047} {\bibfield  {journal} {\bibinfo
  {journal} {Phys.\ Rev.\ D}\ }\textbf {\bibinfo {volume} {96}},\ \bibinfo
  {pages} {024047} (\bibinfo {year} {2017})},\ \Eprint
  {http://arxiv.org/abs/1705.09030} {arXiv:1705.09030 [gr-qc]} \BibitemShut
  {NoStop}%
\bibitem [{\citenamefont {Han}\ \emph {et~al.}(2019)\citenamefont {Han},
  \citenamefont {Huang},\ and\ \citenamefont {Zipfel}}]{Han:2018fmu}%
  \BibitemOpen
  \bibfield  {author} {\bibinfo {author} {\bibfnamefont {Muxin}\ \bibnamefont
  {Han}}, \bibinfo {author} {\bibfnamefont {Zichang}\ \bibnamefont {Huang}}, \
  and\ \bibinfo {author} {\bibfnamefont {Antonia}\ \bibnamefont {Zipfel}},\
  }\bibfield  {title} {\enquote {\bibinfo {title} {{Emergent four-dimensional
  linearized gravity from a spin foam model}},}\ }\href {\doibase
  10.1103/PhysRevD.100.024060} {\bibfield  {journal} {\bibinfo  {journal}
  {Phys.\ Rev.\ D}\ }\textbf {\bibinfo {volume} {100}},\ \bibinfo {pages}
  {024060} (\bibinfo {year} {2019})},\ \Eprint
  {http://arxiv.org/abs/1812.02110} {arXiv:1812.02110 [gr-qc]} \BibitemShut
  {NoStop}%
\bibitem [{\citenamefont {Speziale}(2017)}]{Speziale:2016axj}%
  \BibitemOpen
  \bibfield  {author} {\bibinfo {author} {\bibfnamefont {Simone}\ \bibnamefont
  {Speziale}},\ }\bibfield  {title} {\enquote {\bibinfo {title} {{Boosting
  Wigner's nj-symbols}},}\ }\href {\doibase 10.1063/1.4977752} {\bibfield
  {journal} {\bibinfo  {journal} {J. Math. Phys.}\ }\textbf {\bibinfo {volume}
  {58}},\ \bibinfo {pages} {032501} (\bibinfo {year} {2017})},\ \Eprint
  {http://arxiv.org/abs/1609.01632} {arXiv:1609.01632 [gr-qc]} \BibitemShut
  {NoStop}%
\bibitem [{\citenamefont {Bayle}\ \emph {et~al.}(2016)\citenamefont {Bayle},
  \citenamefont {Collet},\ and\ \citenamefont {Rovelli}}]{Bayle:2016doe}%
  \BibitemOpen
  \bibfield  {author} {\bibinfo {author} {\bibfnamefont {Vincent}\ \bibnamefont
  {Bayle}}, \bibinfo {author} {\bibfnamefont {François}\ \bibnamefont
  {Collet}}, \ and\ \bibinfo {author} {\bibfnamefont {Carlo}\ \bibnamefont
  {Rovelli}},\ }\bibfield  {title} {\enquote {\bibinfo {title} {{Short-scale
  Emergence of Classical Geometry, in Euclidean Loop Quantum Gravity}},}\
  }\href@noop {} {\  (\bibinfo {year} {2016})},\ \Eprint
  {http://arxiv.org/abs/1603.07931} {arXiv:1603.07931 [gr-qc]} \BibitemShut
  {NoStop}%
\bibitem [{\citenamefont {Diener}\ \emph {et~al.}(2014)\citenamefont {Diener},
  \citenamefont {Gupt},\ and\ \citenamefont {Singh}}]{Diener:2013uka}%
  \BibitemOpen
  \bibfield  {author} {\bibinfo {author} {\bibfnamefont {Peter}\ \bibnamefont
  {Diener}}, \bibinfo {author} {\bibfnamefont {Brajesh}\ \bibnamefont {Gupt}},
  \ and\ \bibinfo {author} {\bibfnamefont {Parampreet}\ \bibnamefont {Singh}},\
  }\bibfield  {title} {\enquote {\bibinfo {title} {{Chimera: A hybrid approach
  to numerical loop quantum cosmology}},}\ }\href {\doibase
  10.1088/0264-9381/31/2/025013} {\bibfield  {journal} {\bibinfo  {journal}
  {Class. Quant. Grav.}\ }\textbf {\bibinfo {volume} {31}},\ \bibinfo {pages}
  {025013} (\bibinfo {year} {2014})},\ \Eprint {http://arxiv.org/abs/1310.4795}
  {arXiv:1310.4795 [gr-qc]} \BibitemShut {NoStop}%
\bibitem [{\citenamefont {Assanioussi}\ and\ \citenamefont
  {Bahr}(2020)}]{Assanioussi:2020fml}%
  \BibitemOpen
  \bibfield  {author} {\bibinfo {author} {\bibfnamefont {Mehdi}\ \bibnamefont
  {Assanioussi}}\ and\ \bibinfo {author} {\bibfnamefont {Benjamin}\
  \bibnamefont {Bahr}},\ }\bibfield  {title} {\enquote {\bibinfo {title} {{Hopf
  link volume simplicity constraints in spin foam models}},}\ }\href@noop {} {\
   (\bibinfo {year} {2020})},\ \Eprint {http://arxiv.org/abs/2005.12004}
  {arXiv:2005.12004 [gr-qc]} \BibitemShut {NoStop}%
\bibitem [{\citenamefont {Asante}\ \emph {et~al.}(2020)\citenamefont {Asante},
  \citenamefont {Dittrich},\ and\ \citenamefont {Haggard}}]{Asante:2020qpa}%
  \BibitemOpen
  \bibfield  {author} {\bibinfo {author} {\bibfnamefont {Seth~K.}\ \bibnamefont
  {Asante}}, \bibinfo {author} {\bibfnamefont {Bianca}\ \bibnamefont
  {Dittrich}}, \ and\ \bibinfo {author} {\bibfnamefont {Hal~M.}\ \bibnamefont
  {Haggard}},\ }\bibfield  {title} {\enquote {\bibinfo {title} {{Effective Spin
  Foam Models for Four-Dimensional Quantum Gravity}},}\ }\href@noop {} {\
  (\bibinfo {year} {2020})},\ \Eprint {http://arxiv.org/abs/2004.07013}
  {arXiv:2004.07013 [gr-qc]} \BibitemShut {NoStop}%
\bibitem [{\citenamefont {Delcamp}\ and\ \citenamefont
  {Dittrich}(2017{\natexlab{b}})}]{Delcamp:2016lux}%
  \BibitemOpen
  \bibfield  {author} {\bibinfo {author} {\bibfnamefont {Clement}\ \bibnamefont
  {Delcamp}}\ and\ \bibinfo {author} {\bibfnamefont {Bianca}\ \bibnamefont
  {Dittrich}},\ }\bibfield  {title} {\enquote {\bibinfo {title} {{From 3D
  topological quantum field theories to 4D models with defects}},}\ }\href
  {\doibase 10.1063/1.4989535} {\bibfield  {journal} {\bibinfo  {journal} {J.\
  Math.\ Phys.}\ }\textbf {\bibinfo {volume} {58}},\ \bibinfo {pages} {062302}
  (\bibinfo {year} {2017}{\natexlab{b}})},\ \Eprint
  {http://arxiv.org/abs/1606.02384} {arXiv:1606.02384 [hep-th]} \BibitemShut
  {NoStop}%
\bibitem [{\citenamefont {Dittrich}(2017{\natexlab{b}})}]{Dittrich:2017nmq}%
  \BibitemOpen
  \bibfield  {author} {\bibinfo {author} {\bibfnamefont {Bianca}\ \bibnamefont
  {Dittrich}},\ }\bibfield  {title} {\enquote {\bibinfo {title} {{(3 +
  1)-dimensional topological phases and self-dual quantum geometries encoded on
  Heegaard surfaces}},}\ }\href {\doibase 10.1007/JHEP05(2017)123} {\bibfield
  {journal} {\bibinfo  {journal} {JHEP}\ }\textbf {\bibinfo {volume} {05}},\
  \bibinfo {pages} {123} (\bibinfo {year} {2017}{\natexlab{b}})},\ \Eprint
  {http://arxiv.org/abs/1701.02037} {arXiv:1701.02037 [hep-th]} \BibitemShut
  {NoStop}%
\bibitem [{\citenamefont {Delcamp}\ and\ \citenamefont
  {Tilloy}(2020)}]{Delcamp:2020hzo}%
  \BibitemOpen
  \bibfield  {author} {\bibinfo {author} {\bibfnamefont {Clement}\ \bibnamefont
  {Delcamp}}\ and\ \bibinfo {author} {\bibfnamefont {Antoine}\ \bibnamefont
  {Tilloy}},\ }\bibfield  {title} {\enquote {\bibinfo {title} {{Computing the
  renormalization group flow of two-dimensional $\phi^4$ theory with tensor
  networks}},}\ }\href@noop {} {\  (\bibinfo {year} {2020})},\ \Eprint
  {http://arxiv.org/abs/2003.12993} {arXiv:2003.12993 [cond-mat.str-el]}
  \BibitemShut {NoStop}%
\bibitem [{\citenamefont {Mikovic}(2002)}]{Mikovic:2001xi}%
  \BibitemOpen
  \bibfield  {author} {\bibinfo {author} {\bibfnamefont {Aleksandar~R.}\
  \bibnamefont {Mikovic}},\ }\bibfield  {title} {\enquote {\bibinfo {title}
  {{Spin foam models of matter coupled to gravity}},}\ }\href {\doibase
  10.1088/0264-9381/19/9/301} {\bibfield  {journal} {\bibinfo  {journal}
  {Class.\ Quant.\ Grav.}\ }\textbf {\bibinfo {volume} {19}},\ \bibinfo {pages}
  {2335--2354} (\bibinfo {year} {2002})},\ \Eprint
  {http://arxiv.org/abs/hep-th/0108099} {arXiv:hep-th/0108099} \BibitemShut
  {NoStop}%
\bibitem [{\citenamefont {Oriti}\ and\ \citenamefont
  {Pfeiffer}(2002)}]{Oriti:2002bn}%
  \BibitemOpen
  \bibfield  {author} {\bibinfo {author} {\bibfnamefont {Daniele}\ \bibnamefont
  {Oriti}}\ and\ \bibinfo {author} {\bibfnamefont {Hendryk}\ \bibnamefont
  {Pfeiffer}},\ }\bibfield  {title} {\enquote {\bibinfo {title} {{A Spin foam
  model for pure gauge theory coupled to quantum gravity}},}\ }\href {\doibase
  10.1103/PhysRevD.66.124010} {\bibfield  {journal} {\bibinfo  {journal}
  {Phys.\ Rev.\ D}\ }\textbf {\bibinfo {volume} {66}},\ \bibinfo {pages}
  {124010} (\bibinfo {year} {2002})},\ \Eprint
  {http://arxiv.org/abs/gr-qc/0207041} {arXiv:gr-qc/0207041} \BibitemShut
  {NoStop}%
\bibitem [{\citenamefont {Speziale}(2007)}]{Speziale:2007mt}%
  \BibitemOpen
  \bibfield  {author} {\bibinfo {author} {\bibfnamefont {Simone}\ \bibnamefont
  {Speziale}},\ }\bibfield  {title} {\enquote {\bibinfo {title} {{Coupling
  gauge theory to spinfoam 3d quantum gravity}},}\ }\href {\doibase
  10.1088/0264-9381/24/20/014} {\bibfield  {journal} {\bibinfo  {journal}
  {Class.\ Quant.\ Grav.}\ }\textbf {\bibinfo {volume} {24}},\ \bibinfo {pages}
  {5139--5160} (\bibinfo {year} {2007})},\ \Eprint
  {http://arxiv.org/abs/0706.1534} {arXiv:0706.1534 [gr-qc]} \BibitemShut
  {NoStop}%
\bibitem [{\citenamefont {Smolin}(2009)}]{Smolin:2007rx}%
  \BibitemOpen
  \bibfield  {author} {\bibinfo {author} {\bibfnamefont {Lee}\ \bibnamefont
  {Smolin}},\ }\bibfield  {title} {\enquote {\bibinfo {title} {{The Plebanski
  action extended to a unification of gravity and Yang-Mills theory}},}\ }\href
  {\doibase 10.1103/PhysRevD.80.124017} {\bibfield  {journal} {\bibinfo
  {journal} {Phys.\ Rev.\ D}\ }\textbf {\bibinfo {volume} {80}},\ \bibinfo
  {pages} {124017} (\bibinfo {year} {2009})},\ \Eprint
  {http://arxiv.org/abs/0712.0977} {arXiv:0712.0977 [hep-th]} \BibitemShut
  {NoStop}%
\bibitem [{\citenamefont {Han}\ and\ \citenamefont
  {Rovelli}(2013)}]{Han:2011as}%
  \BibitemOpen
  \bibfield  {author} {\bibinfo {author} {\bibfnamefont {Muxin}\ \bibnamefont
  {Han}}\ and\ \bibinfo {author} {\bibfnamefont {Carlo}\ \bibnamefont
  {Rovelli}},\ }\bibfield  {title} {\enquote {\bibinfo {title} {{Spin-foam
  Fermions: PCT Symmetry, Dirac Determinant, and Correlation Functions}},}\
  }\href {\doibase 10.1088/0264-9381/30/7/075007} {\bibfield  {journal}
  {\bibinfo  {journal} {Class.\ Quant.\ Grav.}\ }\textbf {\bibinfo {volume}
  {30}},\ \bibinfo {pages} {075007} (\bibinfo {year} {2013})},\ \Eprint
  {http://arxiv.org/abs/1101.3264} {arXiv:1101.3264 [gr-qc]} \BibitemShut
  {NoStop}%
\bibitem [{\citenamefont {Bianchi}\ \emph {et~al.}(2013)\citenamefont
  {Bianchi}, \citenamefont {Han}, \citenamefont {Rovelli}, \citenamefont
  {Wieland}, \citenamefont {Magliaro},\ and\ \citenamefont
  {Perini}}]{Bianchi:2010bn}%
  \BibitemOpen
  \bibfield  {author} {\bibinfo {author} {\bibfnamefont {Eugenio}\ \bibnamefont
  {Bianchi}}, \bibinfo {author} {\bibfnamefont {Muxin}\ \bibnamefont {Han}},
  \bibinfo {author} {\bibfnamefont {Carlo}\ \bibnamefont {Rovelli}}, \bibinfo
  {author} {\bibfnamefont {Wolfgang}\ \bibnamefont {Wieland}}, \bibinfo
  {author} {\bibfnamefont {Elena}\ \bibnamefont {Magliaro}}, \ and\ \bibinfo
  {author} {\bibfnamefont {Claudio}\ \bibnamefont {Perini}},\ }\bibfield
  {title} {\enquote {\bibinfo {title} {{Spinfoam fermions}},}\ }\href {\doibase
  10.1088/0264-9381/30/23/235023} {\bibfield  {journal} {\bibinfo  {journal}
  {Class.\ Quant.\ Grav.}\ }\textbf {\bibinfo {volume} {30}},\ \bibinfo {pages}
  {235023} (\bibinfo {year} {2013})},\ \Eprint {http://arxiv.org/abs/1012.4719}
  {arXiv:1012.4719 [gr-qc]} \BibitemShut {NoStop}%
\bibitem [{\citenamefont {Donà}\ \emph {et~al.}(2014)\citenamefont {Donà},
  \citenamefont {Eichhorn},\ and\ \citenamefont {Percacci}}]{Dona:2013qba}%
  \BibitemOpen
  \bibfield  {author} {\bibinfo {author} {\bibfnamefont {Pietro}\ \bibnamefont
  {Donà}}, \bibinfo {author} {\bibfnamefont {Astrid}\ \bibnamefont
  {Eichhorn}}, \ and\ \bibinfo {author} {\bibfnamefont {Roberto}\ \bibnamefont
  {Percacci}},\ }\bibfield  {title} {\enquote {\bibinfo {title} {{Matter
  matters in asymptotically safe quantum gravity}},}\ }\href {\doibase
  10.1103/PhysRevD.89.084035} {\bibfield  {journal} {\bibinfo  {journal}
  {Phys.\ Rev.\ D}\ }\textbf {\bibinfo {volume} {89}},\ \bibinfo {pages}
  {084035} (\bibinfo {year} {2014})},\ \Eprint {http://arxiv.org/abs/1311.2898}
  {arXiv:1311.2898 [hep-th]} \BibitemShut {NoStop}%
\end{thebibliography}%

\end{document}